\documentclass[preprint,11pt,3p]{elsarticle}
\usepackage{amssymb}
\usepackage{lineno}
\usepackage{hyperref}
\hypersetup{
    colorlinks=true,
    linkcolor=blue,
    filecolor=magenta,      
    urlcolor=cyan,
}
\usepackage{setspace}
\usepackage{geometry}
\usepackage{lipsum}
\usepackage{multicol}
\usepackage{balance}       
\usepackage{graphicx}      
\usepackage{svg}
\usepackage{xcolor}
\usepackage{changepage}
\usepackage{caption}    
\usepackage{subfig} 
\usepackage[export]{adjustbox} 
\usepackage{multicol}
\usepackage{bm}
\usepackage{amsmath, array,tabstackengine}
\TABstackMath
\TABstackMathstyle{\displaystyle}
\usepackage{amssymb, mathtools}
\usepackage{pdfpages}
\usepackage{verbatim}  
\usepackage{indentfirst}
\usepackage{nicematrix}
\usepackage{caption}
\usepackage{minitoc}
\usepackage{multirow}
\usepackage{floatrow}
\usepackage{enumitem}
\usepackage[nameinlink]{cleveref}
\usepackage{titletoc}
\usepackage{float}
\hypersetup{urlcolor=blue, colorlinks=true, citecolor = blue}  
\hypersetup{pdfauthor={Name}}
\usepackage{cleveref}
\bibpunct{\textcolor{blue}{[}}{\textcolor{blue}{]}}{,}{a}{,}{,}
\usepackage{siunitx}
\usepackage{etoolbox}
\usepackage{bbold}
\usepackage{booktabs}
\newcommand{\revision}[1]{{\textcolor{black}{#1}}}
\usepackage{pdflscape}

\begin{document}
\begin{frontmatter}
\title{Hydrodynamic Behavior of Non-spherical Particles in Confined Vertical Flows: A Resolved CFD-DEM Study}

\author[inst1]{Amiya Prakash Das}
\ead{apdas@connect.ust.hk}

\affiliation[inst1]{organization={Mechanics of Materials Lab, Department of Mechanical Engineering},
            addressline={Indian Institute of Technology Madras}, 
            city={Chennai},
            postcode={600036}, 
            state={Tamil Nadu},
            country={India}
            }
\affiliation[inst2]{organization={Deep Sea Mining},
            addressline={National Institute of Ocean Technology}, 
            city={Chennai},
            postcode={600100}, 
            state={Tamil Nadu},
            country={India}
            }
\author[inst1]{Shakti Swaroop Choudhury}
\author[inst1]{Sujith Reddy Jaggannagari}
\author[inst2]{Amudha Krishnan}
\author[inst2]{Gopkumar Kuttikrishnan}
\author[inst2]{Balaji Ramakrishnan}
\author[inst1]{Ratna Kumar Annabattula\corref{cor}}
\ead{ratna@iitm.ac.in}
\cortext[cor]{Corresponding author}

\begin{abstract}
We investigate the sedimentation and vertical hydraulic transport of non-spherical polymetallic nodules (PMNs) using resolved computational fluid dynamics-discrete element method (CFD-DEM) with multisphere particles spanning \revision{\(98 < Re_\text{p} < 2904\). Shape effects induce 1.8-2.0 times} drag enhancement relative to volume-equivalent spheres, arising from 50\% larger frontal areas and wake asymmetry, reducing terminal velocities by \revision{27-29\%.} Vertical transport exhibits velocity-driven transitions from intermittent settling to stable convection, as demonstrated by residence-time and drag-force statistics. While PMNs exhibit enhanced rotational-translational coupling and broader force fluctuations, the regime progression qualitatively resembles that of volume-equivalent spherical particles. Drag variance evolution reveals contrasting behavior: small particles \((d/D=0.082)\) show narrow distributions and wake suppression at higher velocities, while large particles \((d/D=0.22)\) exhibit non-monotonic variance. These findings elucidate shape-confinement interactions in vertical transport and establish bounds on the applicability of volume-equivalent spherical particles in reduced-order models.
\end{abstract}


\end{frontmatter}
\section{Introduction}\label{Sec:Introduction}
\noindent The recovery of polymetallic nodules (PMNs) from abyssal plains relies on hydraulic systems to transport coarse non-spherical particles through vertical risers extending several kilometers~\citep{hein2020deep,leng2021brief}. The efficiency of these systems depends on the ability to predict particle entrainment, suspension stability, and transport behavior under strong gravitational forcing and geometric confinement~\citep{van2016stability,dai2024cfd,chen2025deep,yang2025numerical}. In such risers, particles are often comparable in size to the riser diameter and interact strongly with both the carrier fluid and the confining walls, leading to complex multiphase flow behavior~\citep{shen2022resolved,schnorr2022resolved,sun2025transition}.

\revision{Hydraulic collection systems are favored in deep-sea mining for their operational efficiency and mechanical simplicity~\citep{li2024cfd,zhang2024one}, yet their design rests on a multiphase flow problem whose governing physics, i.e., the transport of coarse, irregular, and strongly confined particles in dense slurries, remains incomplete~\citep{zhou2010discrete,elskamp2017strategy,fan2024lattice,chen2025particle,huang2025hydraulic}. Two features distinguish PMN transport from canonical particle-laden flow: irregular particle morphology and strong geometric confinement. Unlike spheres, PMNs exhibit enhanced drag, shape-dependent lift, coupled rotational-translational motion, and altered settling dynamics~\citep{haider1989drag,ganser1993rational,holzer2008new}, all of which directly affect transport efficiency, minimum suspension velocity, and energy consumption. Although shape effects have been characterized extensively in unconfined and weakly confined flows~\citep{diamant2009hydrodynamic,suresh2011effects,bagheri2016drag,cheng2023numerical}, their behavior in narrow risers with confinement ratios \(d/D \ge 0.2\), representative of deep-sea mining systems, remains underexplored. This is a consequential gap: at such confinement ratios, it alters wake development and amplifies particle-wall momentum exchange, mechanisms absent or weak in unconfined studies.}

\revision{Most numerical studies of hydraulic transport employ unresolved computational fluid dynamics-discrete element method (CFD-DEM), in which particle-fluid interactions are modeled using empirical drag correlations~\citep{di1994voidage,zhao2013coupled}. Such correlations are calibrated against spherical particles or parameterized by scalar descriptors such as sphericity~\citep{holzer2008new,malone2008particle,nan2022cfd,cheng2023numerical}. This averaging is reasonable in dilute and unconfined flows where particles tumble freely, but it fails in confined flows: (1) it cannot represent orientation-dependent drag when wall proximity sustains preferred particle orientations, (2) it ignores the asymmetric, wall-deflected wakes responsible for shape-induced lift, and (3) it decouples translation from rotation, suppressing the very mechanism through which non-spherical particles redistribute momentum to the carrier fluid. The limitations are relevant in regimes where particle-fluid and particle-wall interactions dominate and may trigger flow instabilities, intermittent settling, and jamming~\citep{sommerfeld1992modelling,aponte2016simulation,chen2020prediction}.}

\revision{These closure problems are addressed in resolved CFD-DEM, which compute particle-fluid interaction forces directly at the interface, without empirical drag laws~\citep{luo2007modified,lu2018direct,yan2025complex}. Integrating particle representations such as signed distance functions or multisphere approximations extends the approach's ability to model particles with arbitrary shapes~\citep{shen2022resolved,lai2023signed}. Resolved CFD-DEM is well suited for modeling confinement-driven phenomena, including shear jamming, wake-mediated interactions, and regime transitions in narrow geometries~\citep{cunez2020crystallization,schnorr2022resolved}, with broader applicability to engineering-relevant solid-fluid problems~\citep{nan2023high,hu2024resolved}. However, existing studies have largely focused on sedimentation, fluidized beds, or short-bend geometries; sustained vertical hydraulic transport of X-ray computed tomography (CT) derived non-spherical particles at high confinement ratios has not been systematically examined. This gap is the practical motivation for the present study.} 

\revision{Resolved CFD-DEM at industrial-scale riser dimensions remains computationally prohibitive: representing realistic PMN morphology for \(\mathcal{O}(10^3-10^4)\) particles over kilometer-scale domains via SDF or multisphere approaches demands CPU runtime that preclude parametric design. Consequently, large-scale simulations routinely use volume-equivalent spheres, trading morphological fidelity for tractable runtimes. Whether this substitution preserves ensemble-level transport behavior or biases predictions of entrainment thresholds and pressure drop is unknown. Addressing this requires a controlled configuration that isolates shape-induced hydrodynamic effects from confounding collective phenomena such as dense-phase clustering and polydispersity, distinctions that are obscured in fully representative industrial simulations. We therefore adopt a controlled vertical transport configuration with moderate particle ensembles, enabling a direct comparison of CT reconstructed PMNs and volume-equivalent spheres under identical flow and confinement conditions.}

In this work, we employ resolved CFD-DEM to investigate the hydrodynamic behavior of non-spherical PMNs in vertical pipes representative of deep-sea mining risers. The model couples the incompressible Navier-Stokes equations with Newton's equations of motion for discrete particles, whose irregular geometries are represented using multisphere approximations derived from CT scans. We systematically analyze the influence of particle-to-pipe diameter ratio, particle Reynolds number, and flow velocity on particle trajectories, residence time distributions, and drag force statistics. Residence time analysis characterizes the transition from settling-dominated to convection-dominated transport, enabling prediction of minimum suspension velocities~\citep{chen2019cfd,lan2020long}, while drag force statistics reveal underlying force balance mechanisms. The objectives are twofold: first, to elucidate mechanisms by which particle shape influences drag, wake dynamics, and rotational-translational coupling in confined vertical flows; and second, to assess whether ensemble-averaged transport metrics for non-spherical PMNs qualitatively converge toward those of volume-equivalent spheres, thereby establishing when spherical models are suitable for reduced-order simulations of large-scale hydraulic transport systems.

The remainder of this paper is structured as follows.~\Cref{Sec:Methodology} presents the mathematical formulation of the resolved CFD-DEM framework, detailing the Navier-Stokes equations for the fluid phase, Newton's equations for particle dynamics, the immersed boundary method for fluid-solid coupling, and the multisphere representation for complex geometries.~\Cref{sec:numerical_verification} presents numerical verification through comparisons with benchmark experiments for spherical and non-spherical particle sedimentation, establishing the model's accuracy across the Stokes-to-intermediate Reynolds number regime. In addition, we have used the classical DKT problem to resolve multiple particle interactions using CFD-DEM framework.~\Cref{sec:results_discussion} presents results in two parts: sedimentation analysis, which quantifies the effects of shape on drag coefficients and terminal velocities; and vertical transport simulations that analyze entrainment dynamics, residence time distributions, and drag force statistics as functions of flow velocity and confinement ratio.~\Cref{Sec:Summary} summarizes key findings and discusses implications for hydraulic transport in deep-sea mining applications.

\section{Methodology}\label{Sec:Methodology}
\noindent We employ a resolved CFD-DEM framework to investigate the hydraulic transport of coarse, non-spherical PMNs in a vertical cylindrical pipe. The framework couples the Immersed Boundary (IB) method with CFD-DEM, integrating OpenFOAM for fluid dynamics simulation and LIGGGHTS for discrete particle tracking~\citep{klossnew}. This fully resolved, bidirectional coupling framework captures the complex multiphase interactions between fluid flow, particle transport, and sedimentation dynamics by resolving meso-scale physics at the fluid-solid interface through a Lagrangian-Eulerian formulation.

\subsection{Discrete Element Method for Particle Dynamics}
\noindent Individual particles are modeled within a Lagrangian framework using DEM, where Newton's second law governs each particle's motion for both translational and rotational dynamics
\begin{equation}
m_i \frac{d\mathbf{u}_i}{dt} = m_i \mathbf{g} + \sum_{j \neq i}^{N_c} \mathbf{F}_{\text{c},ij} + \sum_{k}^{N_w} \mathbf{F}_{\text{c},ik} + \mathbf{F}_{\text{f},i},
\label{eq:translation}
\end{equation}
\begin{equation}
\mathbb{I}_i \frac{d\mathbf{\omega}_i}{dt} = \sum_{j \neq i}^{N_c} \mathbf{T}_{\text{c},ij} + \sum_{k}^{N_w} \mathbf{T}_{\text{c},ik}
\label{eq:rotation},
\end{equation}
\noindent where \(m_i\) and \(\mathbb{I}_i\) represent the mass and moment of inertia tensor of particle \(i\), \(\mathbf{u}_i\) and \(\mathbf{\omega}_i\) denote linear and angular velocity vectors, \(\mathbf{g}\) is the gravitational acceleration vector, \(\mathbf{F}_{\text{c},ij}\) and \(\mathbf{T}_{\text{c},ij}\) are contact forces and torques between particles \(i\) and \(j\), \(\mathbf{F}_{\text{c},ik}\) and \(\mathbf{T}_{\text{c},ik}\) represent wall-particle interactions with wall \(k\), and \(\mathbf{F}_{\text{f},i}\) is the particle-fluid interaction force computed via the IB method.
The particle-particle and particle-wall contact forces are computed using the Hertz-Mindlin contact model with Coulomb friction, which provides an accurate representation of non-linear deformation and energy dissipation during collisions between non-spherical particles~\citep{mindlin1953elastic,hager2012parallel}. The contact detection algorithm efficiently handles complex particle geometries represented as multisphere assemblies.

\subsection{Computational Fluid Dynamics Formulation}
\noindent The continuous fluid phase is resolved using an Eulerian framework based on the incompressible Navier-Stokes equations within the fluid domain \(\Omega_\text{f}\), using~\Cref{eq:continuity,eq:momentum}
\begin{equation}
\mathbf{\nabla} \cdot \mathbf{u}_\text{f} = 0,
\label{eq:continuity}
\end{equation}
\begin{equation}
\frac{\partial \mathbf{u}_\text{f}}{\partial t} + (\mathbf{u}_\text{f} \cdot \mathbf{\nabla}) \mathbf{u}_\text{f} = -\frac{1}{\rho_\text{f}} \nabla p + \nu \nabla^2 \mathbf{u}_\text{f},
\label{eq:momentum}
\end{equation}
where \(\mathbf{u}_\text{f}\) represents the fluid velocity field, \(p\) is the pressure, \(\rho_\text{f}\) is the fluid density, and \(\nu = \mu_\text{f} / \rho_\text{f}\) is the kinematic viscosity with \(\mu_\text{f}\) being the dynamic viscosity.
The fluid velocity field is initialized as \(\mathbf{u}_\text{f}(\mathbf{x}, t=0) = \mathbf{u}_0(\mathbf{x})\) in \(\Omega_\text{f}\), with Dirichlet boundary conditions \(\mathbf{u}_\text{f} = \mathbf{u}_\Gamma\) imposed at domain boundaries \(\Gamma\). At the fluid-solid interface \(\Gamma_\text{s}\), the no-slip condition enforces velocity continuity
\begin{equation}
\mathbf{u}_\text{f} = \mathbf{u}_\text{s} \quad \text{on} \quad \Gamma_\text{s},
\label{eq:no_slip}
\end{equation}
where \(\mathbf{u}_\text{s}\) represents the local solid velocity. The fluid stress tensor is defined as \(\mathbb{\sigma} = -p \mathbb{1} + \mu_\text{f} (\nabla \mathbf{u}_\text{f} + (\nabla \mathbf{u}_\text{f})^T)\), and the surface traction at the interface is \(\mathbf{t} = \mathbb{\sigma} \cdot \mathbf{n}\) with \(\mathbf{n}\) being the outward unit normal vector. The governing equations are discretized using OpenFOAM's finite-volume method with second-order spatial accuracy, and temporal integration employs the Pressure-Implicit with Splitting of Operators (PISO) algorithm for robust pressure-velocity coupling~\citep{shen2022resolved}.

\subsection{Immersed Boundary Method for Fluid-Solid Coupling}
\noindent The immersed boundary method provides direct fluid-solid coupling by resolving the interface \(\Gamma_s\) without empirical drag correlations, offering significant advantages over the unresolved CFD-DEM framework in terms of accuracy and physical fidelity~\citep{schnorr2022resolved}. The no-slip condition (\Cref{eq:no_slip}) is enforced through a direct forcing approach that modifies the momentum equation. The total hydrodynamic force acting on particle \(i\) is computed by integrating the fluid stress over the particle surface
\begin{equation}
\mathbf{F}_{\text{f},i} = \int_{\Gamma_{\text{s},i}} \mathbb{\sigma} \cdot \mathbf{n} \, dA \approx \sum_{\text{c} \in V_{\Omega_{\text{s},i}}} \left( -\nabla p + \mu_\text{f} \nabla^2 \mathbf{u}_\text{f} \right)_\text{c} V_\text{c},
\label{eq:fluid_force}
\end{equation}

\noindent where \(V_{\Omega_{\text{s},i}}\) represents the set of fluid cells overlapping with solid domain \(\Omega_{\text{s},i}\), \(V_\text{c}\) is the volume of cell \(c\), and the summation extends over all fluid cells influenced by particle \(i\). This formulation captures pressure (buoyancy), viscous drag, and added mass effects without requiring empirical closure relations~\citep{schnorr2022resolved,lai2023signed}. Void fraction weighting ensures accurate force distribution across the fluid-solid interface, accounting for partial cell occupancy by solid particles.

To ensure accurate resolution of boundary layers around complex particle geometries, the computational grid satisfies the criterion \(\Delta x / D < 0.1\), where \(\Delta x\) is the characteristic grid spacing and \(D\) is the characteristic particle dimension~\citep{shen2022resolved}. This resolution requirement ensures that the viscous boundary layer and near-wall velocity gradients are properly captured for non-spherical particles across the range of \(Re_\text{p}\) investigated. The coupling between CFD and DEM solver occurs at a coupling interval of 10 timesteps, with fluid forces from~\Cref{eq:fluid_force} passed to the DEM solver. The updated particle positions and velocities are communicated back to the CFD domain. This bidirectional feedback maintains momentum conservation across the fluid-solid interface while preserving numerical stability through appropriate time step restrictions based on the Courant-Friedrichs-Lewy (Courant number) condition and particle collision time scales.

\subsection{Particle Representation}
\begin{figure}[H]
    \centering
    \includegraphics[width = 0.6\textwidth]{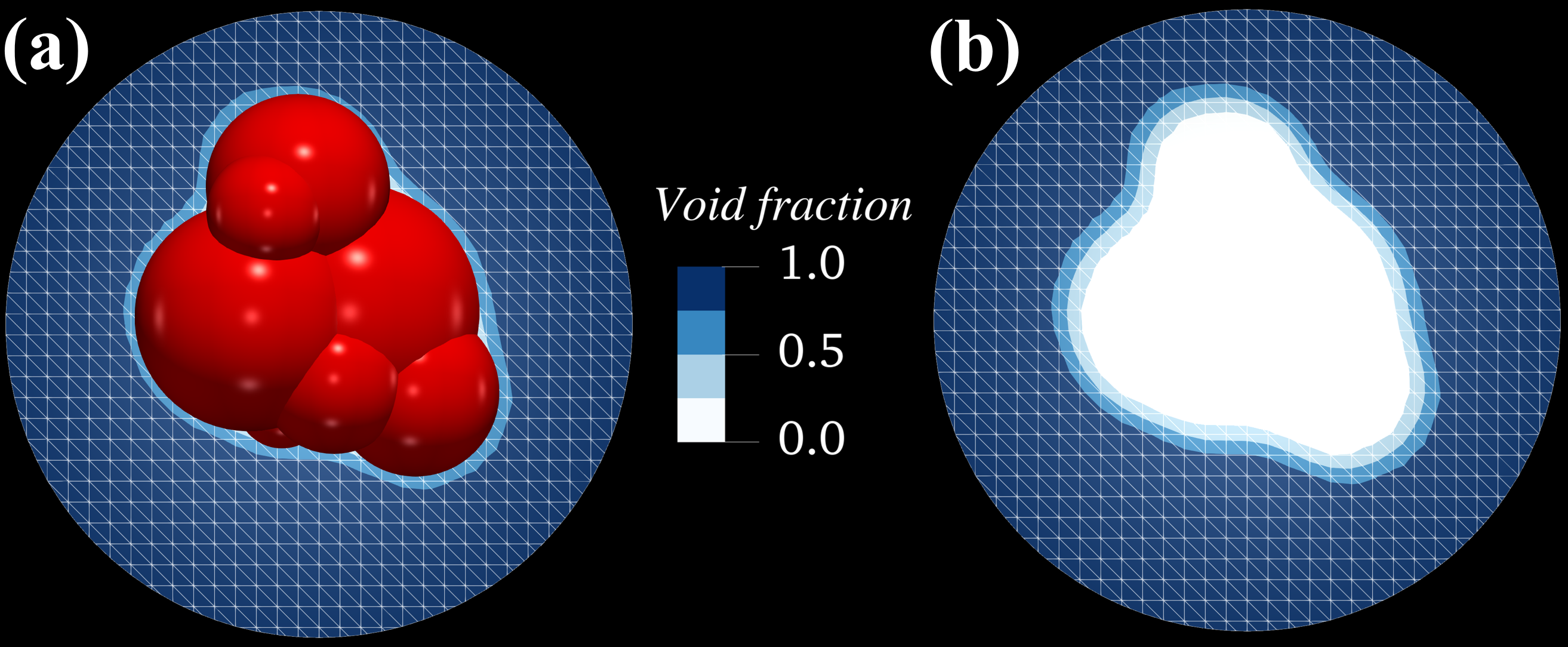}
    \caption{Multisphere representation of non-spherical particles in the resolved CFD-DEM framework. (a) Example particle geometry constructed from 14 overlapping spherical sub-particles, demonstrating the multisphere approach for capturing irregular surface features. (b) Corresponding void fraction field distribution in the computational domain, where white regions indicate solid particle volume and blue regions represent pure fluid. The void fraction field resolves the fluid-solid interface within the Eulerian CFD grid, enabling accurate computation of hydrodynamic forces via the IB method without requiring empirical drag correlations.}
    \label{fig:particle_representation}
\end{figure}

\noindent Complex particle geometries are represented using the multisphere approach, where non-spherical particles are constructed as rigid assemblies of overlapping spherical sub-particles, see~\Cref{fig:particle_representation}(a). This approach is computationally efficient and high-fidelity, capturing the essential geometric features that influence particle-fluid interactions and collision dynamics. The void fraction field visualization, see~\Cref{fig:particle_representation}(b), demonstrates the approach's capability to resolve complex particle-fluid interfaces within the Eulerian CFD grid.

\section{Numerical verification}\label{sec:numerical_verification}
\subsection{Single Sphere Settling}
\noindent The computational framework is verified against experimental data and numerical benchmarks through simulations of a single spherical particle settling in quiescent fluid. We compare our results with the experimental measurements of~\citet{ten2002particle} and numerical predictions from Lattice Boltzmann Method-DEM (LBM-DEM)~\citep{ten2002particle} and CFD-DEM simulations~\citep{lai2023signed} to assess the accuracy of terminal velocity predictions.

The numerical verification configuration follows the experimental setup of~\citet{ten2002particle}, where a submerged sphere with diameter \(d = \SI{15}{\milli\meter}\) and density \(\rho_\text{s}= \SI{1120}{\kilo\gram\per\meter\cubed}\) undergoes free sedimentation within a confined domain of dimensions \(100 \times 100 \times \SI{160}{\milli\meter\cubed}\). \revision{The particle is initially released from a height of \(\SI{120}{\milli\meter}\).} Two test cases with different fluid properties are investigated, as summarized in~\Cref{tab:table_2}, with particles' Reynolds number (\(Re_\text{p} = \rho_\text{f} u_\text{t} d/\mu_\text{f}\)) in the Stokes regime and intermediate regime where inertial effects become significant. \revision{The computational domain is discretized into 1,600,000 hexahedral cells, with a particle diameter-to-cell ratio of \(d/\Delta x = 15\), ensuring adequate boundary-layer resolution around the particle.} Dynamic mesh refinement is applied locally at the particle-fluid interface to maintain accuracy during particle motion. \revision{For this particular case, we are not interested in the collision dynamics, but rather in settling behavior. The time integration employs \(\Delta t_{\text{DEM}} = 1 \times 10^{-5}\ \SI{}{\second}\) for particle dynamics and \(\Delta t_{\text{CFD}} = 1 \times 10^{-4}\ \SI{}{\second}\) for fluid flow, with the Courant number maintained below 0.1 for numerical stability~\citep{schnorr2022resolved}.  Convergence is monitored through pressure and velocity residuals, with a tolerance of \(10^{-6}\).}

\begin{figure}[H]
    \centering
    \includegraphics[width = 0.55\textwidth]{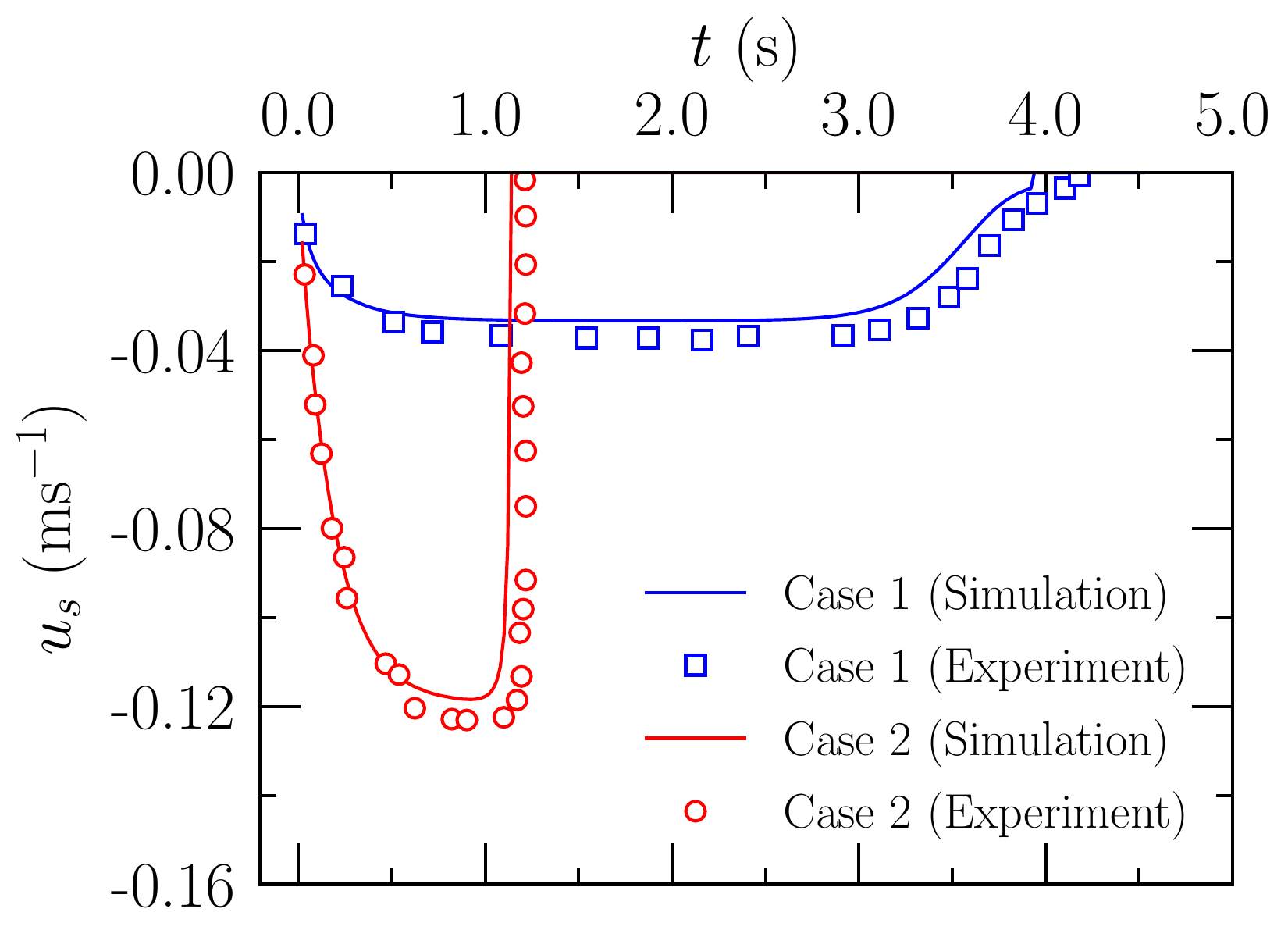}
    \caption{Temporal evolution of settling velocity for verification cases comparing CFD-DEM predictions (solid lines) with experimental measurements from~\citet{ten2002particle} (symbols). The simulations capture both the acceleration phase and the approach to terminal velocity, with an accuracy of 10\%, verifying the IB method for hydrodynamic force calculation across the transitional Reynolds number regime.}
    \label{fig:fig1}
\end{figure}

\revision{\Cref{fig:fig1} presents the temporal evolution of particle velocity for the two verification cases, where solid lines represent the CFD-DEM predictions and markers denote experimental measurements from~\citet{ten2002particle}. The two cases probe distinct
hydrodynamic regimes: Case 1 at \(Re_\text{p} = 1.4\) corresponds to viscous-dominated Stokes flow, while Case 2 at \(Re_\text{p} = 29.8\) corresponds to the intermediate regime where inertial effects are not negligible, and a wake forms behind the particle. The simulated trajectories capture the full settling history in both cases, including the initial acceleration phase, the approach to terminal velocity, and the deceleration upon bottom approach. The acceleration time scale differ, reflecting the different balance between viscous drag and particle inertia: Case 1 reaches terminal velocity within \(\sim \SI{0.8}{\second}\), while Case 2 requires \(\sim \SI{1.2}{\second}\), consistent with the higher \(Re_\text{p}\) and lower fluid viscosity. The CFD-DEM predictions accurately reproduce both time scales. The terminal velocities (\(u_\text{t} = \SI{0.036}{\meter\per\second}\) for Case 1 and \(\SI{0.12}{\meter\per\second}\) for Case 2) are recovered to within \(6\%\) of the experimental values, with the normalized maximum velocity \(u_\text{max}/u_\text{t}\) likewise captured within \(6\%\) (\Cref{tab:table_2}). The deceleration phase as the particle approaches the bottom wall is well reproduced, indicating that near-wall hydrodynamic interactions are correctly resolved by the IB method, even without empirical wall-correction terms.}

\begin{table}[H]
\centering
\caption{Verification metrics comparing CFD-DEM predictions with experimental data from~\citet{ten2002particle}. Terminal velocity \(u_\text{t}\) is the experimental terminal velocity; \(u_{\max}\)/\(u_\text{t}\) is reported from the present simulations; \(Re_\text{p}\) is computed from the simulated \(u_\text{t}\).}
\renewcommand{\arraystretch}{1.5}
\begin{tabular}{c c c c c c}
\hline \hline
\textbf{Case} & \textbf{\(\rho_\text{f} (\SI{}{\kilo\gram\per\meter\cubed})\)} & \textbf{\(\mu_\text{f} (\SI{}{\pascal\cdot\second})\)} & \textbf{\(u_t (\SI{}{\meter\per\second} )\)}\footnote{Experimental terminal velocity from~\citet{ten2002particle}} & \textbf{\(u_{\max}/u_t\)} & \textbf{\(Re_\text{p}\)} \\
\hline 
Case 1 & 970  & 0.373  & 0.038  & 0.95  & 1.4 \\
Case 2 & 960  & 0.058  & 0.128  & 0.94  & 29.8 \\
\hline \hline
\end{tabular}
\label{tab:table_2}
\end{table}

\revision{The flow field snapshots (\Cref{fig:figS2,fig:figS3} in Supplementary material (SM)) demonstrate the transition from viscous-dominated to inertia-dominated settling as \(Re_\text{p}\) increases. Case 1 exhibits the symmetric streamline pattern characteristic of Stokes flow, with the disturbance decaying smoothly into the far field and no wake formation,
consistent with the analytical Stokes solution for creeping flow past a sphere. Case 2 instead shows flow separation behind the particle and a recirculating wake region extending approximately one particle diameter downstream, with a corresponding pressure deficit that contributes to the drag component absent in Case 1. The transition from viscous-dominated to inertia-dominated settling is therefore evident in both the velocity history and the flow-fields, and the agreement with the experimental benchmarks of~\citet{ten2002particle} verifies the numerical method across the range \(1 < Re_\text{p} < 50\) (see movie M1 in the SM).}
\subsection{Drafting, Kissing and Tumbling}
\begin{figure}[ht]
    \centering
    \includegraphics[width = 0.65\textwidth]{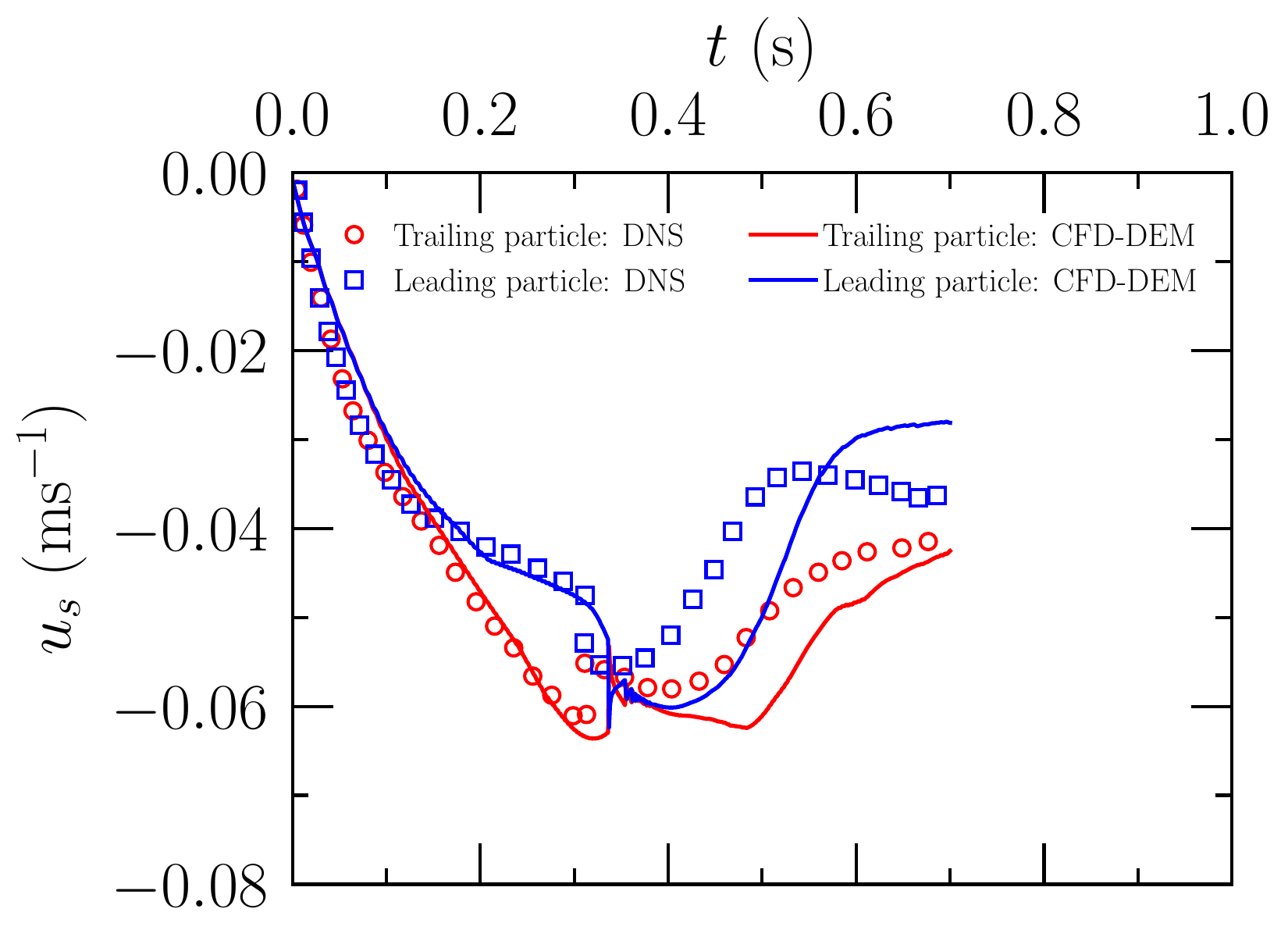}
    \caption{\revision{Settling velocity evolution for the two-particle drafting–kissing–tumbling benchmark. CFD–DEM results (solid) agree well with DNS from~\citet{sharma2005fast} (symbols). The three phases are captured: drafting \((\SI{0.14}{\second} < t < \SI{0.35}{\second})\), kissing \((t \approx \SI{0.35}{\second})\), and tumbling \((t > \SI{0.35}{\second})\). Velocity deviations remain below 5\% during drafting and kissing, confirming accurate multi-particle hydrodynamic interactions.}}
    \label{fig:N2}
\end{figure}

\noindent The CFD-DEM capacity to resolve hydrodynamic interactions between multiple particles is verified through simulations of two settling spheres in quiescent fluid, following the benchmark configuration of \citet{glowinski2001fictitious}. The computational domain is a vertical rectangular box (\SI{10}{\milli\meter} \(\times\) \SI{10}{\milli\meter} \(\times\) \SI{40}{\milli\meter}). Two identical spheres are released vertically, initially separated, and their settling velocities are tracked as they approach, collide, and tumble. \Cref{fig:N2} compares the settling velocity evolution of the two spherical particles with the direct numerical simulations (DNS) from~\citet{sharma2005fast}, revealing three characteristic phases of particle-particle interactions:
\begin{enumerate}[leftmargin=0.5cm,labelsep = 0.2em]
    \item \textbf{Initial Settling Regime (\(t < \SI{0.14}{\second}\))}: The particles settle at identical velocities as the trailing particle remains beyond the wake of the leading particle. Each particle experiences drag from the undisturbed far-field flow, producing independent settling dynamics with a velocity difference of \(<5\%\).
    \item \textbf{Drafting\textemdash Wake Entrainment Regime (\(\SI{0.14}{\second} < t < \SI{0.35}{\second}\))}: The trailing particle enters the leading particle's low-pressure wake, where fluid velocity is elevated relative to the far-field. This reduces the drag force on the trailing particle, causing it to accelerate and rapidly close the gap.
    \item \textbf{Kissing\textemdash Contact and Tumbling Dynamics \((t\ge\SI{0.35}{\second})\)}: At \(t = \SI{0.35}{\second}\), the particles come in contact, known as the kissing phase. The Hertz-Mindlin contact model calculates normal and tangential contact forces, leading to particle tumbling and momentum exchange.
\end{enumerate}

Quantitative comparison from the CFD-DEM shows good agreement with DNS benchmarks: maximum velocity deviation \(<5\%\) throughout all phases, accurate prediction of collision time (\(\SI{0.35}{\second}\), within 5\% of DNS), and correct post-collision behavior. The verification confirms that the IB method accurately resolves: (1) long-range wake-mediated hydrodynamic interactions, (2) transition from independent to coupled settling, (3) contact collision dynamics, and (4) momentum transfer during tumbling. Flow field snapshots in~\Cref{fig:NS3} in SM show particle positions and velocity contours at representative times of (\(t = \SI{0.0}{\second}\), \SI{0.14}{\second}, \SI{0.35}{\second}, \SI{0.7}{\second}), illustrating the progressive wake entrainment mechanism and collision-induced flow reorganization (see movie M2 in the SM).

\subsection{Non-spherical Particle Settling}
\noindent The framework's suitability in handling complex particle morphologies is assessed through a sedimentation simulation of an idealized spherical particle, approximated using a multisphere approach. The test particle comprises 206 overlapping sub-spheres arranged to approximate a spherical geometry while maintaining identical density (\(\rho_\text{s} = \SI{1120}{\kilo\gram\per\cubic\meter}\)) and volume, as the reference sphere from Case 2 (\(d = \SI{15}{\milli\meter}\)). The configuration enables a direct comparison of the hydrodynamic behavior of a spherical particle with that of a non-spherical proxy under identical physical conditions. The computational setup maintains consistency with the verification case 2; fluid properties \(\rho_\text{f} = \SI{960}{\kilo\gram\per\cubic\meter}\) and \(\mu_\text{f} = \SI{0.058}{\pascal\cdot\second}\), rectangular domain dimensions \(100 \times 100 \times \SI{160}{\milli\meter\cubed}\), and particle is released from the same initial position. The mesh resolution is enhanced to \(d/\Delta x = 16\) to ensure adequate representation of the irregular particle surface. Time step remain unchanged at \(\Delta t_\text{DEM} = \SI{1e-5}{\second}\) and \(\Delta t_\text{CFD} = \SI{1e-4}{\second}\) to maintain temporal accuracy and numerical stability.

\begin{figure}[H]
    \centering
    \includegraphics[width = 0.65\textwidth]{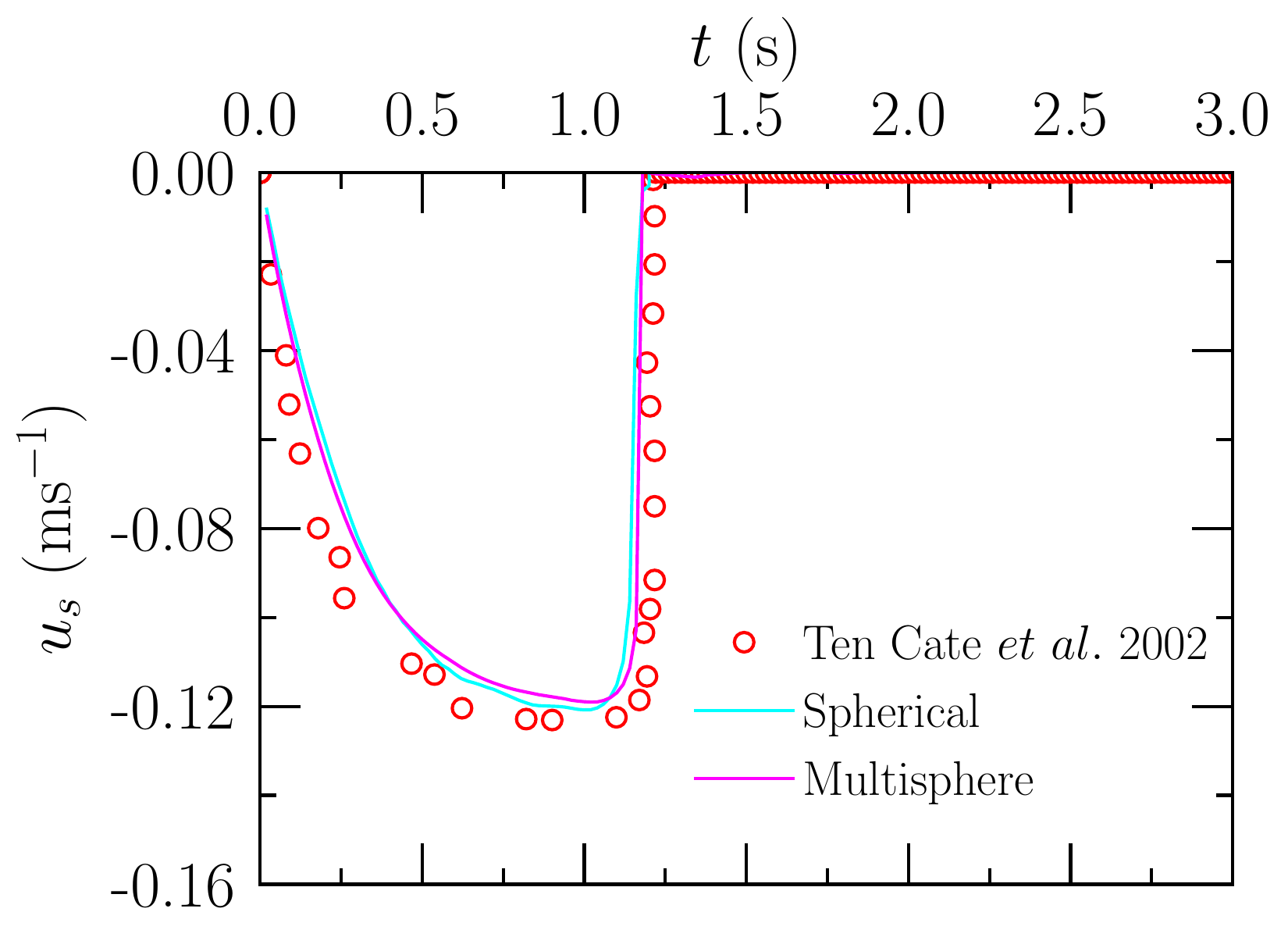}
    \caption{\revision{Settling velocity comparison between spherical and multisphere particles verifying the multisphere approximation. A 206 sub-sphere assembly approximates a sphere with identical density \((\rho_\text{s} = \SI{1120}{\kilo\gram\per\cubic\meter})\), volume, and mass (Case 2), settling in silicon oil \((\rho_\text{f} = \SI{960}{\kilo\gram\per\cubic\meter},\ \mu_\text{f} = \SI{0.058}{\pascal\cdot\second})\). Experimental data from \citet{ten2002particle} (red open symbols) and numerical results for the sphere (cyan solid line) and multisphere (magenta solid line) agree closely, with terminal velocity differences within 6\%.}}
    \label{fig:fig5}
\end{figure}

\Cref{fig:fig5} compares the settling velocity evolution for spherical and multisphere particles. The multisphere particle exhibits terminal velocity \(u_\text{t} = \SI{0.12}{\meter\per\second}\), representing \(\le 6\%\) decrease relative to the sphere (\(u_\text{t} = \SI{0.128}{\meter\per\second}\)). Despite the irregular surface geometry, particles exhibit quantitatively similar acceleration dynamics, with a smooth, monotonic approach to terminal velocity over \(t \approx \SI{0.8}{\second}\). The subtle velocity decrease for the multisphere particle reflects enhanced drag due to its increased surface area, which is greater than that of the equivalent sphere. The flow field around the multisphere particle is qualitatively similar to that of the smooth sphere at the same \(Re_\text{p}\), exhibiting flow separation and a recirculating wake, characteristic of inertia-dominated flow (\Cref{fig:figS4} in SM). The irregular surface introduces additional finer-scale distortions in the near-wall velocity contours, particularly downstream of surface protrusions where local velocity gradients intensify. These features reflect the particle's geometry but do not significantly alter the global force balance or settling dynamics in this intermediate Reynolds number regime.

The close agreement between spherical and multisphere terminal velocities (within 6\%) validates two critical aspects of the computational framework: (1) the multisphere approximation accurately represents the hydrodynamic resistance of irregular particles while maintaining computational efficiency, and (2) the IB method accurately resolves fluid-solid interactions around complex geometries with accuracy comparable to smooth surfaces. \revision{This verification establishes confidence in the numerical framework that distinguishes the non-spherical simulations from the spherical baseline: the multisphere representation and the IB force integration over complex geometries. Because the resolved CFD-DEM framework computes hydrodynamic forces by directly integrating the fluid stress tensor over the particle volume, the accuracy of this integration\textemdash rather than agreement with empirical drag correlations\textemdash constitutes the appropriate verification target.}

\revision{\textbf{Remark}: We note that established non-spherical drag correlations (e.g.,~\citet{haider1989drag,holzer2008new}) are calibrated for geometrically regular shapes parameterized by well-defined sphericity values, and their applicability to the highly irregular, non-convex PMN morphologies considered here is itself uncertain. The present verification strategy, therefore, focuses on validating the numerical framework at the level of the governing equations, providing a robust basis for PMN transport investigations.}

\section{Results and Discussion}\label{sec:results_discussion}
\noindent We systematically investigate the hydrodynamic behavior of non-spherical PMNs in confined vertical flows, which is directly relevant to hydraulic lifting systems used in deep-sea mining risers. The analysis is structured in two parts. First, sedimentation in quiescent fluid is examined to quantify shape-induced drag enhancement and establish reference settling velocities for irregular PMNs relative to volume-equivalent spheres. These results provide calibration data for drag models commonly used in riser-scale simulations. Second, vertical hydraulic transport is analyzed using residence-time statistics and drag-force distributions to identify entrainment behavior and transport stability as flow velocity increases.

\revision{All transport simulations are performed with 40 particles to provide statistically meaningful ensemble measures while preserving fully resolved particle-fluid interactions. This particle number captures particle-to-particle variability without introducing collective effects such as sustained clustering or plug formation, which are beyond the scope of the present study. Statistical convergence of the drag-force distribution \(P(\hat{f})\) is verified by comparing ensemble sizes of \(N_\text{p} = 20\), \(40\), and \(80\) for the large sphere case at \(u_\text{f} = 3.0u_\text{t}\) (\Cref{fig:NS4} in the SM); all the distributions agree closely in both peak location and tail behavior, confirming that \(N_\text{p} = 40\) is sufficient for the first- and second-order statistics reported here. Continuous particle injection is not considered here; instead, the focus is on isolating the intrinsic hydrodynamic response of PMNs under controlled confinement. This approach enables clear physical interpretation and directly supports the reduced-order modeling strategies for riser-scale transport.}

\subsection{Settling Analysis}
\begin{figure}[H]
    \centering
    \includegraphics[width=0.6\linewidth]{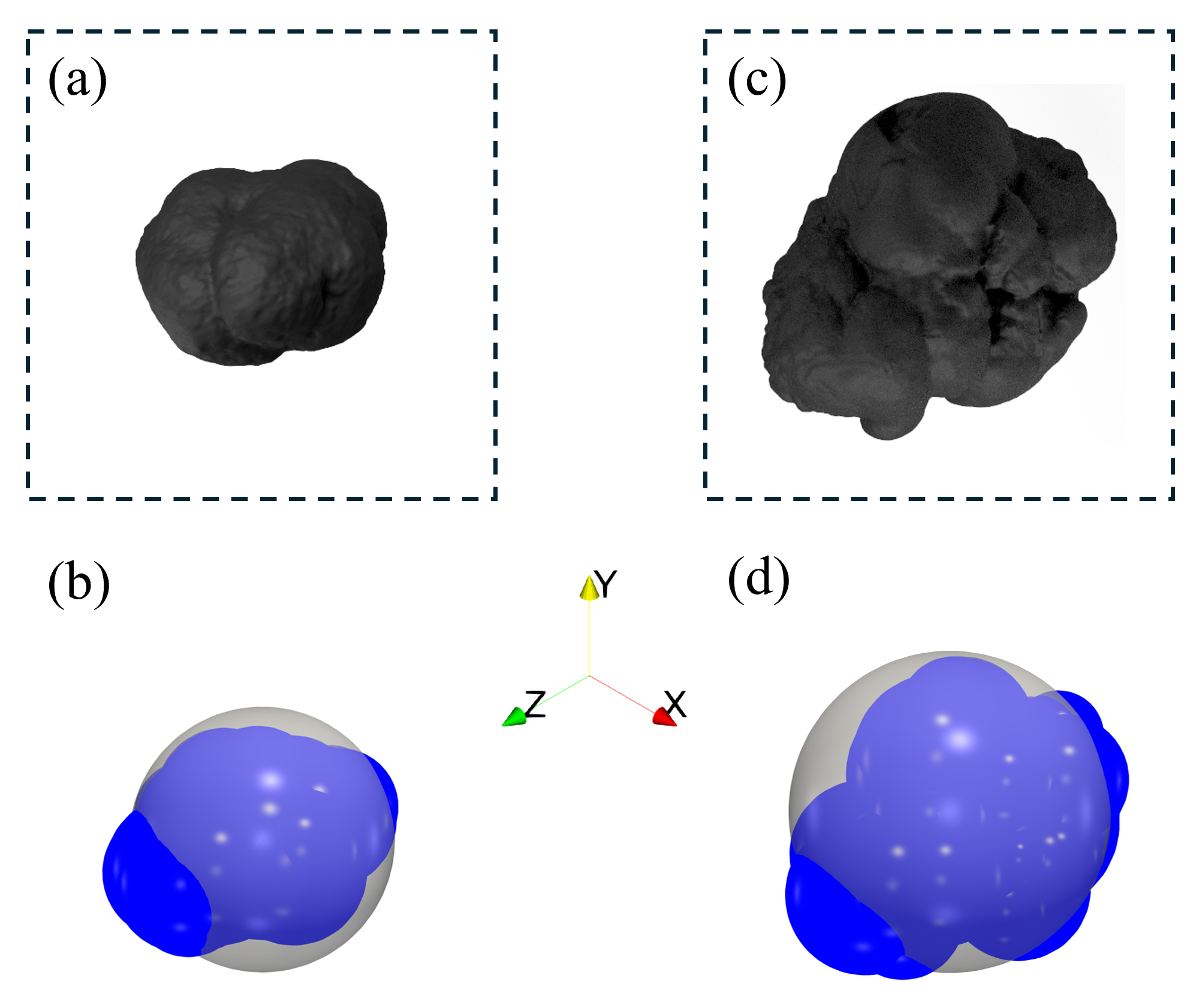}
    \caption{PMN morphology and multisphere approximations. (a) \& (c) CT-reconstructed surfaces of PMNs showing irregular geometry characteristic of natural PMNs. (b) \& (d) Multisphere representations (blue) comprising 28 and 45 sub-spheres for effective diameters \(d_\text{eff} = \SI{20}{\milli\meter}\) and \SI{54}{\milli\meter}, respectively, superimposed on volume-equivalent spheres (grey, \(d_\text{v} = \SI{16.4}{\milli\meter}\) and \(\SI{44}{\milli\meter}\)). The volume ratio \((d_\text{v}/d_\text{eff})^3 \approx 0.55\) indicates that PMNs occupy only 55\% of their circumscribed sphere volume due to non-convex morphology.}
    \label{fig:fig6}
\end{figure}

\noindent \Cref{fig:fig6} illustrates the complex and irregular morphology of PMNs represented using a multisphere approximation based on high-resolution CT scans. This approach employs 28 and 45 sub-spheres for PMNs with effective diameters of \(d_\text{eff} = \SI{20}{\milli\meter}\) and \(d_\text{eff} = \SI{54}{\milli\meter}\), respectively, where \(d_\text{eff}\) represents the diameter of the minimum circumscribed sphere that fully encloses the PMN. The volume-equivalent sphere diameter \(d_\text{v} = (6V_\text{p}/\pi)^{1/3}\), where \(V_\text{p}\) is the PMN volume, yields \(d_\text{v} = \SI{16.4}{\milli\meter}\) and \(d_\text{v} = \SI{44}{\milli\meter}\) for the small and large PMN, respectively. The ratio \((d_\text{v}/d_\text{eff})^3 \approx 0.55\) indicates that PMNs occupy only 55\% of their circumscribed volume, reflecting their irregular, non-convex morphology with surface protrusions and concavities characteristic of natural PMNs. The multisphere representations in \Cref{fig:fig6}b and d show the PMN geometries superimposed with their volume-equivalent spheres, demonstrating that the irregular PMN surfaces extend significantly beyond the volume-equivalent sphere while remaining contained within the circumscribed sphere envelope. This geometric distinction is crucial for understanding hydrodynamic behavior, as the effective frontal area and surface morphology of PMNs differ substantially from volume-equivalent spheres.

\begin{figure}[H]
    \centering
    \includegraphics[width=0.55\linewidth]{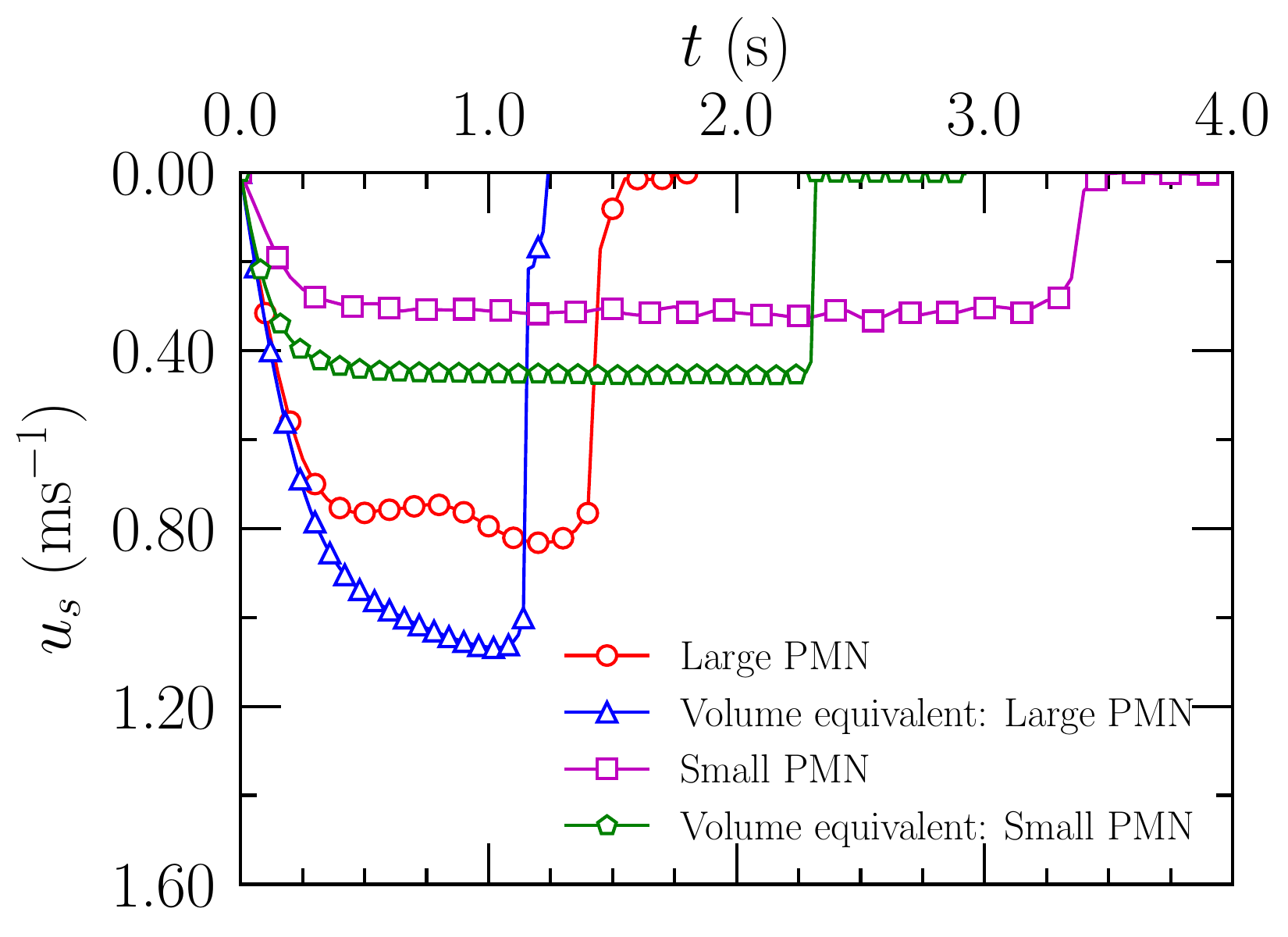}
    \caption{Temporal evolution of settling velocity for PMNs and volume-equivalent spherical particles. PMNs (squares and circles) reach terminal velocities 29\textendash 33\% lower than volume-equivalent spheres (triangles and pentagons) despite identical mass and volume, reflecting enhanced drag coefficients due to increased surface area.}
    \label{fig:fig7}
\end{figure}

\Cref{fig:fig7} compares the temporal evolution of settling velocities for PMNs and their volume-equivalent spherical particles in fluid. The numerical simulations are performed within a cylindrical computational domain with diameter \(D = \SI{200}{\milli\meter}\) and length \(L = \SI{1200}{\milli\meter}\). The mesh is constructed with resolution requirements based on the particle characteristic dimension: for PMNs, we use \(d_\text{eff}/\Delta x \geq 8\) to resolve surface irregularities and wake structures; for volume-equivalent spheres, we use \(d_\text{v}/\Delta x \geq 8\). This ensures at least 8 grid cells across the particle dimension in all cases. Both PMNs and spherical particles have density \(\rho_\text{p} = \SI{2000}{\kilo\gram\per\meter\cubed}\) and identical volumes. They are released from rest at the domain centerline at height \(z = \SI{1000}{\milli\meter}\) in quiescent fluid with density \(\rho_\text{f} = \SI{1000}{\kilo\gram\per\meter\cubed}\) and dynamic viscosity \(\mu_\text{f} = \SI{0.05}{\pascal\cdot\second}\). The terminal velocity \(u_\text{t}\) is determined from the steady-state settling velocity achieved after initial transients decay.~\Cref{tab:settling} lists the details of the settling characteristic for the PMNs and volume-equivalent spheres. \revision{The elastic properties prescribed for the simulations are \(\nu_\text{p} = 0.2\) and \(Y_\text{p} = \SI{1}{\giga\pascal}\). This yields a time step of \(\Delta t_\text{DEM} = \SI{5e-6}{\second}\) (\(\approx 10\%\) of the Rayleigh time step for smaller spheres sets the time step,~\cref{eq:time}) and \(\Delta t_\text{CFD} = \SI{5e-5}{\second}\) to maintain temporal accuracy and numerical stability.}
\revision{\begin{equation}
    t_\text{DEM} = \dfrac{0.5\pi d_\text{p}}{0.163\nu_\text{p} + 0.8766} \times \sqrt{\dfrac{2\rho_\text{p}(1+\nu_\text{p})}{Y_\text{p}}},
    \label{eq:time}
\end{equation}
where \(d_\text{p}\) and \(\rho_\text{p}\) are the particle diameter and density, respectively. \(\nu_\text{p}\) and \(Y_\text{p}\) are the elastic properties.} 

\begin{table}[h]
\centering
\caption{Comparison of settling characteristics for PMNs and volume-equivalent spheres}
\label{tab:settling}
\begin{tabular}{lccccc}
\hline \hline 
Particle & \(d_\text{v}\) (\SI{}{\milli\meter}) & \(d_\text{eff}\) (\SI{}{\milli\meter}) & \(u_t\) (\SI{}{\meter\per\second}) & \(Re_\text{p}\) & \(C_\text{D}\) \vspace{0.05em}\\
\hline
Small sphere & 16.4 & 16.4 & 0.45 & 148 & 1.01 \\
Small PMN & 16.4 & 20.0 & 0.30 & 98 & 2.28 \\
Large sphere & 44.0 & 44.0 & 1.05 & 924 & 0.44 \\
Large PMN & 44.0 & 54.0 & 0.75 & 660 & 0.86 \\
\hline \hline
\multicolumn{6}{l}{\small Note: \(Re_\text{p}\) based on \(d_\text{v}\); \(C_\text{D}\) based on \(A_\text{proj} = \pi d_\text{v}^2/4\)}
\end{tabular}
\end{table}

\begin{figure}[H]
    \centering
    \subfloat[]{\includegraphics[width=0.425\linewidth]{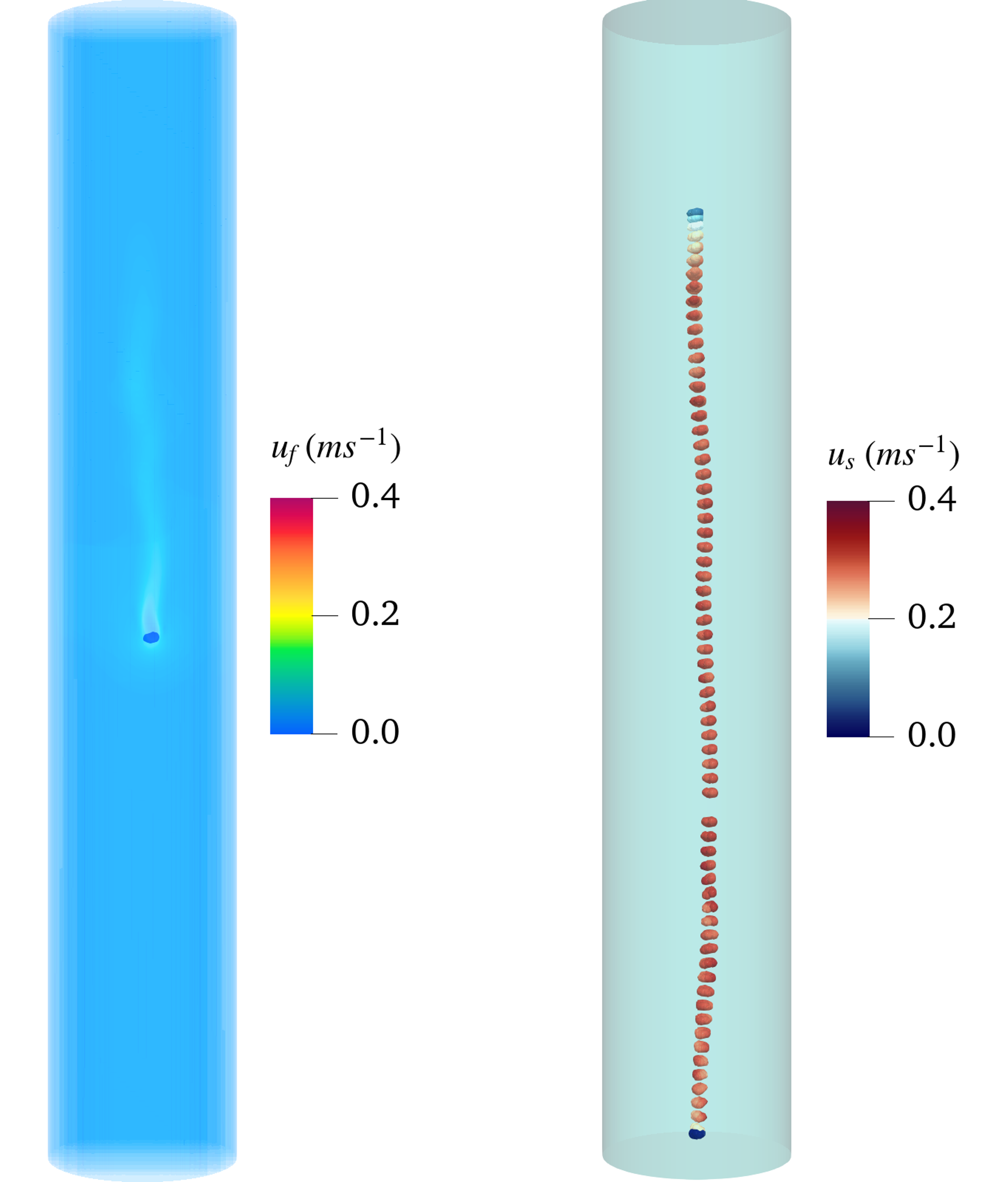}}\hspace{0.4em}
    \subfloat[]{\includegraphics[width=0.425\linewidth]{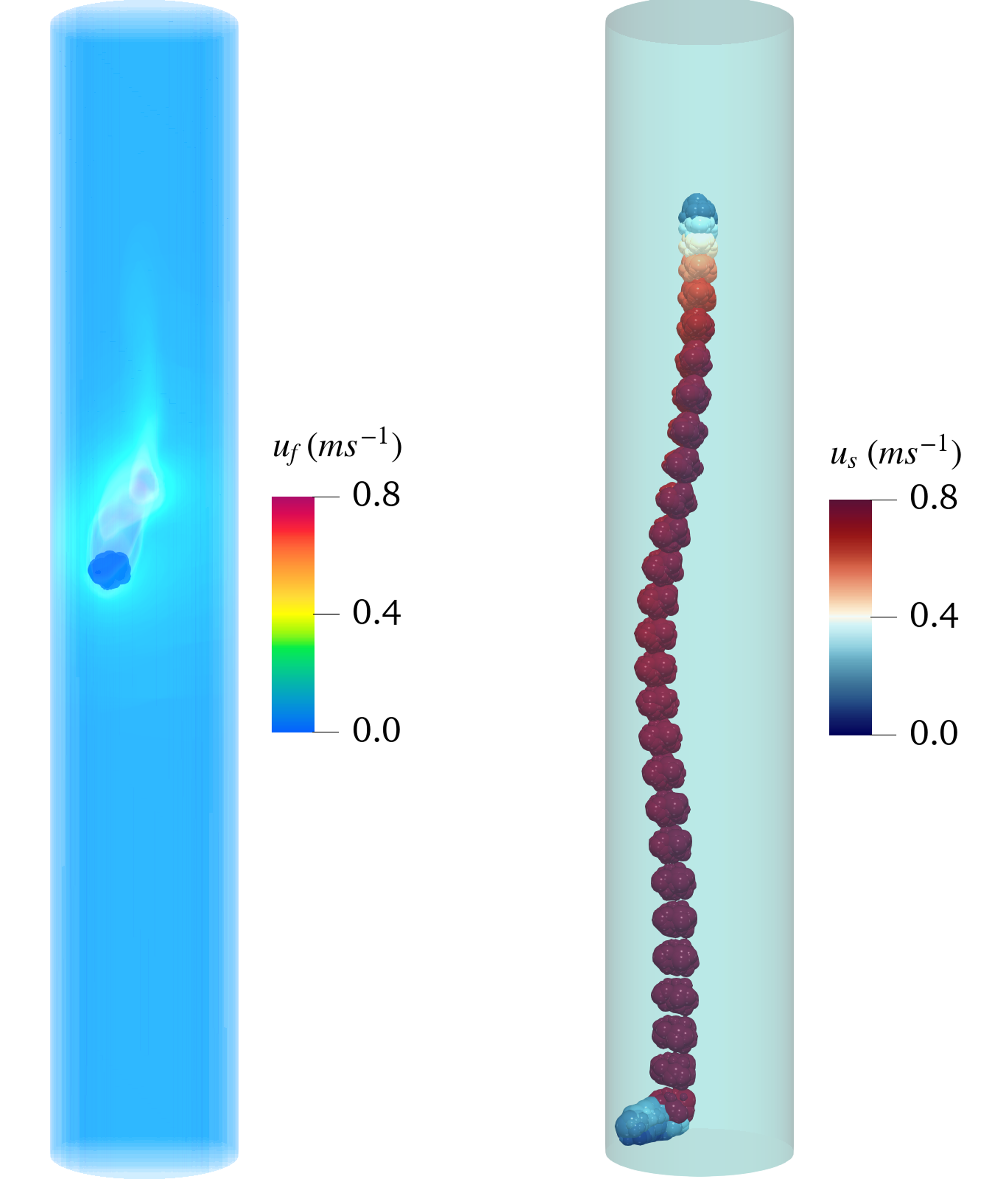}}
    \caption{Instantaneous flow fields and particle trajectories during sedimentation of PMNs in quiescent fluid. (a) Small PMN at terminal velocity. (b) Large PMN. The fluid velocity magnitude contours show asymmetric wake structures resulting from irregular particle geometry (left). Three-dimensional particle trajectories colored by solid velocity magnitude, demonstrating vertical settling with coupled rotational motion (right).}
    \label{fig:fig9}
\end{figure}

\Cref{fig:fig7} reveals that PMNs settle significantly slower than volume-equivalent spheres despite identical mass and buoyancy. The small PMN reaches terminal velocity \(u_\text{t}^\text{PMN} = \SI{0.30}{\meter\per\second}\) compared to \(u_\text{t}^\text{sphere} = \SI{0.42}{\meter\per\second}\) (29\% reduction), while the large PMN achieves \(u_\text{t}^\text{PMN} = \SI{0.8}{\meter\per\second}\) versus \(u_\text{t}^\text{sphere} = \SI{1.1}{\meter\per\second}\) (27\% reduction). These velocity differences are purely shape effects and correspond to \(Re_\text{p}\) of 98 and 138 (small PMN and sphere) and 704 and 968 (large PMN and sphere). The drag coefficient, defined from terminal force balance as \(C_\text{D} = 4(\rho_\text{p} - \rho_\text{f})gd_\text{v}/(3\rho_\text{f} u_\text{t}^2)\) using the projected area \(A_\text{proj} = \pi d_\text{v}^2/4\), yields \(C_\text{D}^\text{PMN} = 2.38\) and 0.9 compared to \(C_\text{D}^\text{sphere} = 1.21\) and 0.48 for small and large particles, respectively\textemdash representing 2.0 and 1.8 times enhancements~\citep{vergara2024drag}. \revision{This drag increase reflects two distinct contributions that are inherently coupled for freely settling non-spherical particles. The first is the larger instantaneous projected area\textemdash the PMNs rotate during settling\textemdash where the frontal area varies continuously with orientation. The maximum orientational projected area exceeds the volume-equivalent value \(\pi d_\text{v}^2/4\) by a factor of approximately \((d_\text{eff}/d_\text{v})^2 \approx 1.5\), accounting for a substantial portion of the observed \(C_\text{D}\) enhancement. The second contribution arises from morphology-induced effects that persist independently of projected-area differences: asymmetric wake structures generated by surface protrusions and concavities (see~\Cref{fig:fig9}), orientation-dependent pressure differences along the irregular surface, and rotational-translational coupling that continuously alters the angle of attack. Accounting for the projected-area difference through normalized drag coefficients \(C_\text{D}^\text{norm} = C_\text{D}^\text{PMN} (d_\text{v}/d_\text{eff})^2\) yields values still 20\textendash 30\% above spherical equivalents, confirming that these morphological effects contribute substantially beyond the geometric area increase alone.} 

\begin{figure}[H]
    \centering
    \includegraphics[width=0.55\linewidth]{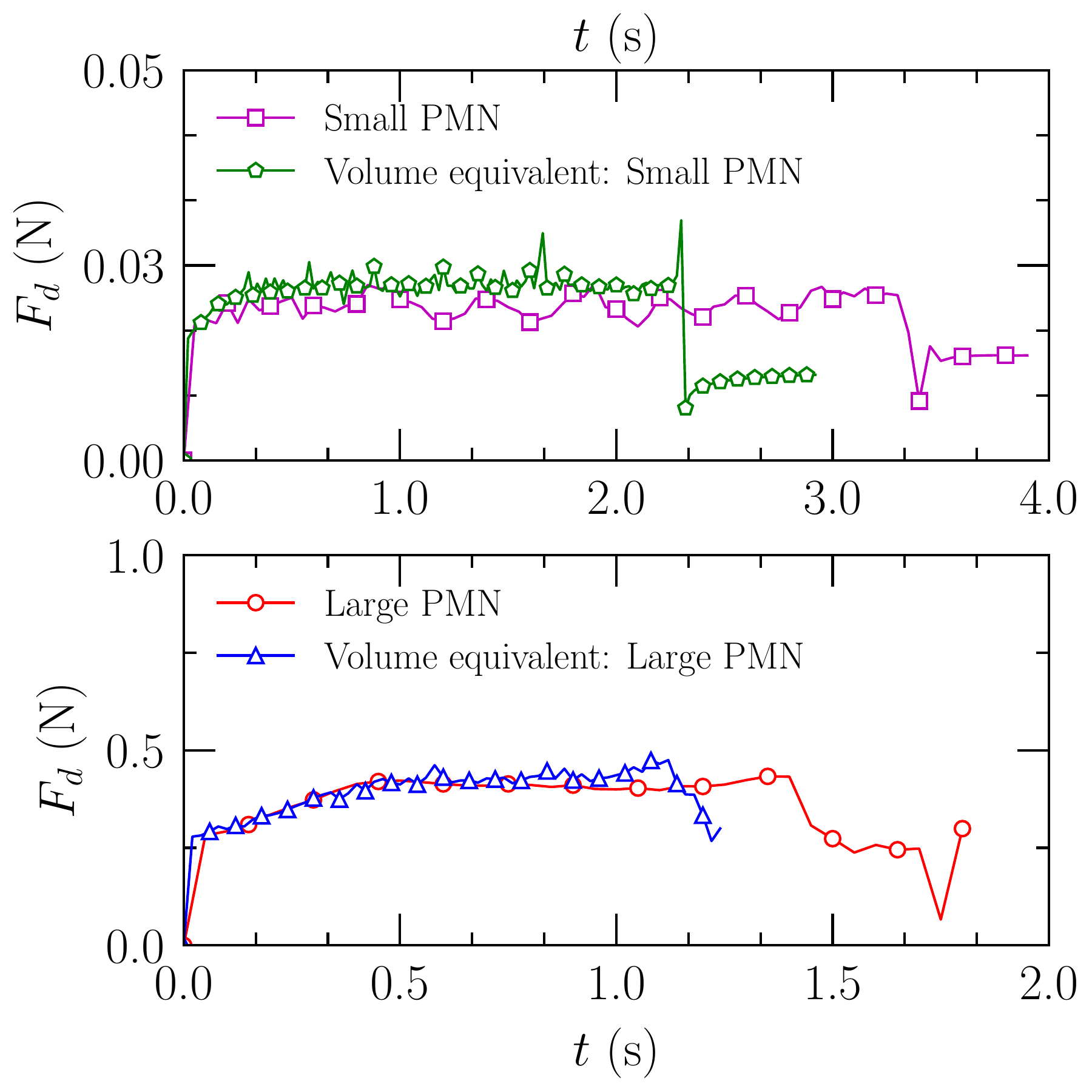}
    \caption{Temporal evolution of vertical drag force during sedimentation for PMNs and volume-equivalent spheres. Small particles reach steady-state drag force \(F_\text{d} \approx \SI{0.023}{\newton}\) (top). Large particles achieve \(F_\text{d} \approx \SI{0.43}{\newton}\) (bottom).}
    \label{fig:fig8}
\end{figure}

\revision{The use of the volume-equivalent projected area \(\pi d_\text{v}^2/4\) as the reference in the definition of \(C_\text{D}\) is a deliberate choice: it is orientation-invariant and enables direct comparison with volume-equivalent spheres, which is key to the settling analysis. The orientation-dependent contribution to drag is thus absorbed into the reported \(C_\text{D}^\text{PMN}\) values. For applications requiring orientation-resolved drag models, the instantaneous projected area must be tracked alongside particle angular dynamics, an approach that is beyond the scope of the present study but naturally accommodated within the resolved CFD-DEM framework.} \Cref{fig:fig8} shows the temporal evolution of vertical drag force during sedimentation, revealing distinct transient dynamics. The drag force initially increases during particle acceleration as the relative fluid-particle velocity and unsteady inertial effects develop. The drag forces reach steady-state values of \(F_\text{d} = (\rho_\text{p} - \rho_\text{f})V_\text{p}g \approx \SI{0.023}{\newton}\) (small) and \(\SI{0.43}{\newton}\) (large), confirming force balance to within 5\%. Notably, the PMNs and spheres experience identical terminal drag forces since they have equal volumes and densities\textemdash the 27-29\% lower settling velocities of PMNs arise because their enhanced drag coefficients allow them to generate the required hydrodynamic resistance at reduced slip velocities. The sharp decrease in \(F_\text{d}\) after \(t \approx 2.8\) s for the small PMN indicates bottom contact, where wall reaction forces partially support the particle weight.  

\subsection{Vertical transport}
\noindent To investigate PMN entrainment in vertical risers, we simulate the transport of mono-sized spherical particles and PMNs (\(d_\text{v} = \SI{16.4}{\milli\meter}\) and \SI{44}{\milli\meter}) through a vertical cylindrical pipe with diameter \(D = \SI{200}{\milli\meter}\) and length \(L = \SI{1200}{\milli\meter}\) at three fluid velocities: \(u_\text{f} = 1.0u_\text{t}\), \(2.0u_\text{t}\), and \(3.0u_\text{t}\), where \(u_\text{t}\) is the terminal settling velocity,~\Cref{tab:tab4} lists the \(Re_\text{p}\) values. Particles are initially distributed randomly in the lower section of the pipe (\(0 < z < \SI{200}{\milli\meter}\)) and allowed to reach quasi-steady transport conditions. \revision{The local hydrodynamic regime around individual particles is more appropriately characterized by the \(Re_\text{p}\), which spans 98-2904, with the confinement ratio \(d/D = 0.082\) and 0.22 further modulating wake behavior.} We analyze transport characteristics through residence time distributions \(P(t_\text{r})\) and normalized drag force distributions \(P(\hat{f})\), where \(t_\text{r}\) is the time required for a particle to traverse one meter vertically and \(\hat{f} = F_{\text{d}}/(\rho_\text{p} - \rho_\text{f})V_\text{p} g\) is the ratio of instantaneous vertical drag to submerged particle weight.
\begin{table}[h]
\centering
\caption{Particles' Reynolds numbers \((Re_\text{p})\) for spherical particles and PMNs during vertical transport at varying fluid velocities. Terminal velocities \(u_\text{t}\) from sedimentation analysis with \(\rho_\text{f} = \SI{1000}{\kilo\gram\per\meter\cubed}\) and \(\mu_\text{f} = \SI{0.05}{\pascal\cdot\second}\).}
\label{tab:tab4}
\begin{tabular}{lccccc}
\hline \hline
\multirow{2}{*}{Particle Type} & \multirow{2}{*}{\(d_\text{v}\) (\SI{}{\milli\meter})} & \multirow{2}{*}{\(u_\text{t}\) (\SI{}{\meter\per\second})} & \multicolumn{3}{c}{\(Re_\text{p} = \rho_\text{f} u_\text{f} d_\text{v}/\mu_\text{f}\)} \\
\cmidrule(lr){4-6}
& & & \(u_\text{f} = 1.0u_\text{t}\) & \(u_\text{f} = 2.0u_\text{t}\) & \(u_\text{f} = 3.0u_\text{t}\) \\
\hline
Small sphere & 16.4 & 0.42 & 138  & 276  & 414  \\
Small PMN    & 16.4 & 0.30 & 98   & 196  & 294  \\
Large sphere & 44.0 & 1.10 & 968  & 1936 & 2904 \\
Large PMN    & 44.0 & 0.80 & 704  & 1408 & 2112 \\
\hline \hline
\multicolumn{6}{l}{\small Note: \(Re_\text{p}\) calculated using volume-equivalent diameter \(d_\text{v}\) and fluid velocity \(u_\text{f}\).} \\
\end{tabular}
\end{table}
\subsubsection{Residence Time Analysis}
\begin{figure}[H]
    \centering
    \subfloat[Small spheres]{\includegraphics[width=0.4\linewidth]{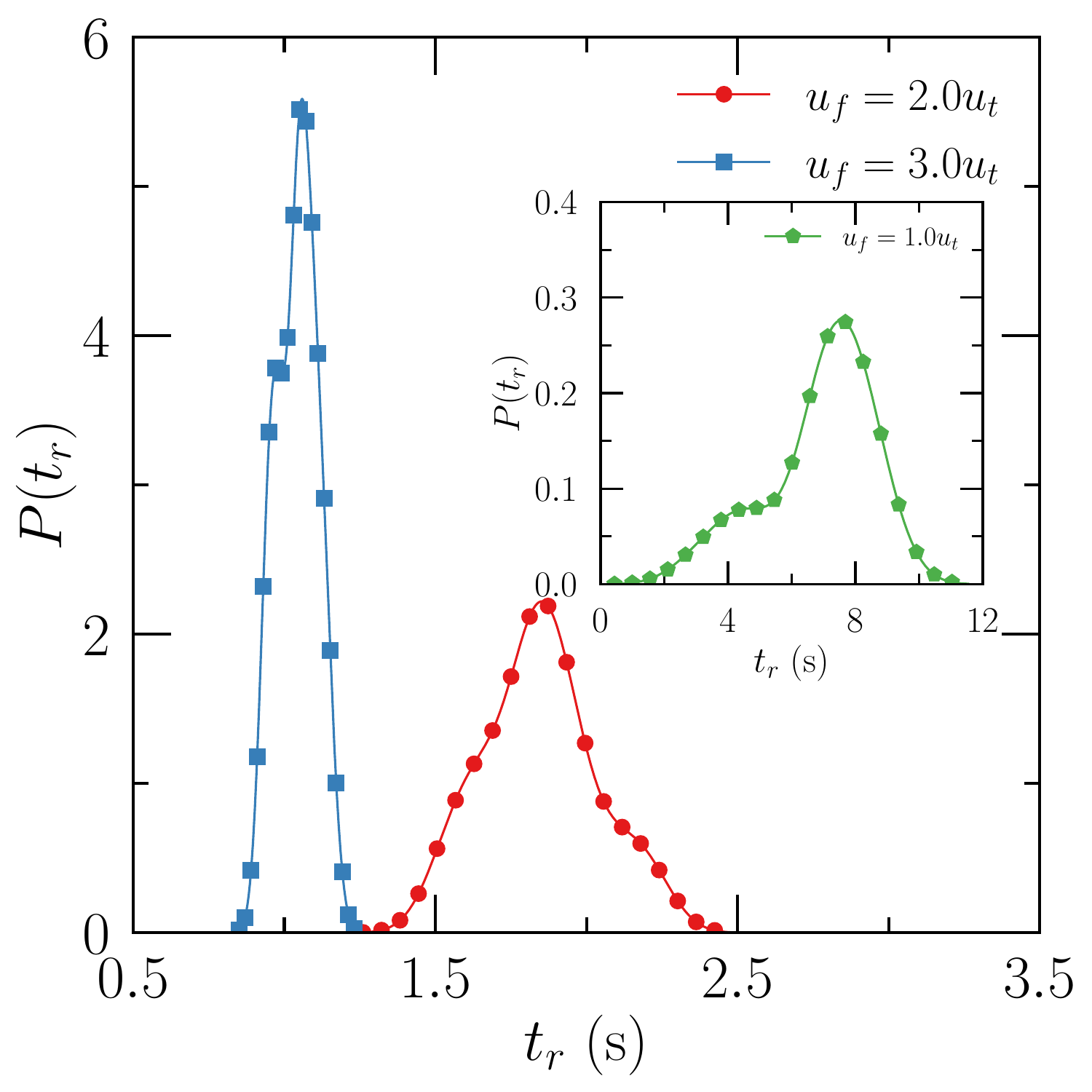}}
    \subfloat[Large spheres]{\includegraphics[width=0.4\linewidth]{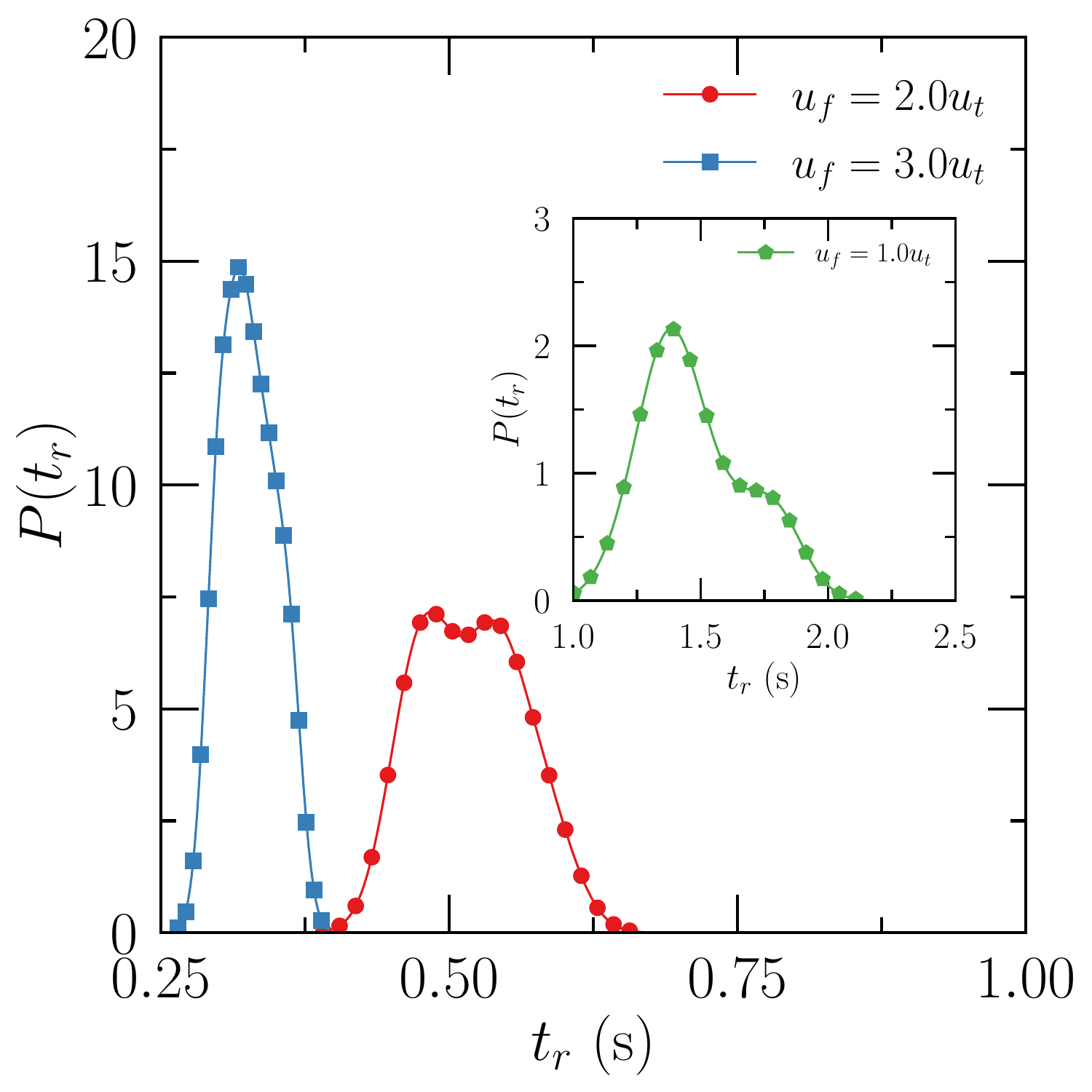}}\\
    \subfloat[Small PMNs]{\includegraphics[width=0.4\linewidth]{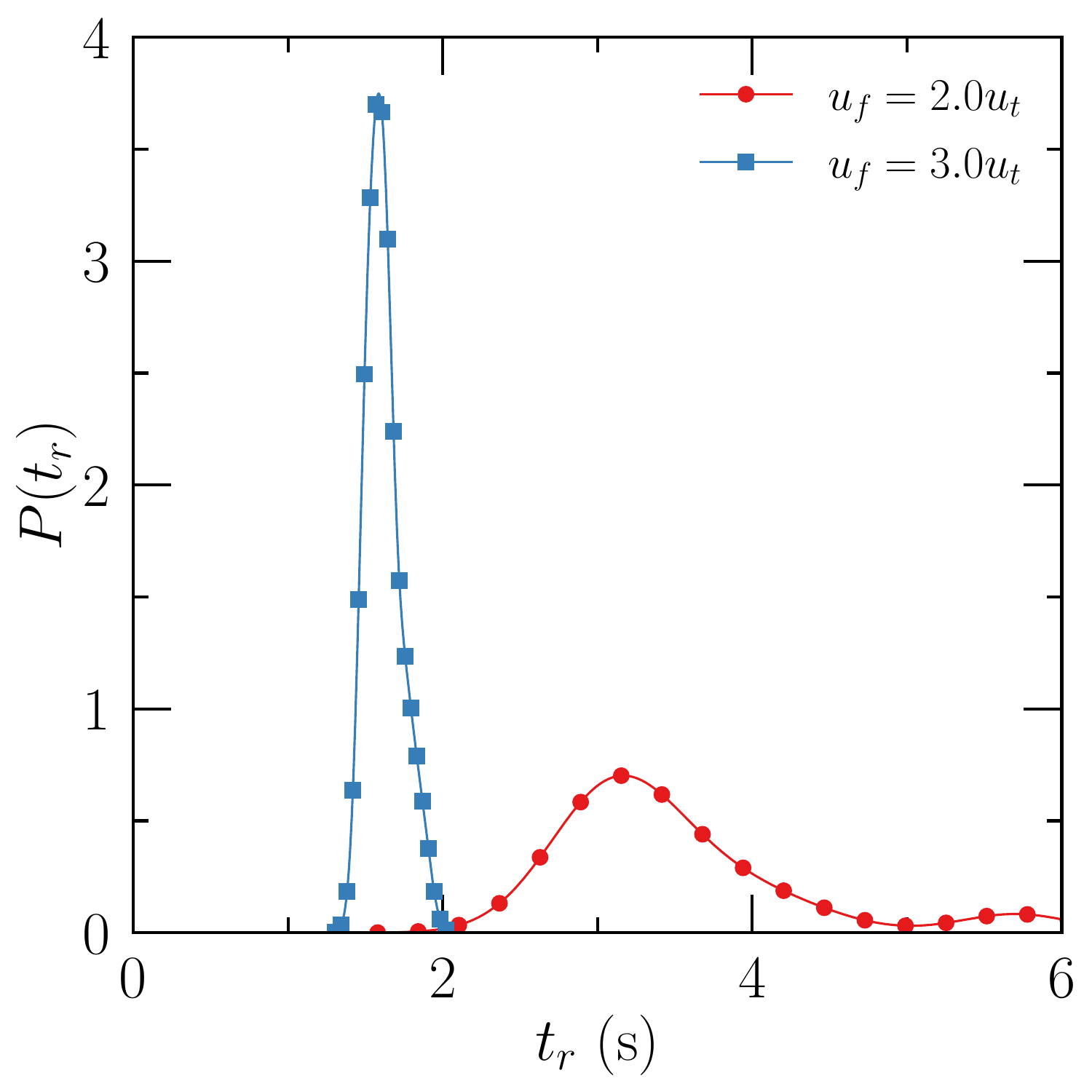}}
    \subfloat[Large PMNs]{\includegraphics[width=0.4\linewidth]{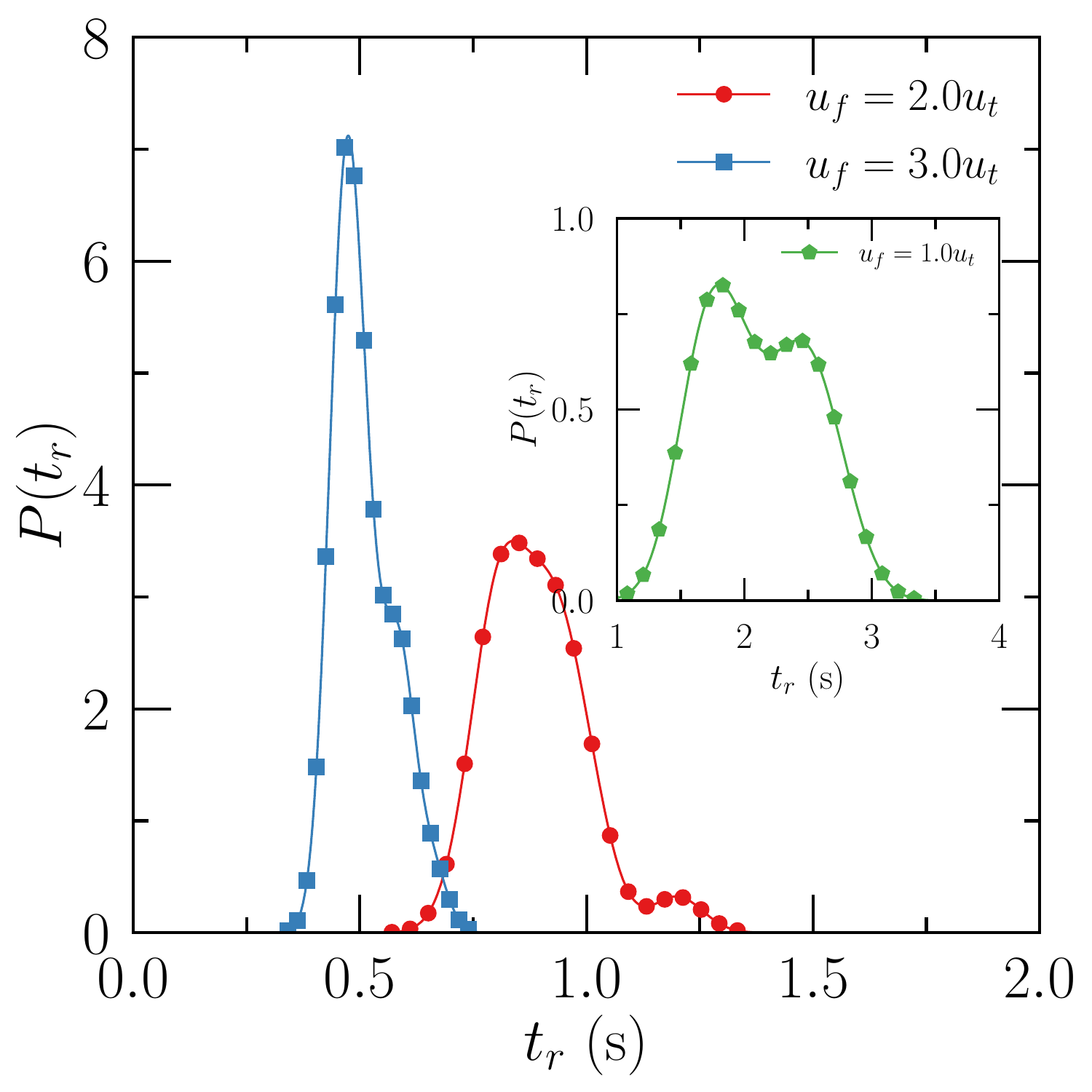}}
    \caption{\revision{Probability distributions of residence time \((t_\text{r})\) at \(u_\text{f} = 1.0u_\text{t}\), \(2.0u_\text{t}\), and \(3.0u_\text{t}\). Panels (a-d) show small spheres, large spheres, small PMNs, and large PMNs, respectively. Distributions narrow with increasing velocity, transitioning from broad, variance-dominated profiles at low flow rates (insets) to sharp, convection-dominated distributions at high rates. Note: Small PMNs remain suspended at \((u_\text{f} = 1.0u_\text{t})\) without net upward transport.}}
    \label{fig:fig10}
\end{figure}

\begin{figure}[H]
    \centering
    \subfloat[]{\includegraphics[width=0.2\linewidth]{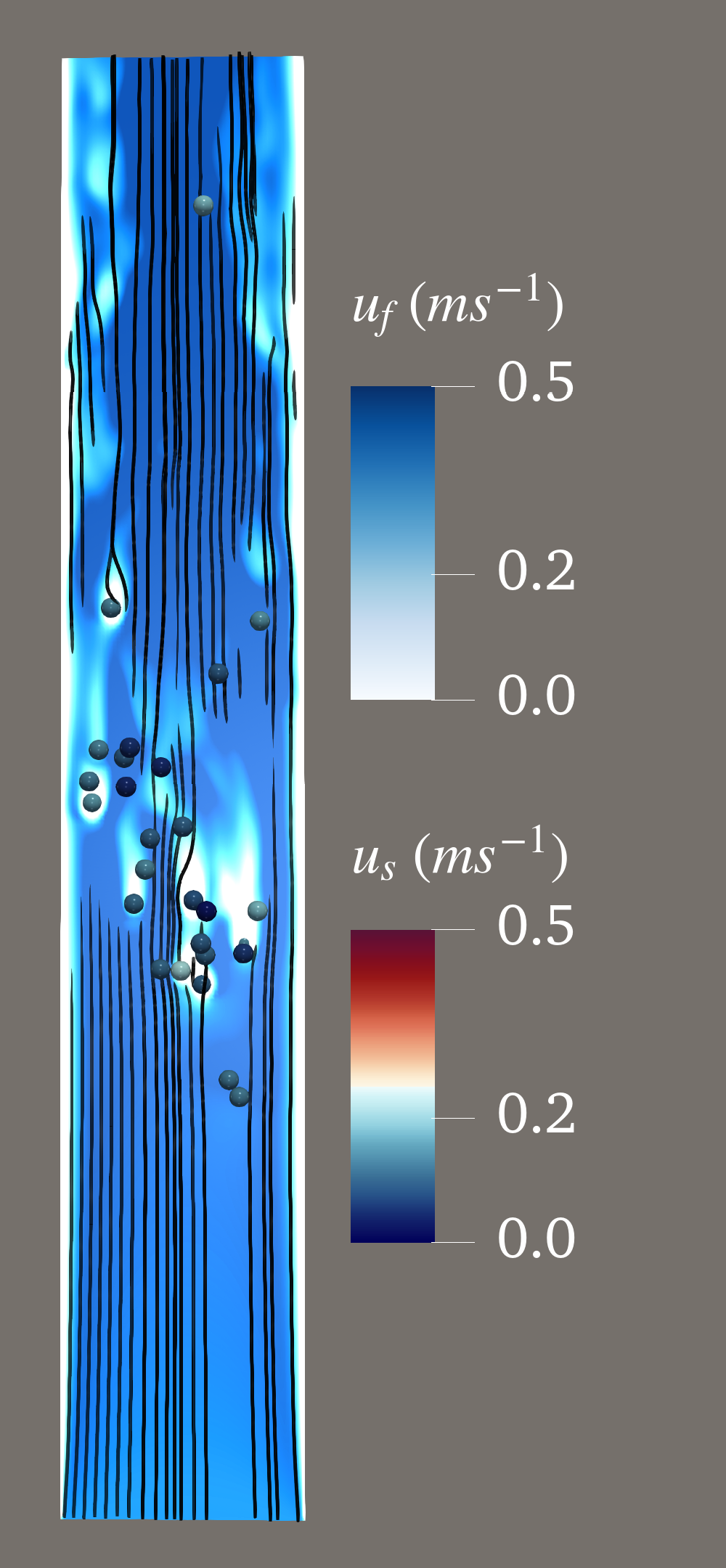}}\hspace{1em}
    \subfloat[]{\includegraphics[width=0.2\linewidth]{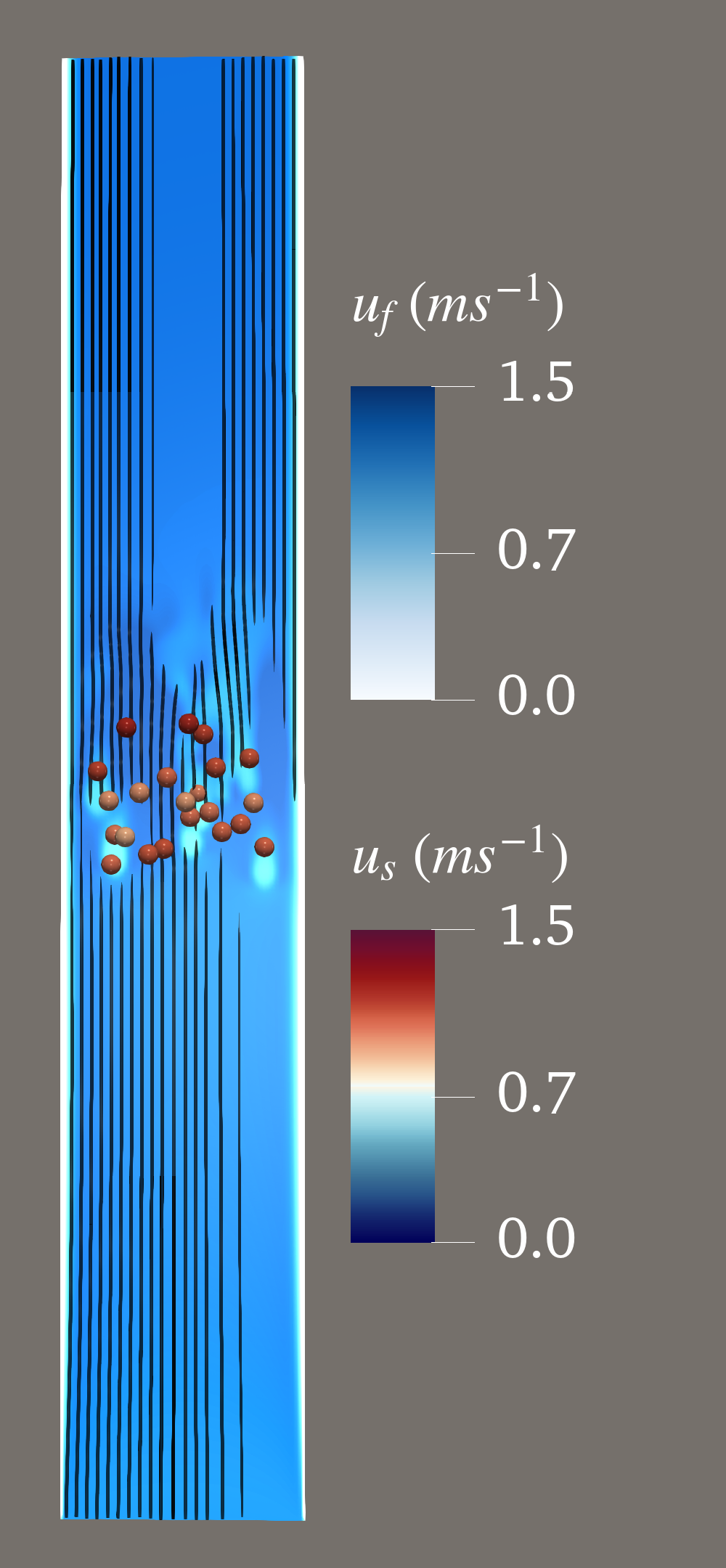}}\hspace{1em}
    \subfloat[]{\includegraphics[width=0.2\linewidth]{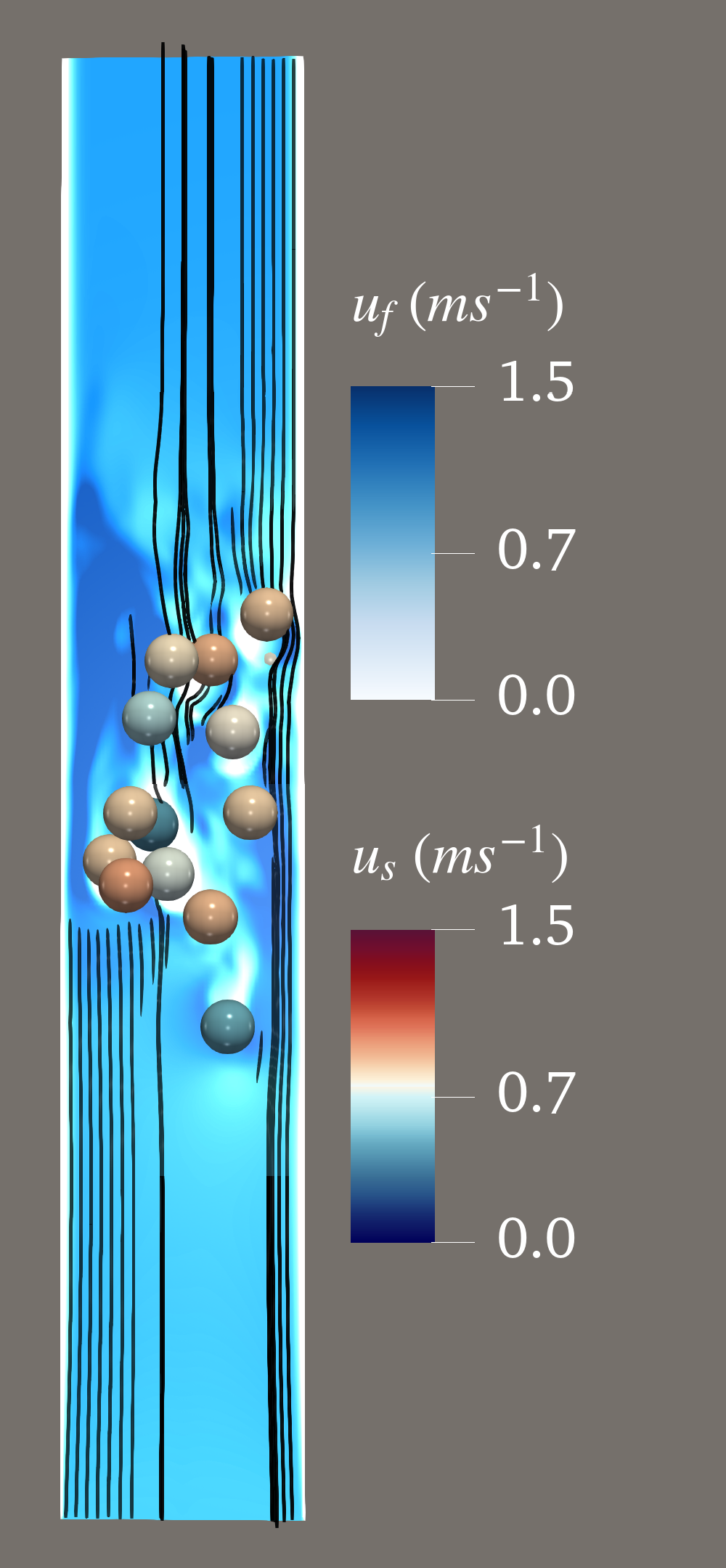}}\hspace{1em}
    \subfloat[]{\includegraphics[width=0.2\linewidth]
    {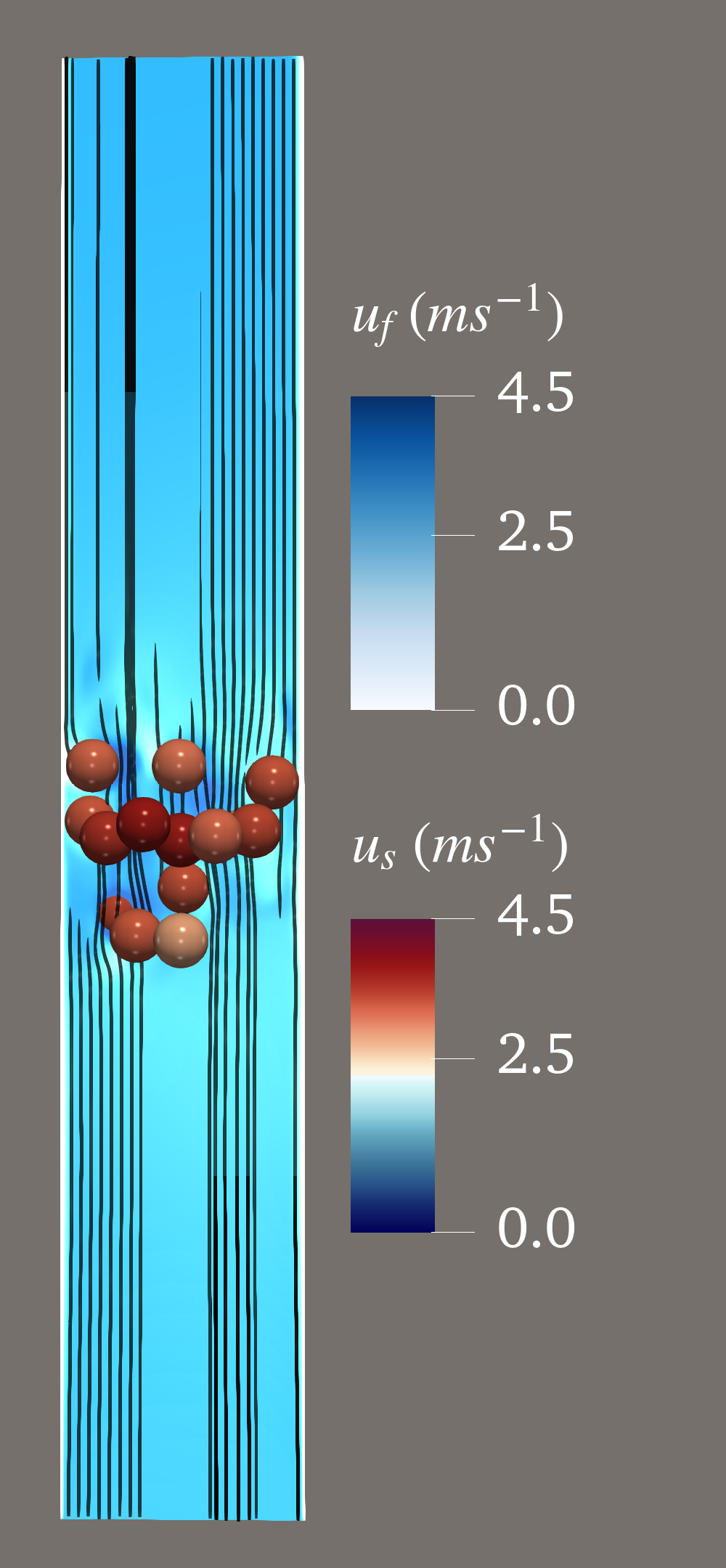}}\\
    \subfloat[]{\includegraphics[width=0.2\linewidth]{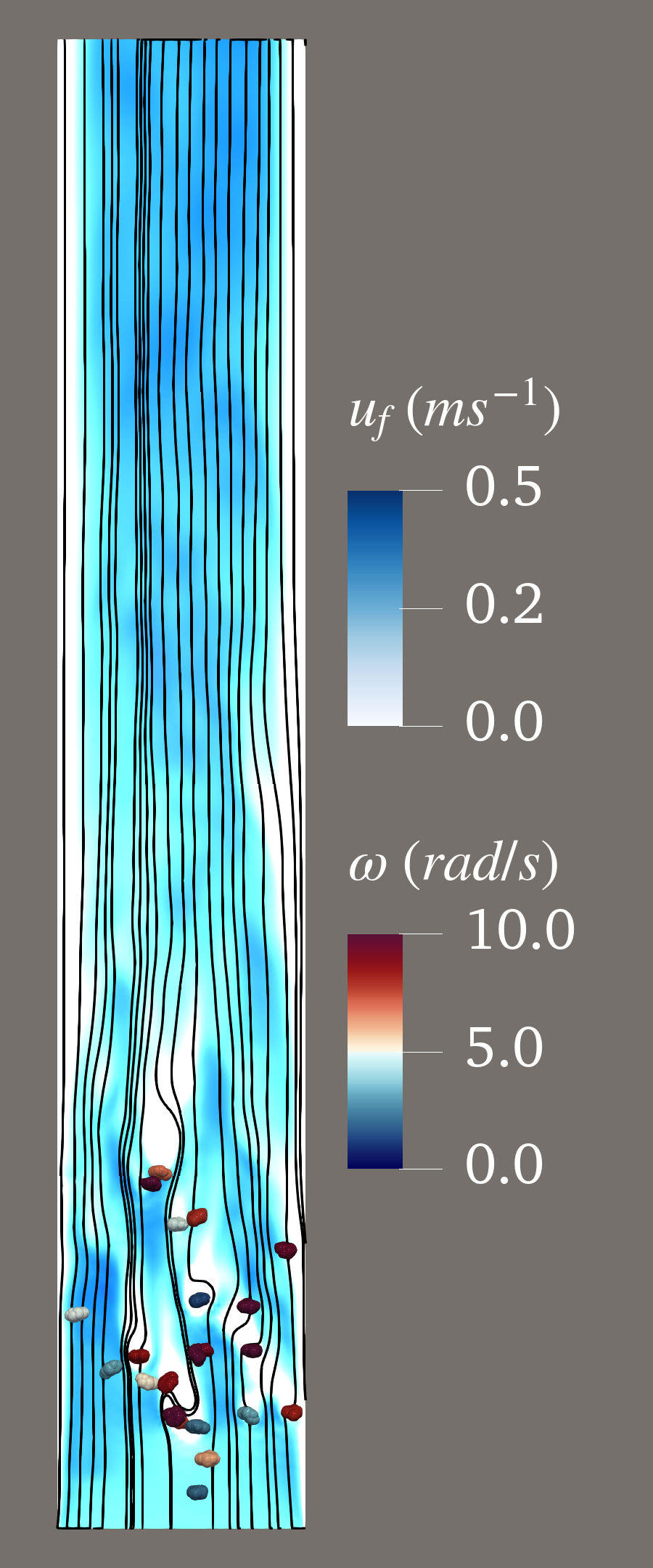}}\hspace{1em}
    \subfloat[]{\includegraphics[width=0.2\linewidth]{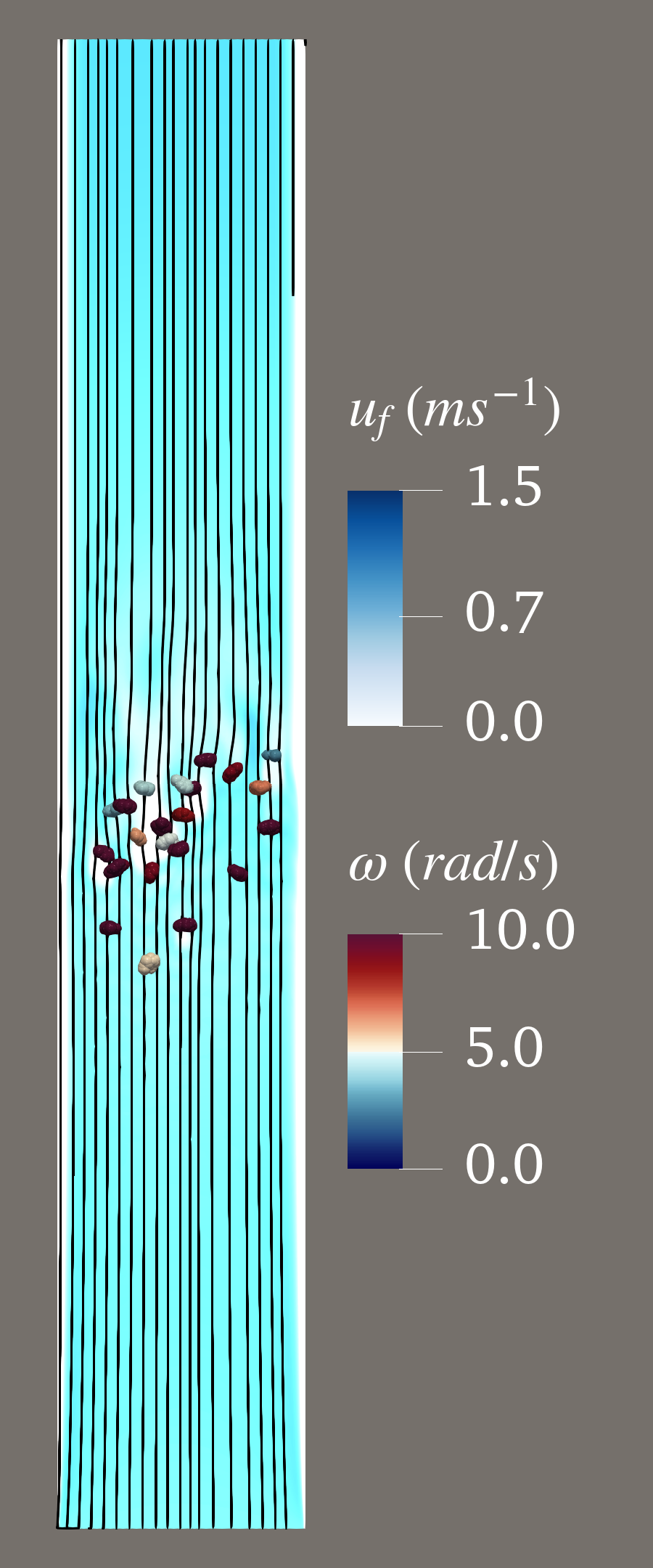}}\hspace{1em}
    \subfloat[]{\includegraphics[width=0.2\linewidth]{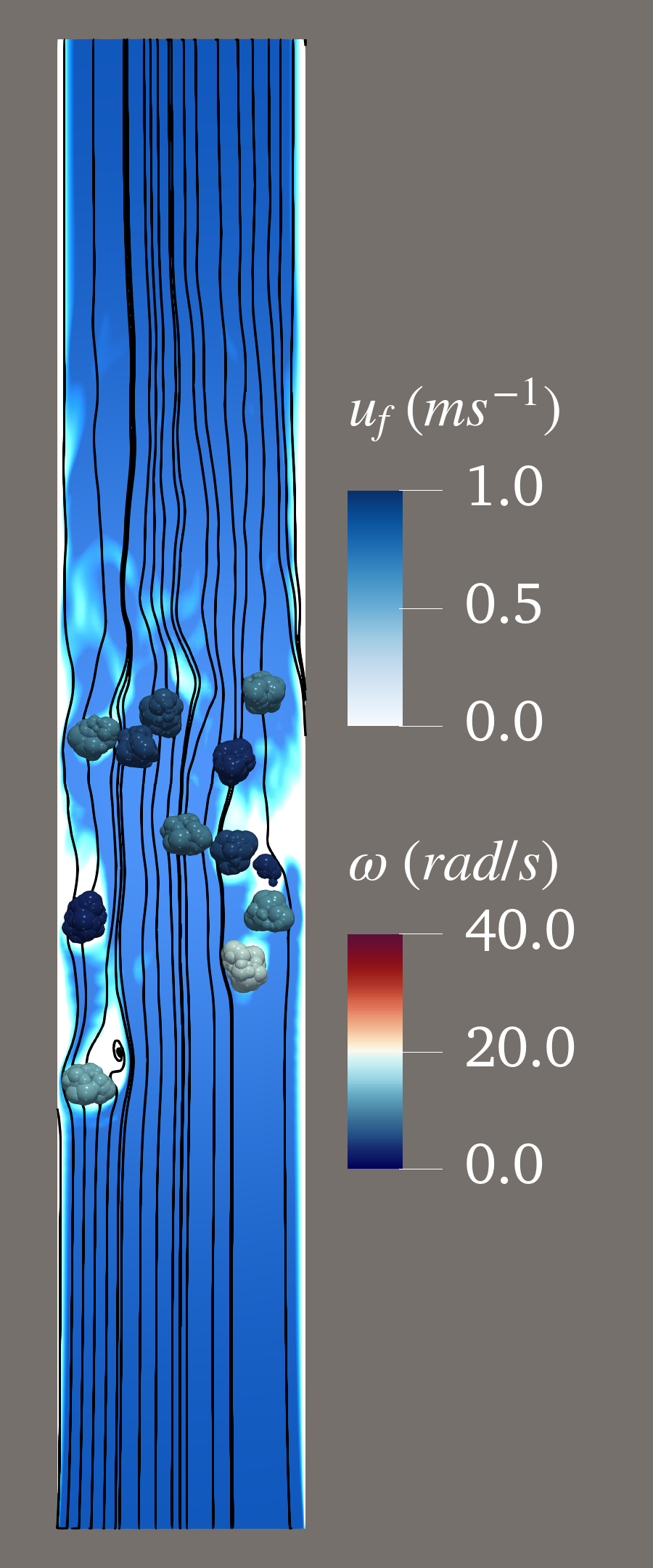}}\hspace{1em}
    \subfloat[]{\includegraphics[width=0.2\linewidth]
    {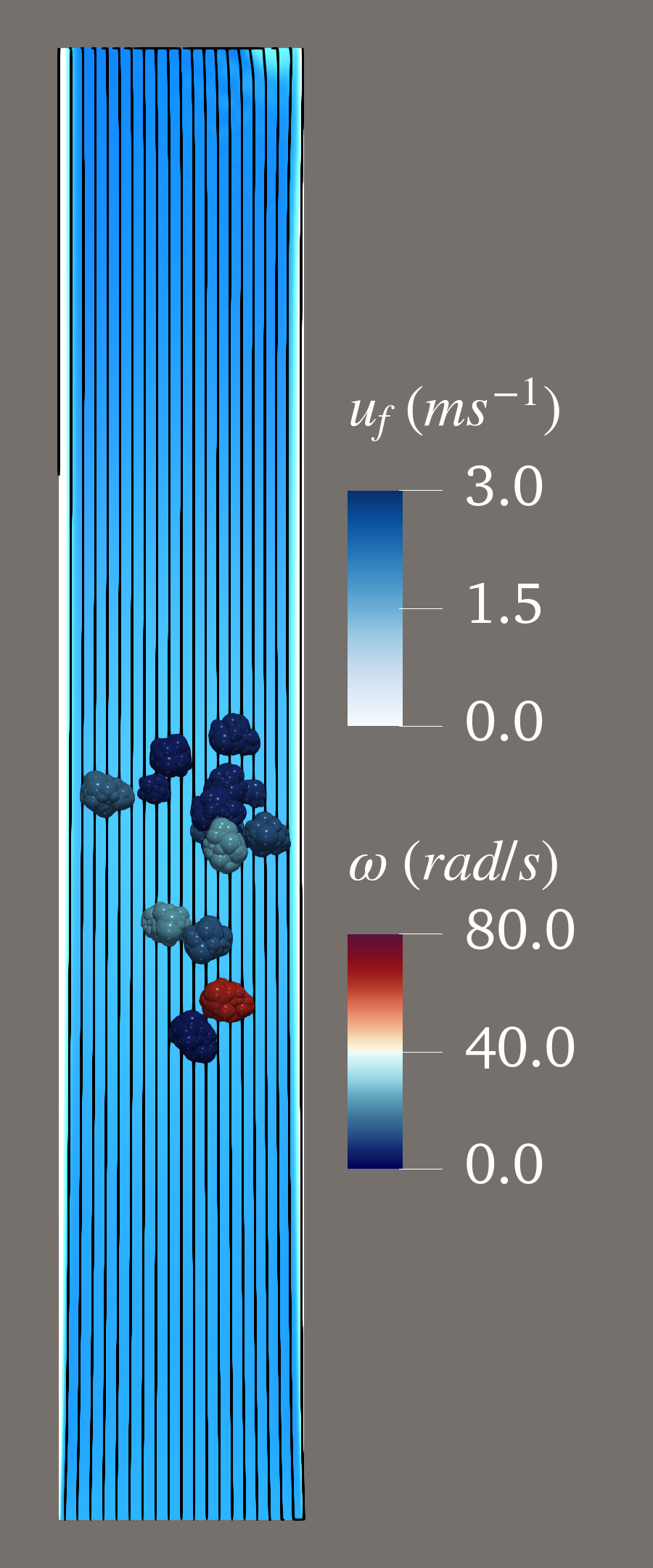}}
    \caption{Particle configurations and flow fields during vertical transport. Small spheres \((d_v = \SI{16.4}{\milli\meter})\): (a) \(u_f = 1.0u_t\) shows vertical dispersion and lateral spread; (b) \(u_f = 3.0u_t\) shows homogeneous distribution with rectilinear motion. 
    Large spheres \((d_v = \SI{44}{\milli\meter})\): (c) \(u_f = 1.0u_t\) shows moderate dispersion; (d) \(u_f = 3.0u_t\) shows homogeneous distribution. 
    Small PMNs: (e) \(u_f = 1.0u_t\) shows asymmetric wakes and helical trajectories; (f) \(u_f = 3.0u_t\) shows reduced dispersion with continued rotational motion. 
    Large PMNs: (g) \(u_f = 1.0u_t\) shows complex wake structures; (h) \(u_f = 3.0u_t\) shows linear trajectories with angular velocity up to \SI{80}{\radian\per\second} and intermittent wall-proximity events. Fluid velocity is colored by magnitude, and streamlines are shown as tube elements (left). Particle trajectories colored by velocity (top right rows) or angular velocity (bottom right rows).}
    \label{fig:fig11}
\end{figure}

\begin{figure}[H]
    \centering
    \sidesubfloat[]{\includegraphics[width=0.2\linewidth]{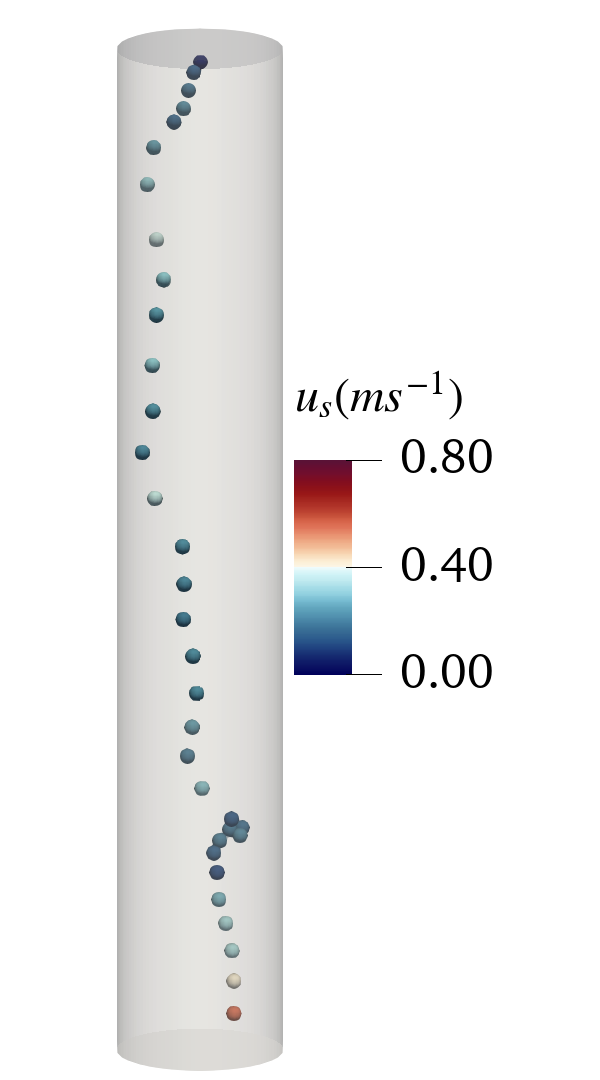}}
    \sidesubfloat[]{\includegraphics[width=0.2\linewidth]{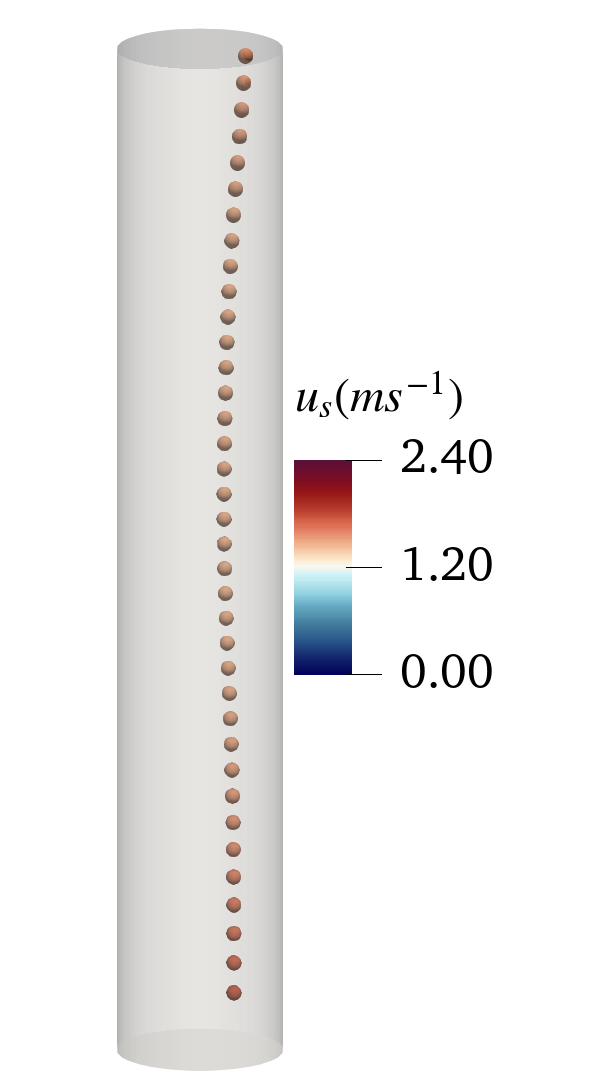}}
    \sidesubfloat[]{\includegraphics[width=0.2\linewidth]{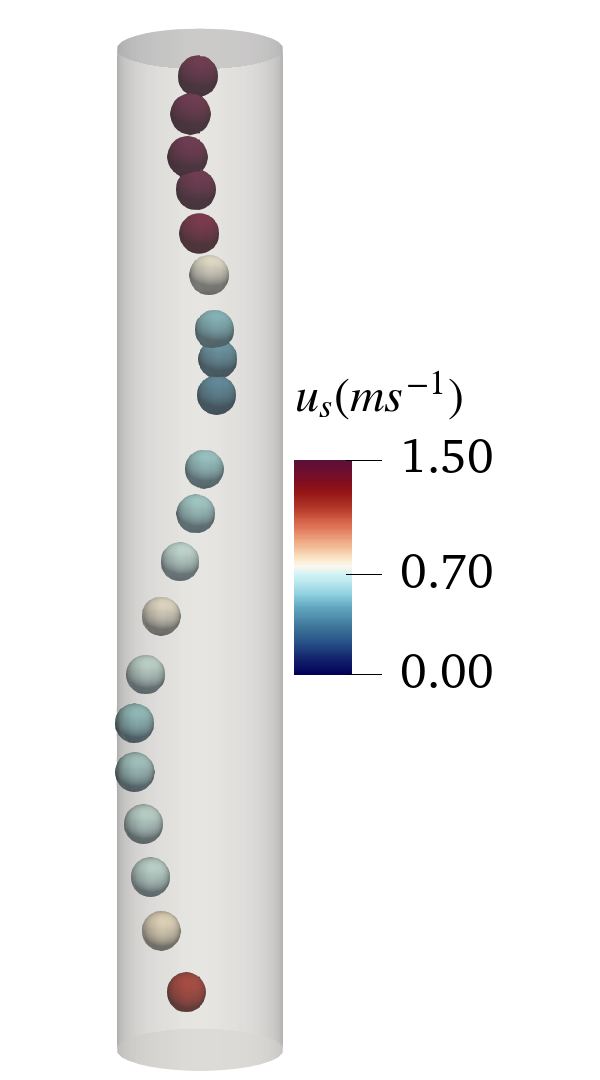}}
    \sidesubfloat[]{\includegraphics[width=0.2\linewidth]{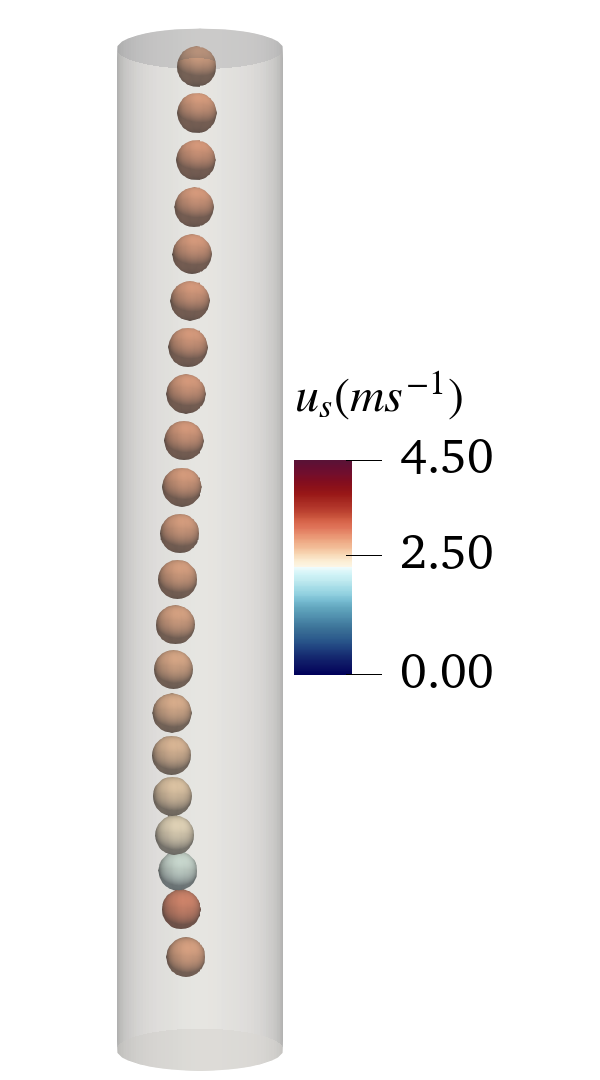}}\vspace{1em}
    \sidesubfloat[]{\includegraphics[width=0.2\linewidth]{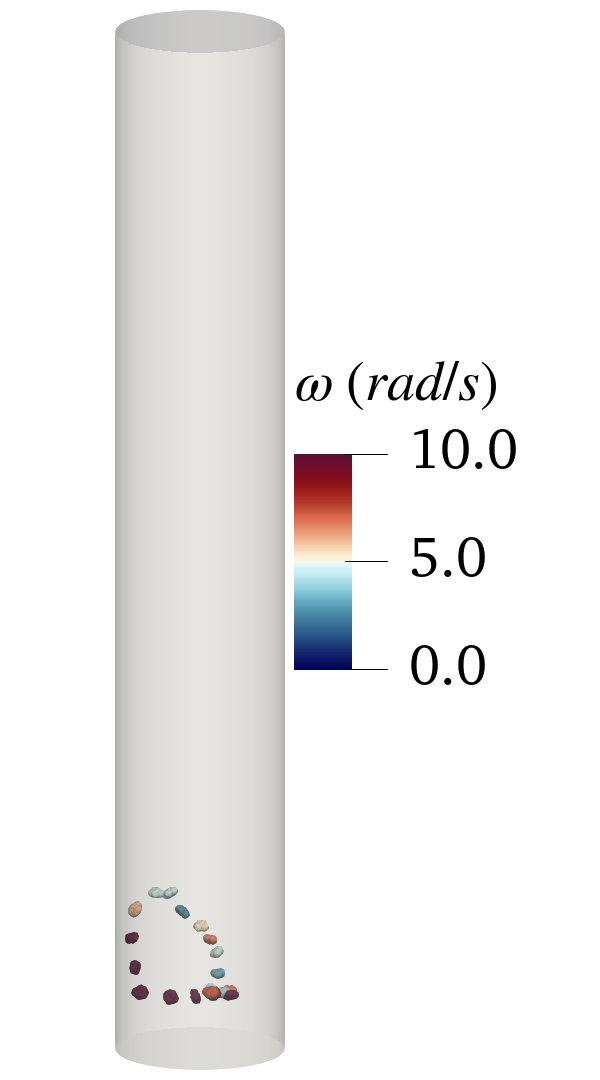}}
    \sidesubfloat[]{\includegraphics[width=0.2\linewidth]{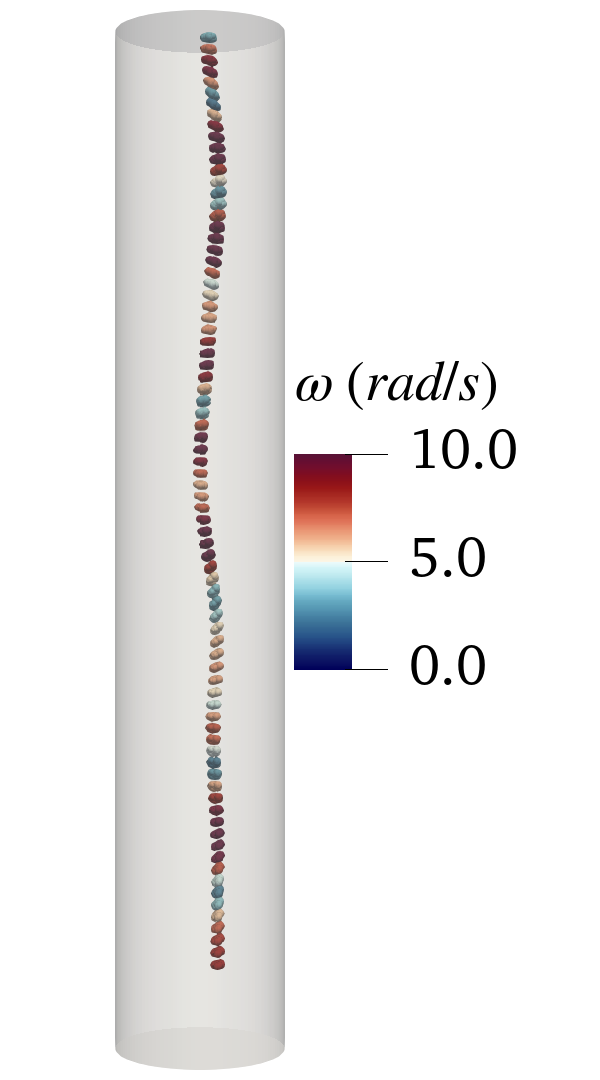}}
    \sidesubfloat[]{\includegraphics[width=0.2\linewidth]{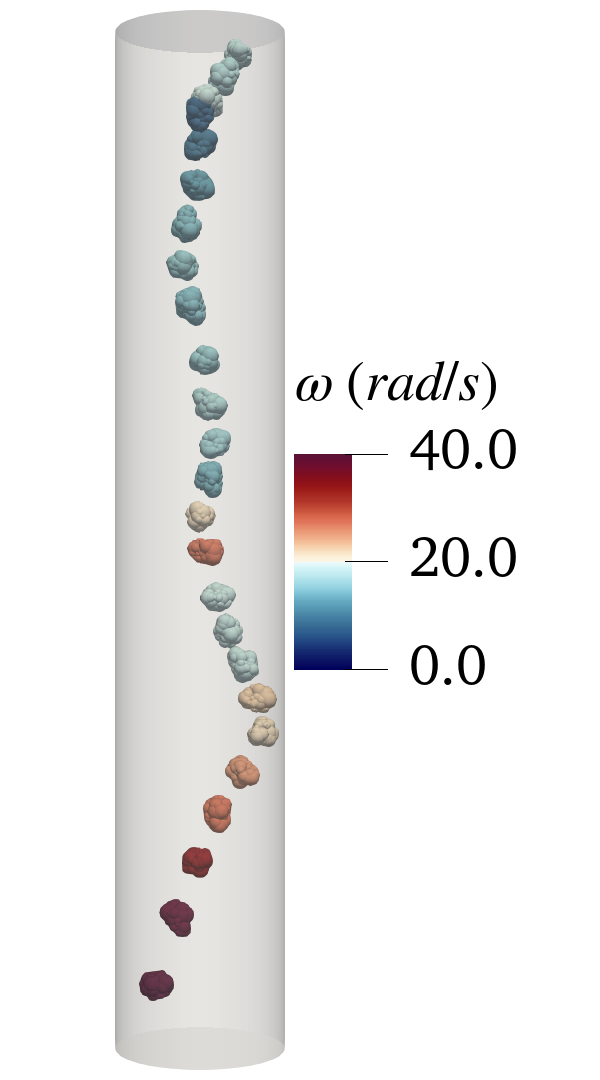}}
    \sidesubfloat[]{\includegraphics[width=0.2\linewidth]{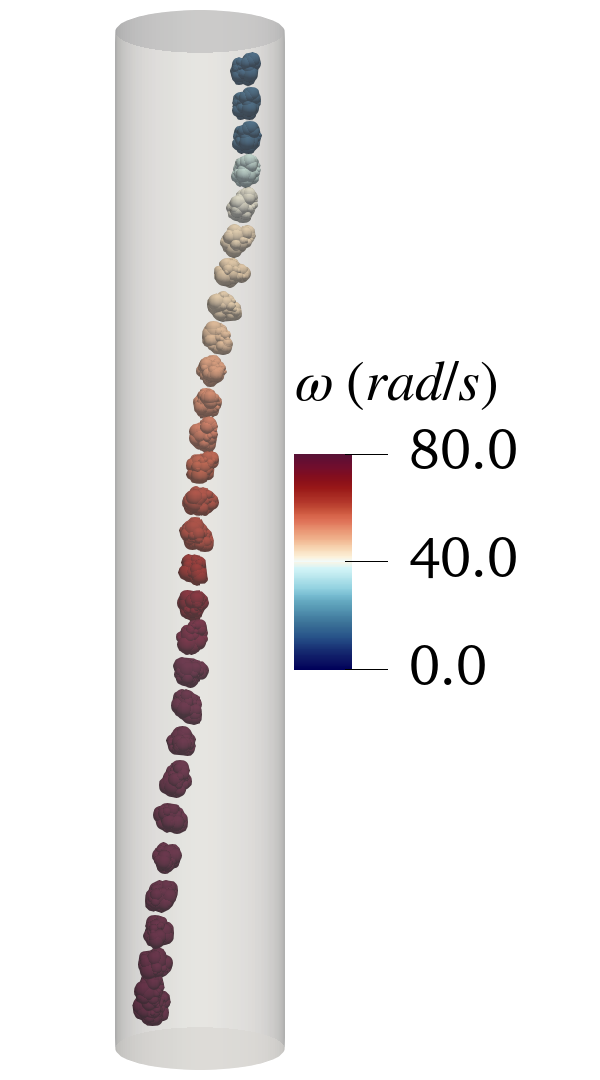}}
    \caption{Particle trajectory for a random particle during vertical hydraulic transport, colored by instantaneous translational \((u_s)\) or angular velocity \((\omega)\).  Small particles with (a) \& (b) as spheres and (e) \& (f) as PMNs at \(u_f = 1.0u_t\) and \(3.0u_t\), respectively, colored by translational \((u_s)\) or angular \((\omega)\) velocity. The small PMNs remain in a state of marginal suspension near the pipe inlet, exhibiting oscillatory motion without net upward transport\textemdash the fluid velocity is insufficient to overcome gravitational settling, resulting in a dynamic equilibrium in which the particles hover with continuous rotational adjustments. Large particles with (c) \& (d) as spheres and (g) \& (h) as PMNs at \(u_f = 1.0u_t\) and \(3.0u_t\), respectively. At low flow velocity \((u_f = 1.0u_t)\), trajectories exhibit significant lateral dispersion and velocity fluctuations reflecting intermittent transport. At high flow velocity \((u_f = 3.0u_t)\), trajectories become more vertically aligned with reduced dispersion, indicating stable convection-dominated transport.  The PMNs exhibit enhanced rotational motion due to shape-induced torque than volume-equivalent spheres.}
    \label{fig:fig13}
\end{figure}

\noindent \Cref{fig:fig10} presents the probability distributions of residence time \((t_\text{r})\) for spherical and non-spherical particles at varying flow velocities. The distributions characterize the transition from intermittent, settling-dominated transport at low velocities to steady, convection-dominated entrainment at high velocities. In~\Cref{fig:fig10}a and c, the small spherical particles and PMNs, respectively, exhibit broad distributions with extended tails at \(u_\text{f} = 1.0u_\text{t}\) and \(u_f = 2.0u_\text{t}\), indicating high variability in individual particle transport velocities. For spheres, the standard deviation decreases from \(\sigma_\text{tr} = \SI{1.55}{\second}\) at \(1.0u_\text{t}\) to \(\sigma = \SI{0.06}{\second}\)  at \(3.0u_\text{t}\). At \(1.0u_\text{t}\), the small PMNs exhibit marginal suspension states\textemdash particles oscillate near the inlet without achieving consistent upward motion. This is illustrated in particle trajectory visualizations (see~\Cref{fig:fig11}e and ~\Cref{fig:fig13}e), where small PMNs remain confined to the lower pipe section at low velocity, exhibiting continuous rotational motion without net vertical displacement. Large spherical particles,~\Cref{fig:fig10}b, show systematic variance reduction from \(\sigma = \SI{0.19}{\second}\) to \(\SI{0.02}{\second}\) as the fluid velocity increases. Large PMNs,~\Cref{fig:fig10}d, exhibit broader distributions at all velocities compared to spherical particles, with more pronounced tails at \(2.0u_\text{t}\) and \(3.0u_\text{t}\). 

The mean residence times \((\mu_{t_\text{r}})\) for all particles follow the expected inverse relationship \(\mu_{t_\text{r}} \propto 1/(u_\text{f} - u_\text{t})\), with measured values of \(\mu_{t_\text{r}} = \SI{6.87}{\second}\), \(\SI{1.85}{\second}\), and \(\SI{1.04}{\second}\) for small spherical particles at \(1.0u_\text{t}\), \(2.0u_\text{t}\), and \(3.0u_\text{t}\), respectively.~\Cref{tab:tab5} delineates the \(\mu_{t_\text{r}}\) and \(\sigma_{t_\text{r}}\) for various particle groups. Particle configuration snapshots in~\Cref{fig:fig11} reveal the spatial distributions underlying these statistical behaviors. At \(1.0u_\text{t}\), small spherical particles,~\Cref{fig:fig11}a exhibit pronounced vertical dispersion with particles spanning the half pipe length, consistent with the broad residence time distribution and intermittent settling events. \revision{The small PMNs (see~\Cref{fig:fig11}e) remain in a state of marginal suspension near the pipe inlet, exhibiting oscillatory motion without net upward transport\textemdash the imposed fluid velocity matches the terminal settling velocity, providing hovering equilibrium in which the time-mean drag balances submerged weight without sustained net upward transport. Particles undergo continuous rotational adjustments and lateral excursions, but accumulate no measurable axial displacement over the simulation window, which is insufficient to overcome gravitational settling, resulting in a dynamic equilibrium in which the particles hover with continuous rotational adjustments.} Large particles,~\Cref{fig:fig11}c and g, show less dispersion due to higher inertia, which reduces susceptibility to local flow fluctuations. 

At \(3.0u_\text{t}\), all particle types achieve spatially homogeneous distributions with minimal vertical dispersion, confirming efficient entrainment and stable transport (see~\Cref{fig:fig11}b, d, f, and h). The absence of extended residence time tails at this velocity indicates that all particles maintain consistent upward motion without settling reversals. Flow field visualizations show that small PMNs develop subtle asymmetric wakes and helical trajectories with variable angular velocity even at high velocities,~\Cref{fig:fig11}f, while large PMNs exhibit no spatially coherent wake structures in instantaneous streamline plots, despite angular velocity \SI{80}{\radian\per\second}, indicative of strong rotational-translational coupling,~\Cref{fig:fig11}h. Individual particle trajectories in~\Cref{fig:fig13} provide a direct visualization of these transport regimes. Small spherical particles (\Cref{fig:fig13}a and b) transition from oscillatory motion with lateral wandering at \(1.0u_\text{t}\) to nearly rectilinear trajectories at \(3.0u_\text{t}\). However, the Small PMNs (\Cref{fig:fig13}e) remain in marginal suspension at low velocity. Large PMNs,~\Cref{fig:fig13}g and h, exhibit enhanced rotational motion, with PMNs achieving significantly higher angular velocities, reflecting shape-induced torques and coupled translational-rotational dynamics.
\begin{table}[h]
\centering
\caption{Mean residence time ($\mu_{tr}$) and standard deviation ($\sigma_{tr}$) for spherical particles and PMNs during vertical transport at varying fluid velocities.}
\label{tab:tab5}
\begin{tabular}{lcccccc}
\hline\hline
\multirow{2}{*}{Particle} & \multicolumn{3}{c}{$\mu_{tr}$ (s)} & \multicolumn{3}{c}{$\sigma_{tr}$ (s)} \\
\cline{2-7}
& $1.0u_t$ & $2.0u_t$ & $3.0u_t$ & $1.0u_t$ & $2.0u_t$ & $3.0u_t$ \\
\hline
Small sphere & 6.87 & 1.85 & 1.04 & 1.55 & 0.18 & 0.06 \\
Small PMN & \textemdash & 3.51 & 1.62 & \textemdash & 0.75 & 0.11 \\
Large sphere & 1.48 & 0.52 & 0.33 & 0.19 & 0.04 & 0.02 \\
Large PMN & 2.11 & 0.89 & 0.51 & 0.39 & 0.10 & 0.06 \\
\hline\hline
\multicolumn{7}{l}{Note: Small PMNs remain in a suspended state at \(1.0u_t\) } \end{tabular}
\end{table}

\subsubsection{Drag Force Analysis}
\begin{figure}[H]
    \centering
    \subfloat[Small spheres]{\includegraphics[width=0.4\linewidth]{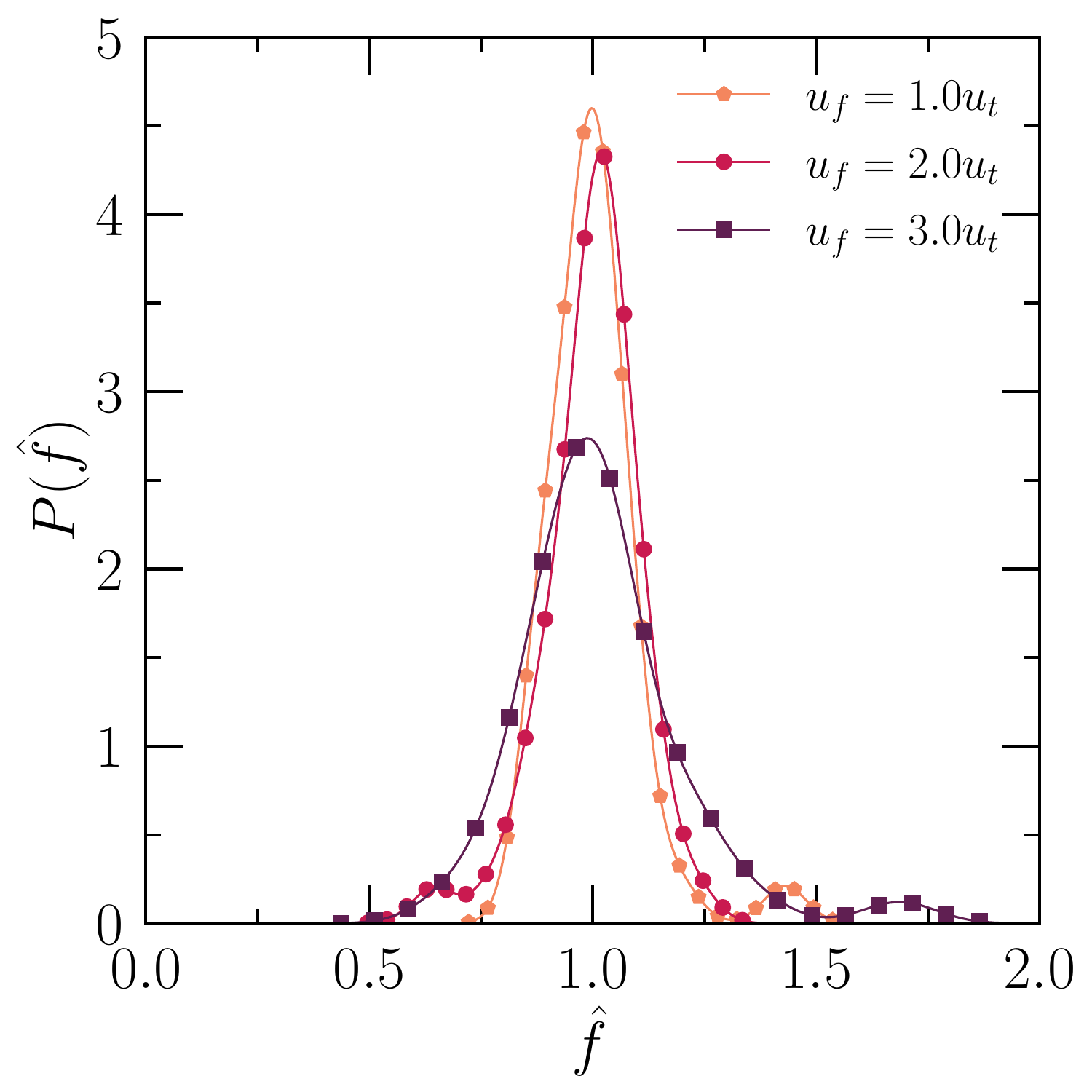}}
    \subfloat[Small PMNs]{\includegraphics[width=0.4\linewidth]{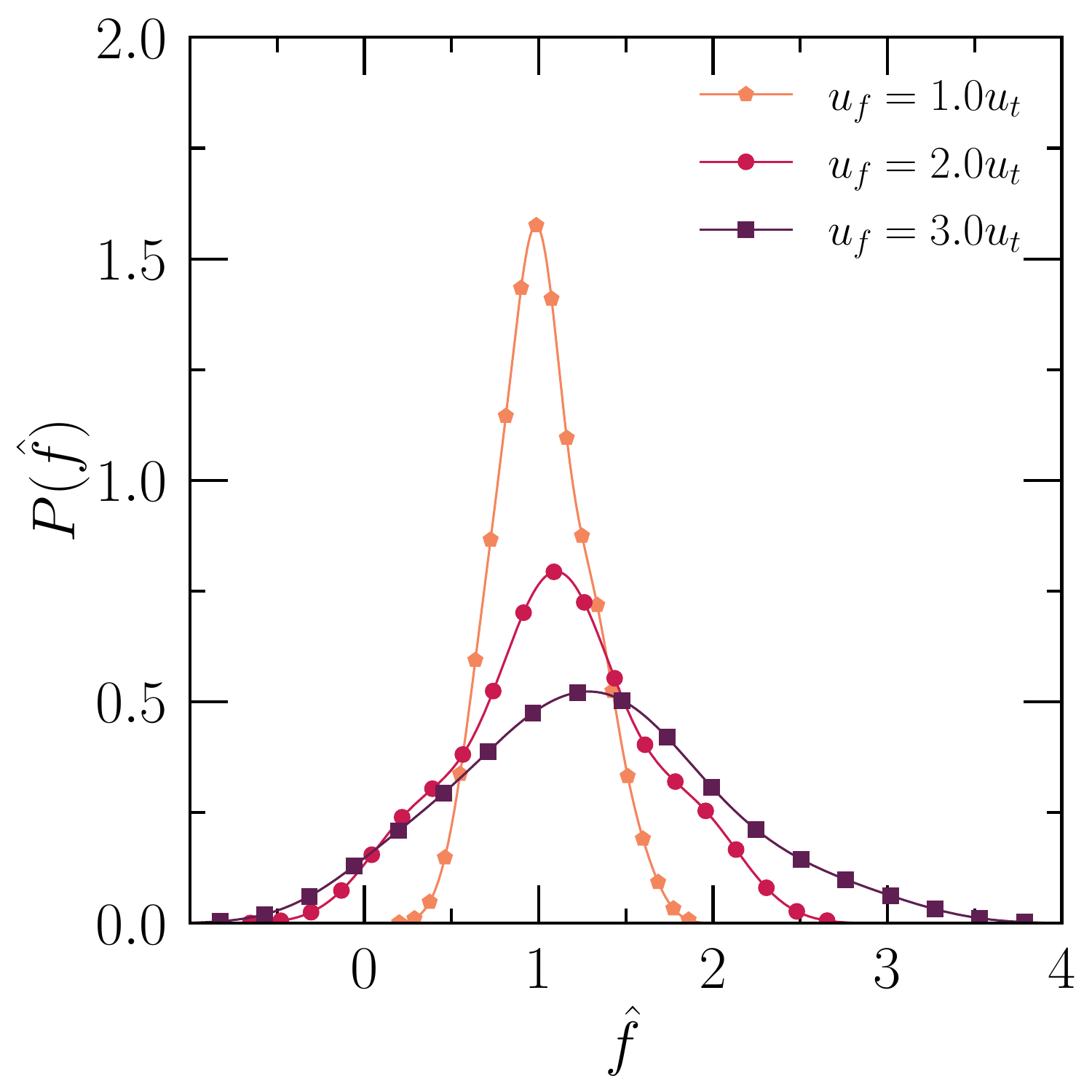}}\\
    \subfloat[Large spheres]{\includegraphics[width=0.4\linewidth]{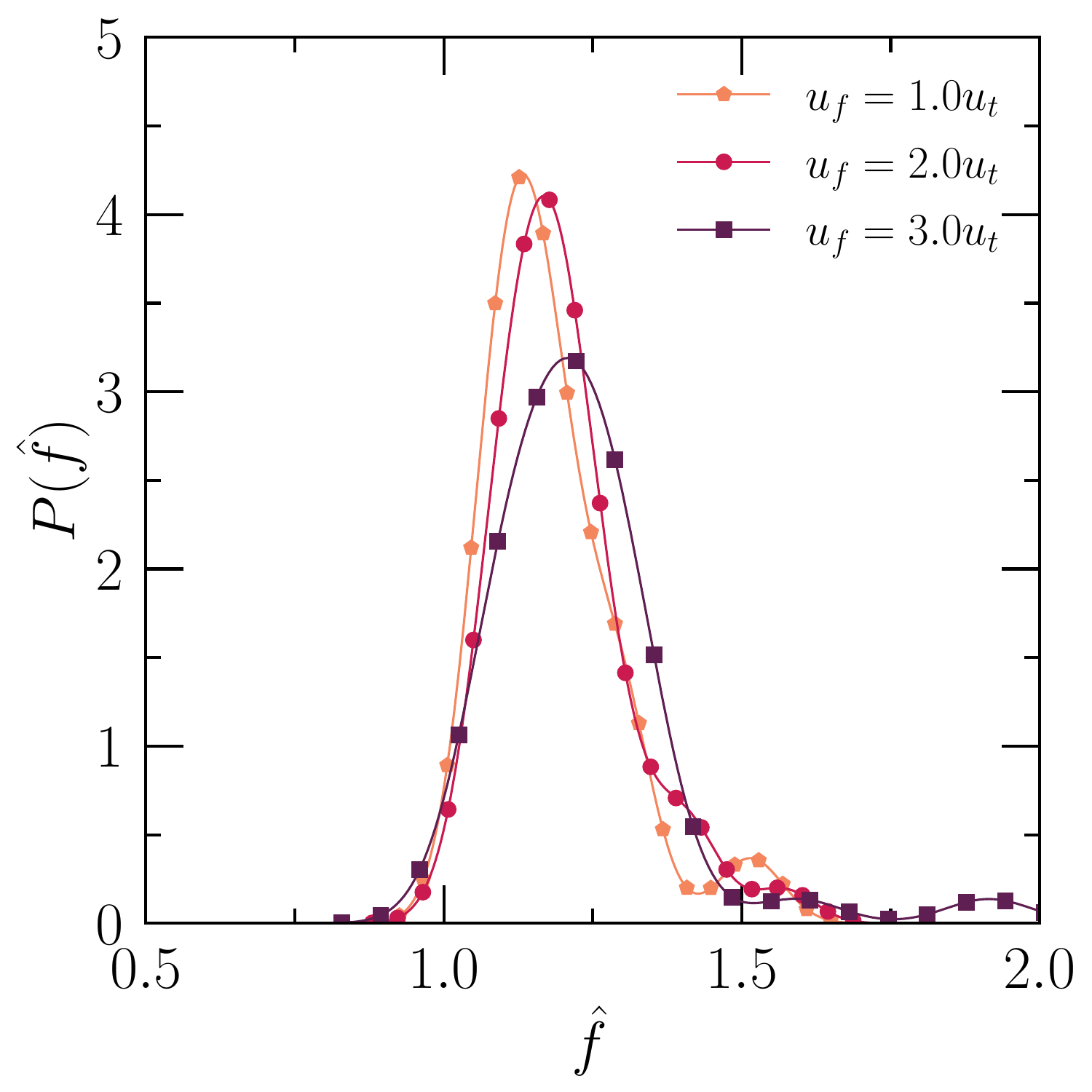}}
    \subfloat[Large PMNs]{\includegraphics[width=0.4\linewidth]{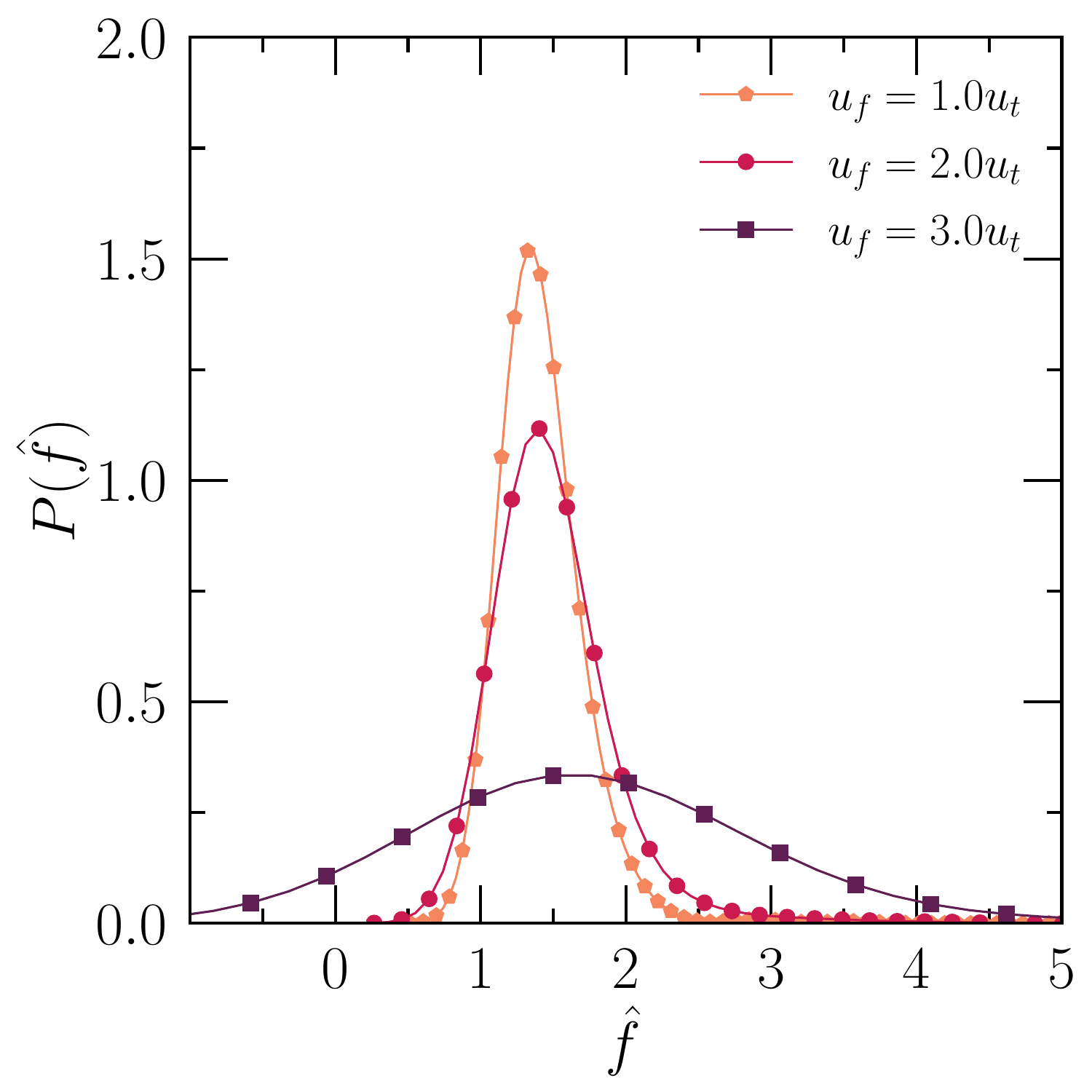}}
    \caption{\revision{Probability distributions of the normalized drag force \((\hat{f} = F_\text{d}/(\rho_\text{p}-\rho_\text{f})V_\text{p} g)\) for spherical and non-spherical particles at \(u_\text{f} = 1.0u_\text{t}\), \(2.0u_\text{t}\), and \(3.0u_\text{t}\). Panels (a-d) show small spheres, large spheres, small PMNs, and large PMNs, respectively. Small particles exhibit narrow, weakly varying distributions across velocities, whereas large spheres show progressive broadening with increasing \(u_\text{f}\), reflecting enhanced wake unsteadiness. Large PMNs display broader, skewed distributions with pronounced high-drag tails, highlighting intermittent, orientation-dependent forcing.}}
    \label{fig:fig12}
\end{figure}

\noindent \Cref{fig:fig12}a and b illustrate the probability distributions of the normalized axial drag force, \(\hat{f} = F_\text{d}/(\rho_\text{p}-\rho_\text{f})V_\text{p} g\), providing insight into the force balance mechanisms governing particle suspension. \revision{The distributions are obtained by sampling \(\hat{f}\) for each particle at every output step and aggregating the resulting realizations across the ensemble and realization window. It characterizes the fluctuating drag experienced by individual particles during transport, rather than their ensemble- or time-averaged mean.} Across all fluid velocities examined, the mean normalized drag force remains approximately near one (\(\langle\hat{f}\rangle \approx 1\)), confirming that drag effectively balances particle weight under steady-state conditions. Particles continuously adjust their slip velocity to maintain this equilibrium, regardless of the imposed fluid velocity. This behavior aligns with classical multiphase flow theory, where terminal velocity defines the threshold for suspension, and higher entrainment velocities enhance stability without significantly altering the mean force balance. While gravitational and hydrodynamic forces remain dynamically balanced on average, the distinction between different operating conditions arises from the shape of the distributions, which encapsulate the fluctuating components of the drag force and their dependence on turbulence, inertia, and confinement.
The evolution of drag-force statistics can be directly interpreted in light of the resolved flow fields shown in~\Cref{fig:fig11}. For the smaller spherical particles, at \(u_\text{f} = u_\text{t}\), the flow exhibits clear local velocity deficits and moderate streamline curvature around individual particles, indicating the presence of short, attached wakes, see~\Cref{fig:fig11}a. In this regime, the particle inertia is moderate \((St \approx 1.26)\), placing the particles in a transitional response regime where they partially track the carrier flow while retaining a measurable lag to the velocity fluctuations. The Stokes number \(St\) is defined as
\revision{
\begin{equation}
   St = \dfrac{\rho_\text{p} d_\text{v}^2 u_\text{f}}{18\mu_\text{f}D}.
\end{equation}}
As a result, drag fluctuations remain weak, and the corresponding force distribution is narrow,~\Cref{fig:fig12}a. As the flow velocity increases to \(u_\text{f} = 3u_\text{t}\), the streamline patterns appear straighter and individual wake structures become less visually distinct,~\Cref{fig:fig11}b. This does not imply the absence of wake dynamics. Instead, the increase in \(Re_\text{p}\) leads to wake structures that are rapidly convected downstream, while axial shear and inter-particle interactions inhibit the formation of spatially coherent recirculation zones~\citep{crowe1998multiphase}. In this regime, wake-induced disturbances persist but are temporally short-lived and not spatially localized, resulting in a modest broadening of the drag-force distribution without significant intermittency.

A qualitatively different behavior is observed for the larger spherical particles \((d_\text{v} = \SI{44}{\milli\meter})\). At \(u_\text{f} = u_\text{t}\), the flow field reveals strong streamline distortion and a pronounced velocity deficit downstream of the particle cluster. The wake remains coherent and visibly asymmetric due to lateral confinement by the pipe wall \((d_\text{v}/D = 0.22)\), but adjusts quasi-steadily to particle motion. Such confinement-induced wake asymmetry and deflection are well documented for bluff bodies in pipes and channels~\citep{zdravkovich1997flow,sahin2004numerical}.  Despite the higher particle inertia \((St \approx 23.7)\), the quasi-steady wake adjustment yields relatively narrow distributions of drag force, see~\Cref{fig:fig12}b. At \(u_\text{f} = 3u_\text{t}\), the instantaneous flow visualizations show predominantly axial streamlines with no clearly identifiable recirculation~\Cref{fig:fig11}d. This apparent suppression of wake structures does not indicate steady drag conditions. At high inertia \((St \approx 71)\) and elevated \(Re_\text{p}\), wake dynamics transition to a regime dominated by rapid downstream convection, confinement-induced suppression of lateral wake expansion, and strong interactions with neighboring particles and the confining wall~\citep{zdravkovich1997flow}. Under these conditions, the particle response time becomes comparable to the characteristic time scales of wake evolution, resulting in history-dependent drag forces that are not uniquely determined by the instantaneous slip velocity \citep{balachandar2010turbulent,leskovec2024turbulent}. Consequently, drag fluctuations become intermittent and heavy-tailed, even though wake structures are not readily apparent in instantaneous streamline plots.

\revision{The persistence of wake unsteadiness at high \(u_\text{f}\) is confirmed quantitatively by the streamwise velocity fluctuation intensity \(I = u^{\prime}_\text{rms}/u_\text{f}\), computed from fluid probes on the pipe centerline at \(z = \SI{0.1}{\meter}\), \(z = \SI{0.6}{\meter}\) and \(z = \SI{1}{\meter}\). For the large-sphere case (\(d/D = 0.22\)), \(I\) grows from \(\approx 11\%\) at mid-pipe to \(\approx 23\%\) near the exit at \(u_\text{f} = u_\text{t}\), indicating that wake disturbances accumulate downstream when the convective time scale is long. At \(u_\text{f} = 3u_\text{t}\), \(I \approx 13\%\) at both locations, indicating a statistically homogeneous fluctuating state in which disturbances advect out of the domain rather than accumulating, while \(u^{\prime}_\text{rms}\) in absolute terms remains substantial. The transition from spatially developing (\(z\)-dependent) to spatially
developed (uniform) fluctuation intensity parallels the residence time
narrowing, and indicates that wake unsteadiness persists at high \(u_\text{f}\) even where instantaneous streamline plots show predominantly axial flow. Several factors contribute to this apparent contrast between the streamline visualizations and the fluctuation statistics, including the short convective time of the wake, confinement-induced suppression of lateral wake expansion, and inter-particle wake interference. The fluctuation intensities reported here arise predominantly from particle-induced disturbances rather than from classical wall-bounded turbulence in the absence of particles (see~\Cref{fig:NS5} in the SM). Probe measurements at \(z\) show that the \(u_z\) relaxes to the imposed value once particles have advected out.}

The drag force distribution for small PMNs in \Cref{fig:fig12}c exhibits behavior quantitatively similar to that of spherical particles, with sharp peaks centered around \(\hat{f} \approx 1.1-1.3\) across all flow velocities. This indicates that, for small particles, geometric irregularity plays a secondary role relative to inertia and confinement. Owing to their low effective inertia and rapid rotational dynamics, small PMNs remain strongly coupled to the carrier flow, such that orientation-dependent variations in projected area are rapidly averaged out and drag fluctuations are primarily governed by local flow variability. In contrast, the drag-force distributions for large PMNs, \Cref{fig:fig12}d, exhibit distinct features associated with orientation-dependent forcing. At \(1.0u_\text{t}\), the distribution is relatively narrow but shifted toward higher values, with a peak at \(\hat{f} \approx 1.5\) and a tail extending to \(\hat{f} \approx 2.0\), reflecting enhanced form drag associated with non-spherical geometry. As the flow velocity increases to \(2.0u_\text{t}\) and \(3.0u_\text{t}\), the distributions broaden and shift toward \(\hat{f} \approx 1.7\), accompanied by extended high-drag tails. At \(3.0u_\text{t}\), the distribution becomes more sharply peaked than at intermediate velocities, yet remains substantially broader than that of spherical particles and retains a long tail extending to \(\hat{f} \approx 4.0\).

\revision{The contribution of the orientation-dependent projected area to these distributions deserves explicit consideration. Although \(\hat{f}\) is normalized by submerged weight and therefore does not contain \(A_\text{proj}\) explicitly, the instantaneous drag force \(F_\text{d}\) is sensitive to the time-varying orientation of non-spherical particles. This orientation-dependent forcing is directly resolved in the present simulations, as the CFD-DEM framework integrates the fluid stress over the instantaneous particle volume and does not invoke any orientation-dependent drag closure or explicit parameterization. The orientation modulation observed in \(P(\hat{f})\) is therefore an emergent feature of the resolved solid-fluid coupling. For the PMN morphologies considered here, \(A_\text{proj}\) varies between bounds set by the volume-equivalent and circumscribed spheres, with \((d_\text{eff}/d_\text{v})^2 \approx 1.5\) representing the ratio of the maximum instantaneous value to the orientation-averaged value. Whether this orientation modulation manifests in \(P(\hat{f})\) depends on the ratio of the rotational time scale to the wake-shedding and transit time scales, which is controlled by the particle Stokes number. For small PMNs (\(St \approx 0.9\) at \(u_\text{t}\)), the projected-area variability is geometrically bounded, and the resulting orientation-induced force fluctuations are small, comparable to carrier flow-driven fluctuations. Consequently, orientation modulation does not produce a distinguishable broadening, which remains quantitatively similar to that of small spheres (\Cref{fig:fig12}c). For large PMNs at higher flow velocity (\(St \approx 51.6\)), the rotational time scales are comparable to wake evolution time scales. The angular velocities reaching \(80~\text{rad}\,\text{s}^{-1}\) (\Cref{fig:fig13}h) sample a wide range of projected areas without complete averaging. The resulting orientation-dependent forcing is transmitted into the instantaneous drag force, broadening \(P(\hat{f})\) and generating the extended tails (see~\Cref{fig:fig12}d). This mechanism operates in addition to, and is distinguishable from, the wake-history effect governing large spheres or PMNs: large spheres exhibit broadening without orientation modulation, whereas large PMNs exhibit both contributions superimposed.}

\revision{Across all velocities, small spheres and small PMNs maintain narrow, peaked distributions, whereas pronounced differences between spherical and non-spherical particles emerge for the larger size class. Large spherical particles exhibit a monotonic broadening of the drag distribution with increasing velocity, consistent with wake unsteadiness modified by confinement. For large PMNs, drag variability is further amplified by orientation-dependent forcing, which introduces an additional source of intermittency beyond wake dynamics alone. At \(1.0u_t\), large PMNs experience irregular translational and rotational motion, including intermittent settling, wall-proximity events, and re-entrainment, leading to strong drag fluctuations. Flow-field visualizations in~\Cref{fig:fig11}g reveal complex three-dimensional wake structures with multiple separation points arising from the non-convex particle morphology. Particle trajectory snapshots (see~\Cref{fig:fig13}) confirm irregular lateral motion and variable angular velocity with intermittent wall-proximity events at low velocities, indicative of short-lived radial excursions. At higher velocities, increased angular velocity promotes more rapid reorientation and sustained transport upward through a more uniform mean flow, suppressing large radial excursions while promoting more persistent near-wall localization.}

\revision{To quantify wall-proximity behavior, we define the instantaneous near-wall fraction \(\phi_\text{w}(t) = \dfrac{1}{N_\text{p}}\sum_i I(\delta_i(t) < 0.1\,d_\text{v})\), where \(\delta_i = R - r_i - d_\text{v}/2\) is the gap between the surface of particle \(i\) and the pipe wall, \(R\) is the pipe radius, and \(r_i\) is the radial distance of the particle center from the pipe axis. For consistency in defining the surface-wall separation, this metric is evaluated for spherical particles, although trajectory visualizations indicate qualitatively similar radial dynamics for non-spherical counterparts. Time-averaged over the stationary transport window, \(\langle \phi_\text{w} \rangle\) increases from \(0.13\) at \(u_\text{f} = 1.0\,u_\text{t}\) to \(0.24\) at \(u_\text{f} = 3.0\,u_\text{t}\), indicating that the near-wall layer is sustained at both velocities but becomes denser at higher flow rates.}

\section{Summary and Conclusions}\label{Sec:Summary}
\noindent This study employed a fully resolved CFD-DEM approach to investigate the hydrodynamic behavior of non-spherical polymetallic nodules (PMNs) during sedimentation and vertical hydraulic transport in confined risers, which are relevant to deep-sea mining operations. Particle-fluid interactions are explicitly resolved without empirical drag closures. The simulations provide mechanistic insight into how particle shape, confinement, and flow velocity collectively govern settling dynamics, residence time statistics, and fluctuations in drag force across low to moderate \(Re_\text{p}\). The key findings are outlined below.
\begin{enumerate}[leftmargin=0.5cm,labelsep = 0.2em]
    \item Shape-induced drag enhancement governs PMN sedimentation, leading to reduced terminal velocities without altering the terminal force balance. Sedimentation simulations demonstrate that PMNs settle 27-29\% slower than volume-equivalent spheres, irrespective of having the same mass and buoyancy. This reduction arises from an increase in the drag coefficient and morphology-induced wake asymmetry. Remarkably, the PMNs and spheres experience identical terminal drag forces equal to their submerged weight. The reduced settling velocity of PMNs reflects that the irregular particles generate the required hydrodynamic resistance at lower slip velocities, rather than an alteration in the equilibrium force balance.
    \item The residence time distributions \((t_\text{r})\) highlight a velocity-driven transition from intermittent to convection-dominated transport, particularly showing that non-spherical particles experience delayed entrainment. At low flow velocities, the spherical particles and PMNs display broad \(t_\text{r}\) indicative of intermittent motion and partial settling, with small PMNs remaining marginally suspended and exhibiting oscillatory trajectories. As flow velocity increases, these distributions narrow and peak, signifying a shift to stable convection-dominated transport. The PMNs demonstrate larger mean residence times and variances compared to spherical particles under similar flow conditions, which is attributable to the greater drag \((\hat{f} > 1)\) and rotational-translational coupling that contribute to the delayed entrainment.
    \item Drag force statistics are significantly influenced by particle morphology and confinement, with non-spherical particles amplifying unsteady force fluctuations. The mean normalized drag force fluctuates around \(\hat{f} = 1-1.5\) across configurations, indicating a steady transport equilibrium between hydrodynamic drag and submerged weight. Distinct transport regimes are characterized by variations in drag fluctuations. Small particles exhibit narrow distributions in drag fluctuations, which aligns with their rapid response to flow changes and low Stokes number dynamics. \revision{In contrast, large particles exhibit monotonic broadening of \(P(\hat{f})\) with increasing \(u_\text{f}\), consistent with wake-history effects. Large PMNs exhibit broadening \(P(\hat{f})\) reflecting the competition between orientation-dependent forcing and increasingly homogeneous wake fluctuations. For small PMNs, rapid reorientation effectively averages out projected-area fluctuations, and the distributions remain comparable to those of equivalent spheres.}
\end{enumerate}

While PMNs exhibit 40-90\% longer residence times and 2 times higher drag coefficients compared to volume-equivalent spheres, the underlying transport physics remains qualitatively similar. Both particle types undergo the same progression from a settling-dominated regime, through a transitional regime, to a convection-dominated regime as flow velocity increases. This suggests that first-order transport behavior can be captured using volume-equivalent spherical particles with appropriately calibrated drag laws, though quantitative predictions require shape-specific corrections. Future work will extend this framework to polydisperse systems with continuous particle injection and quantify pressure drop and transport efficiency under realistic operating conditions. The elevated drag coefficient has direct implications for pressure drop: at fixed input solid concentration for deep sea mining risers, the submerged weight \((\rho_\text{p}-\rho_\text{f})\mathbf{g}\phi_\text{s}\) is shape-independent, but the 27–29\% lower settling velocity raises solid fraction \((\phi_\text{s})\) at a given feed rate and increases the minimum input fluid velocity due to solid-fluid and solid-solid interactions. The results presented in this work apply to the \(Re_\text{p}\) ranges corresponding to dilute ensembles and are intended as calibration input for reduced-order models. Extrapolation of the results to dense slurries characterized by hindered settling, plug flow, and clustering is beyond the scope of the present study. Furthermore, the direct experimental validation of drag coefficients for CT-reconstructed PMN geometries, through controlled settling experiments with simultaneous orientation tracking, would provide complementary verification of the resolved simulations and enable assessment of the applicability of existing non-spherical drag correlations to highly irregular natural particle morphologies.

\section*{Acknowledgements}
\nolinenumbers
\noindent RKA acknowledges the financial support from the National Institute of Ocean Technology, Ministry of Earth Sciences, Government of India.

\section*{Conflict of Interest}
\noindent The authors declare that they have no conflict of interest.

\bibliographystyle{elsarticle-num-names}
\biboptions{comma,round} 
\bibliography{Ref}

\newpage
\setcounter{section}{0}
\setcounter{page}{1}
\setcounter{figure}{0}
\setcounter{equation}{0}
\renewcommand{\thesection}{S\arabic{section}}
\renewcommand{\thepage}{S\arabic{page}}
\renewcommand{\thetable}{S\arabic{table}}
\renewcommand{\thefigure}{S\arabic{figure}}
\renewcommand{\theequation}{S\arabic{equation}}

\setcounter{affn}{0}
\resetTitleCounters

\makeatletter
\let\@title\@empty
\makeatother

\title{\textbf{\Large Supplementary Material}\\ \vspace{2em} \large Hydrodynamic Behavior of Non-spherical Particles in Confined Vertical Flows: A Resolved CFD-DEM Study}

\makeatletter
\renewenvironment{abstract}{\global\setbox\absbox=\vbox\bgroup
  \hsize=\textwidth\def\baselinestretch{1}%
  \noindent\unskip\textbf{Contents}
 \par\medskip\noindent\unskip}
 {\egroup}


\startlist{toc}
\begin{abstract}
\vspace{-48pt}
\printlist{toc}{}{\section*{}}
\end{abstract}
\maketitle
\section*{}
\parindent0pt

\setcounter{table}{0}
\newpage
\section{Movies}
\noindent Detailed movie captions are provided below.
\begin{enumerate}[label=M\arabic*:~]
    \item A sphere settling in the viscous fluid: transition from a Stokes regime \((Re_\text{p} = 1.4)\) to an intermediate regime \((Re_\text{p} = 29.8)\), where inertial effects become significant. \\
    \href{https://youtu.be/8Z_YzIGVHtc}{Click on the link for the Movie}
    
    \item Particle positions and fluid velocity magnitude contours during particle sedimentation, illustrating the drafting-kissing-tumbling phenomenon.\\
    \href{https://youtu.be/YoOdlxFwlpE}{Click on the link for the Movie}
    
    \item Particle configurations and stream lines during vertical transport of polymetallic nodules entrained in a carrier fluid where \(u_\text{f} = \SI{0.8}{\meter\per\second}\) (i.e., 1.0 times terminal velocity \(u_\text{t}\)) and \(\omega\) is the angular velocity.   
    \item Particle configurations and stream lines during vertical transport of polymetallic nodules entrained in a carrier fluid where \(u_\text{f} = \SI{2.4}{\meter\per\second}\) (i.e., 3.0 times terminal velocity \(u_\text{t}\)) and \(\omega\) is the angular velocity.
\end{enumerate}

\newpage
\section{Single sphere settling}
\begin{figure}[H]
    \centering
    \subfloat[]{\includegraphics[width = 0.25\textwidth]{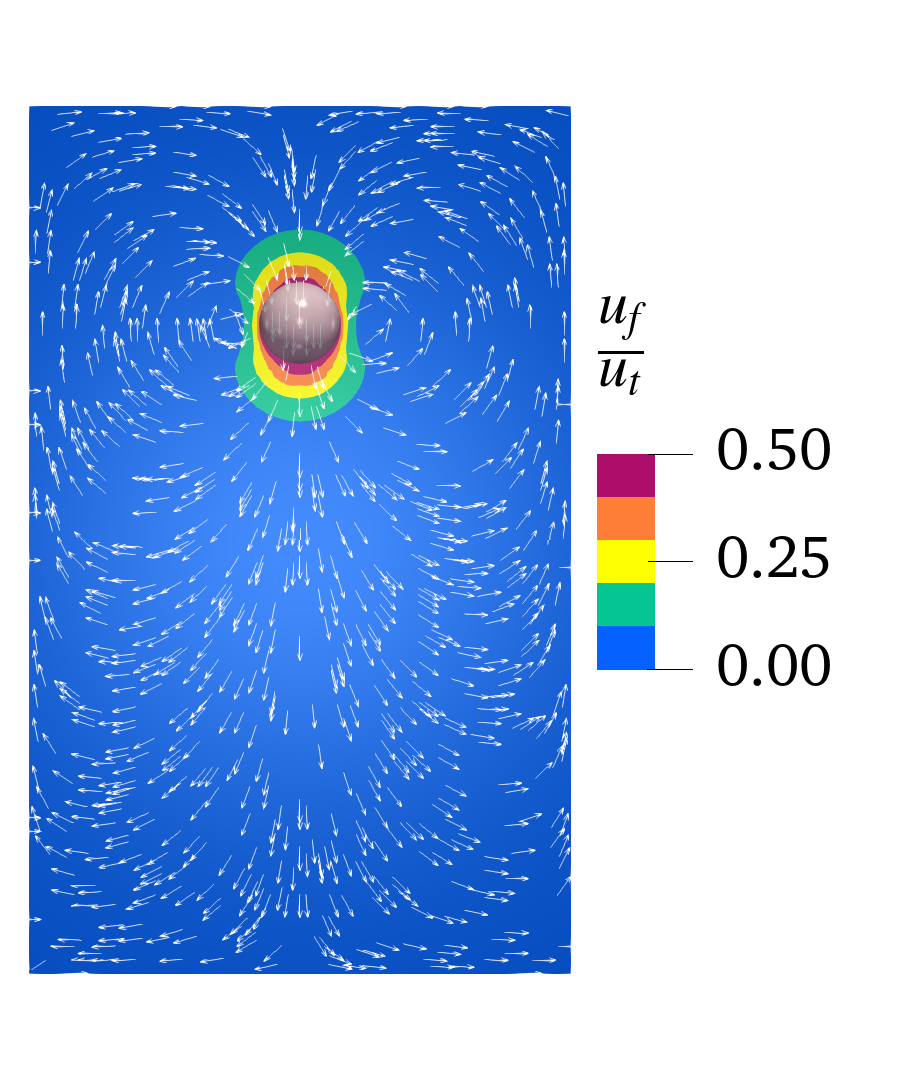}}
    \subfloat[]{\includegraphics[width = 0.25\textwidth]{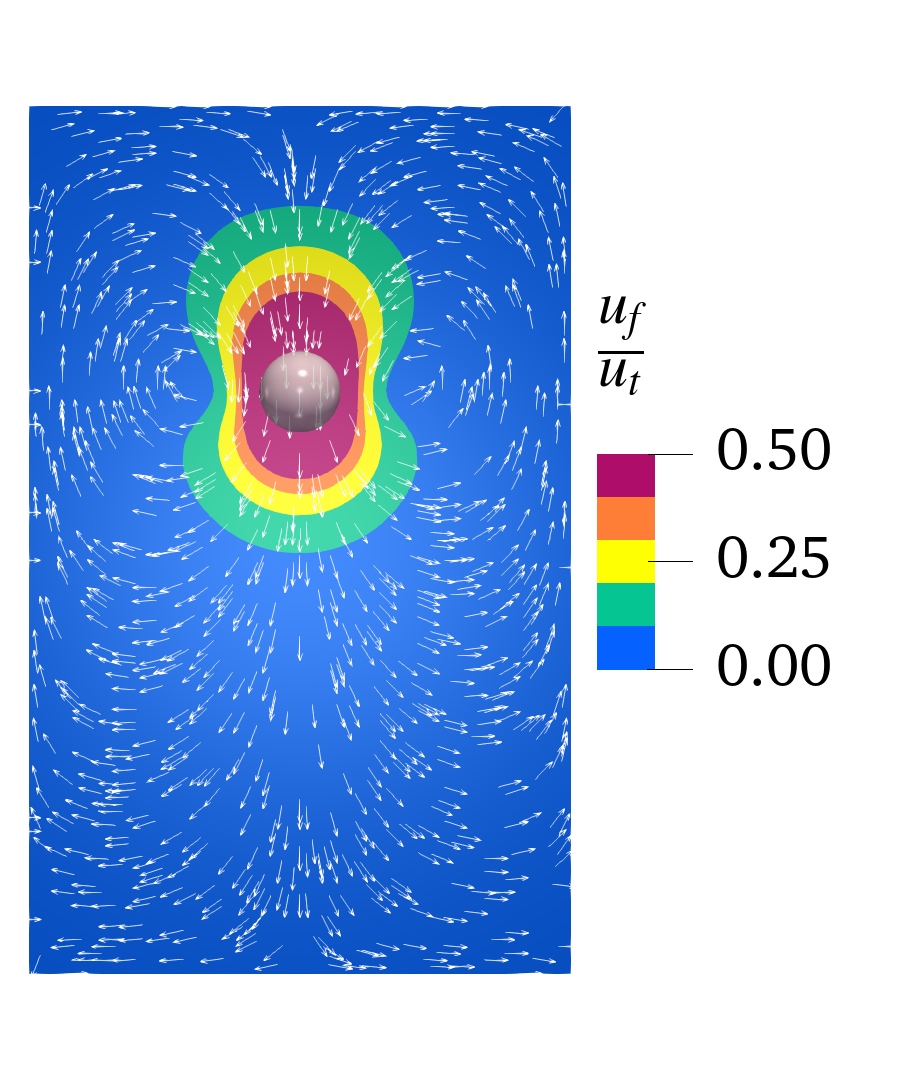}}
    \subfloat[]{\includegraphics[width = 0.25\textwidth]{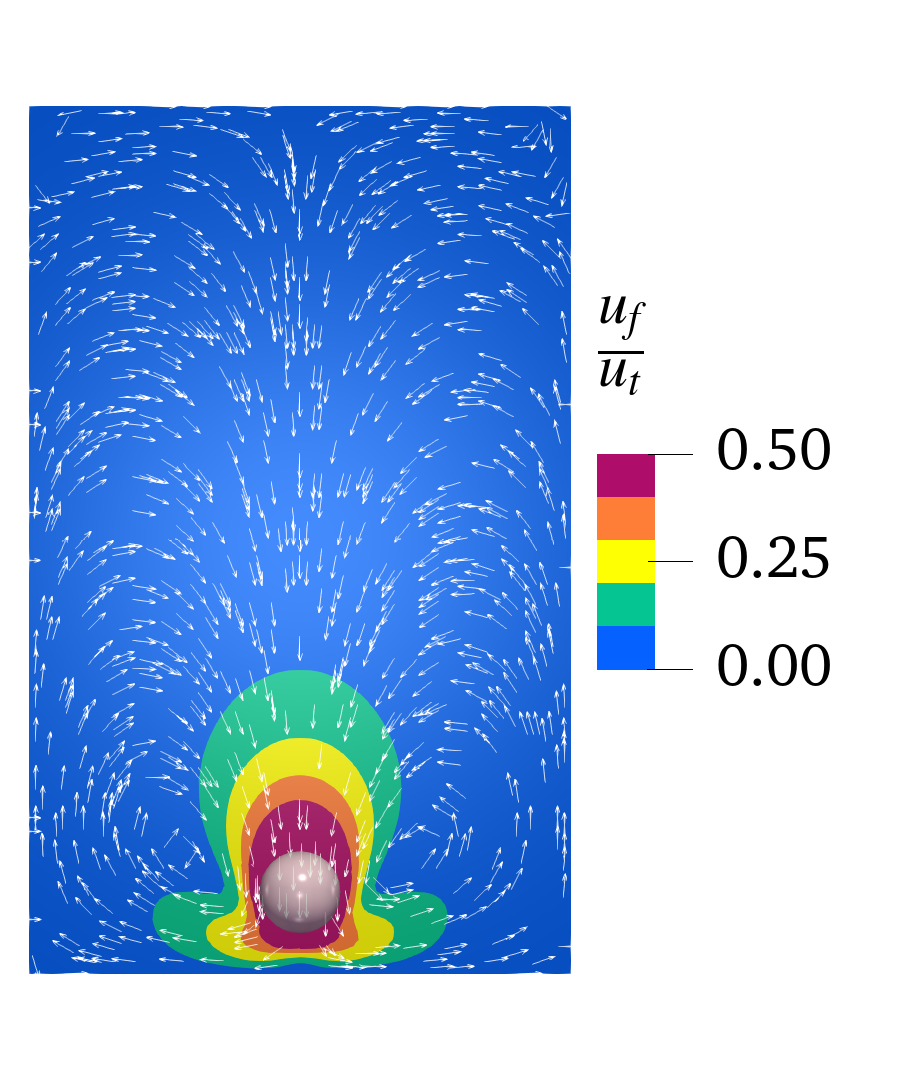}}
    \caption{Velocity contours and flow fields for \(Re_\text{p} = 1.4\) (case 1) showing symmetric flow patterns characteristic of the viscous regime. (a) Initial acceleration phase with developing boundary layer. (b) Approach to terminal velocity exhibiting symmetric streamlines that are consistent with Stokes flow past a sphere. (c) Particle at \(0.5d\) from the domain bottom showing a fully developed symmetric flow field with smooth velocity decay to far-field values.}
    \label{fig:figS3}
\end{figure}

\begin{figure}[H]
    \centering
    \subfloat[]{\includegraphics[width = 0.25\textwidth]{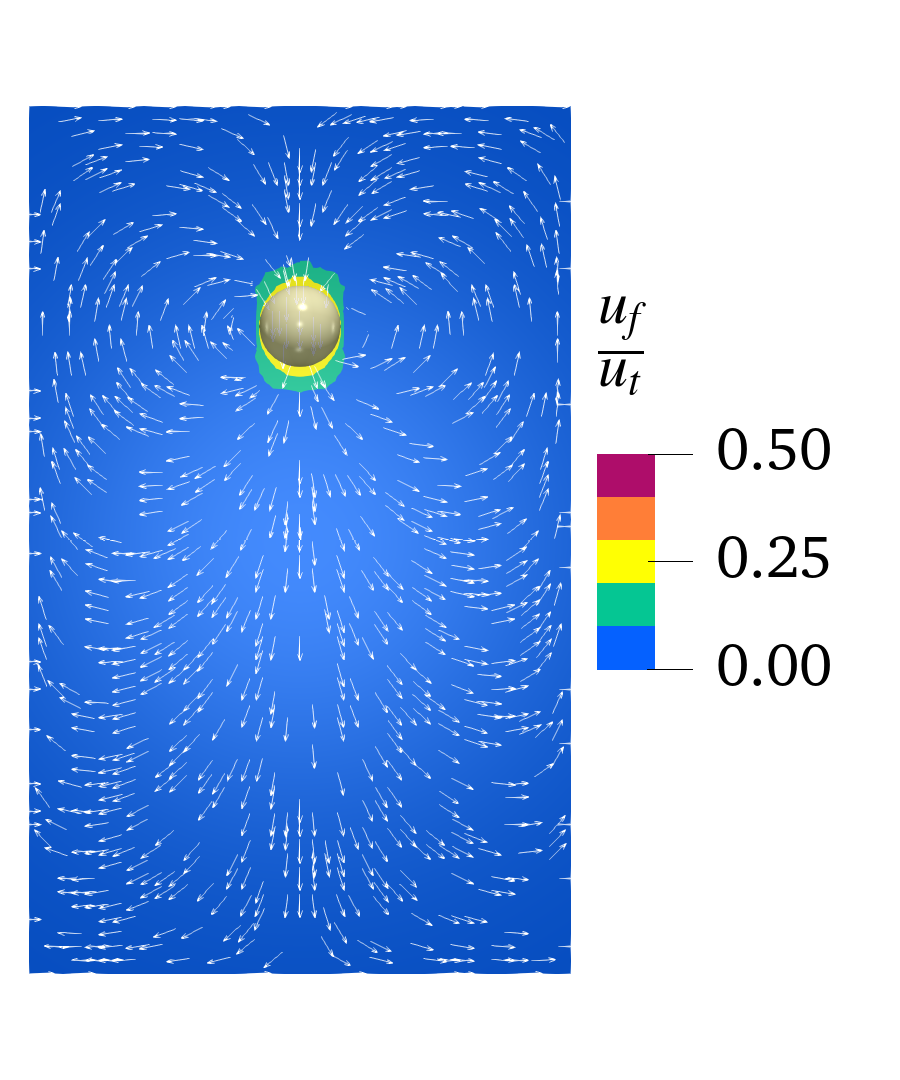}}
    \subfloat[]{\includegraphics[width = 0.25\textwidth]{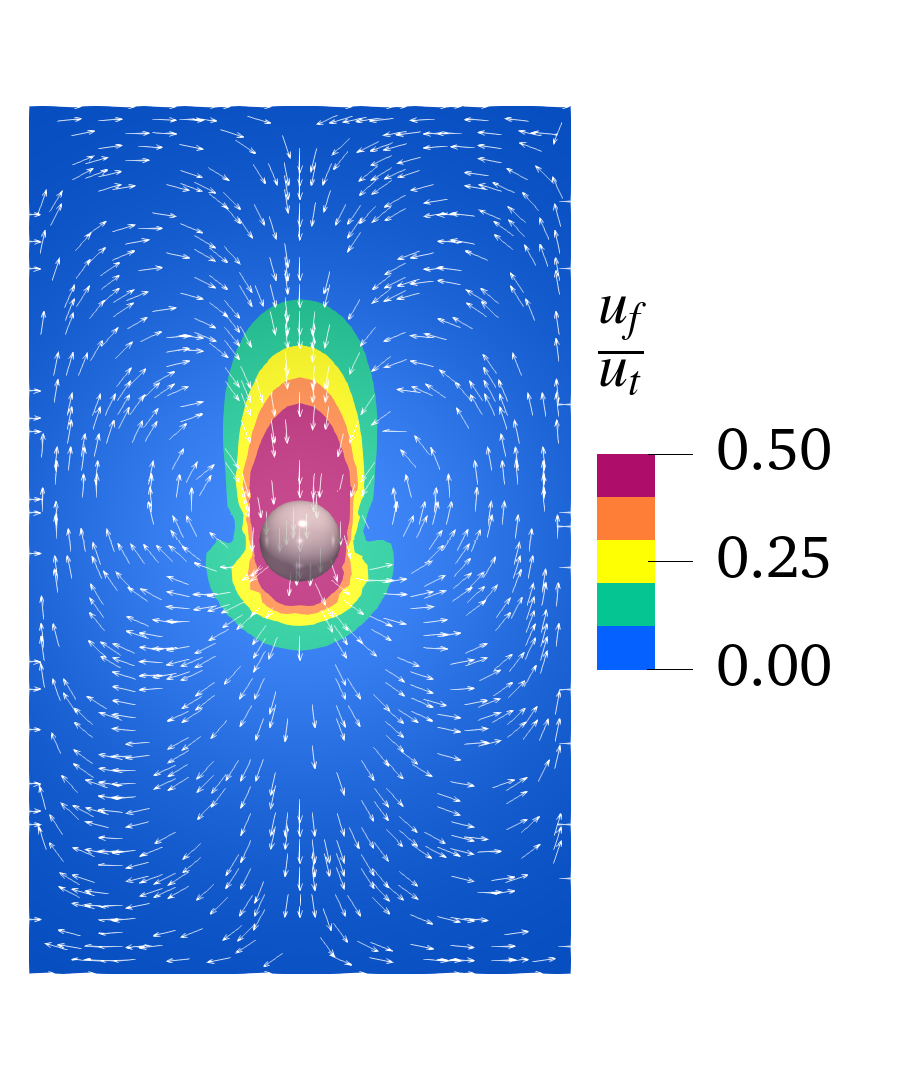}}
    \subfloat[]{\includegraphics[width = 0.25\textwidth]{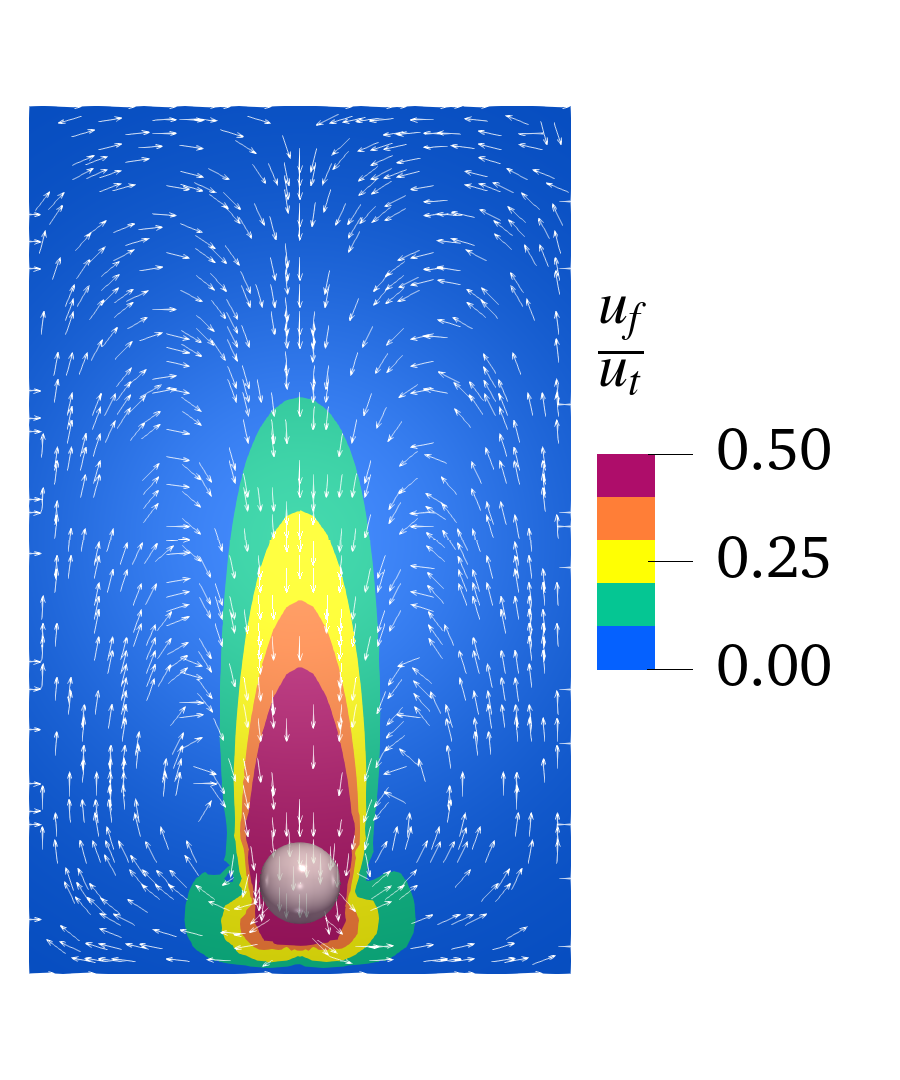}}
    \caption{Velocity contours and flow fields for \(Re_\text{p} = 29.8\) (case 2) showing wake formation and flow separation characteristic of the inertial regime. (a) Initial acceleration phase. (b) Approach to terminal velocity with clear flow separation and wake region extending approximately \(1.0d\) downstream, indicating transition from viscous-dominated to inertia-dominated settling. (c) Particle at \(0.5d\) from the domain bottom, showing an established recirculating wake structure and an asymmetric pressure distribution that contributes to enhanced form drag.}
    \label{fig:figS2}
\end{figure}

\Cref{fig:figS3,fig:figS2} illustrate the flow fields (vector glyphs) and normalized velocity contours\((u_f/u_t)\) for \(Re_\text{p} =\) 1.4 and 29.8, respectively. Where \(u_f\) and \(u_t\) are the instantaneous fluid velocity and terminal velocity, respectively.

\begin{figure}[H]
    \centering
    \subfloat[]{\includegraphics[width = 0.25\textwidth]{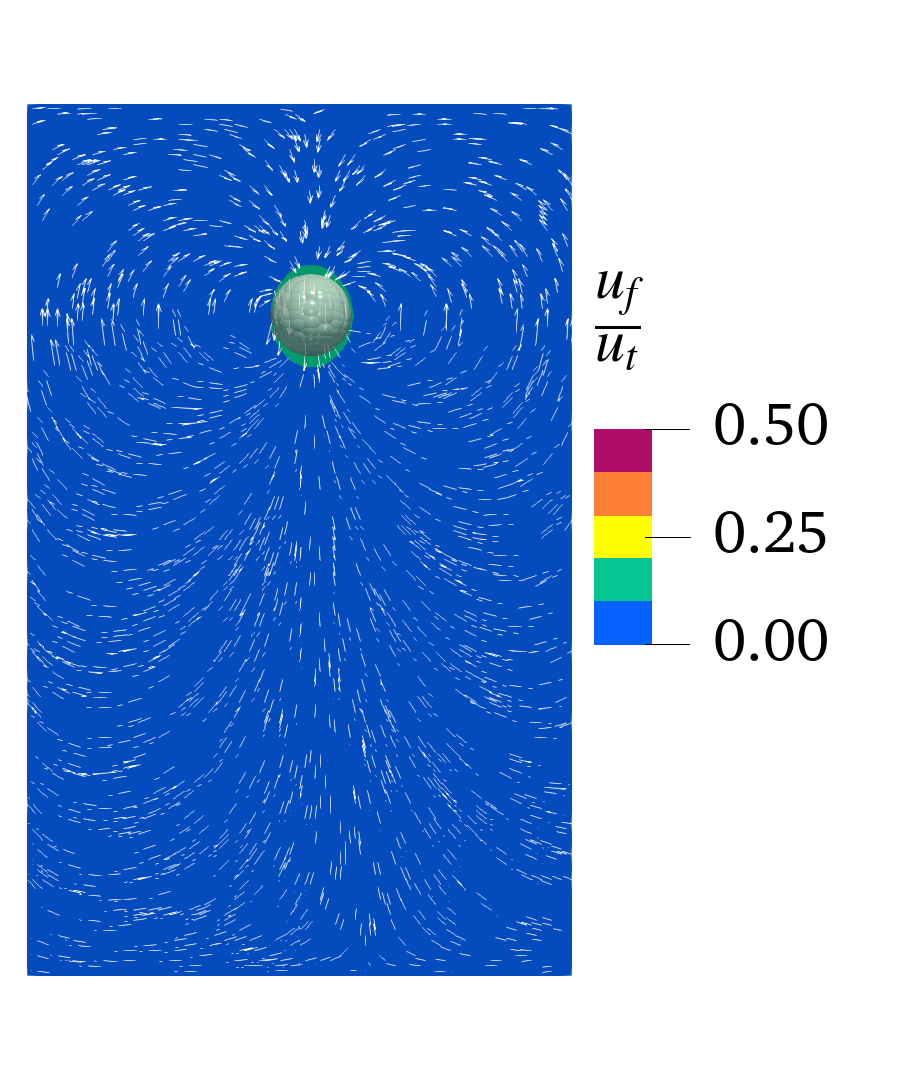}}
    \subfloat[]{\includegraphics[width = 0.25\textwidth]{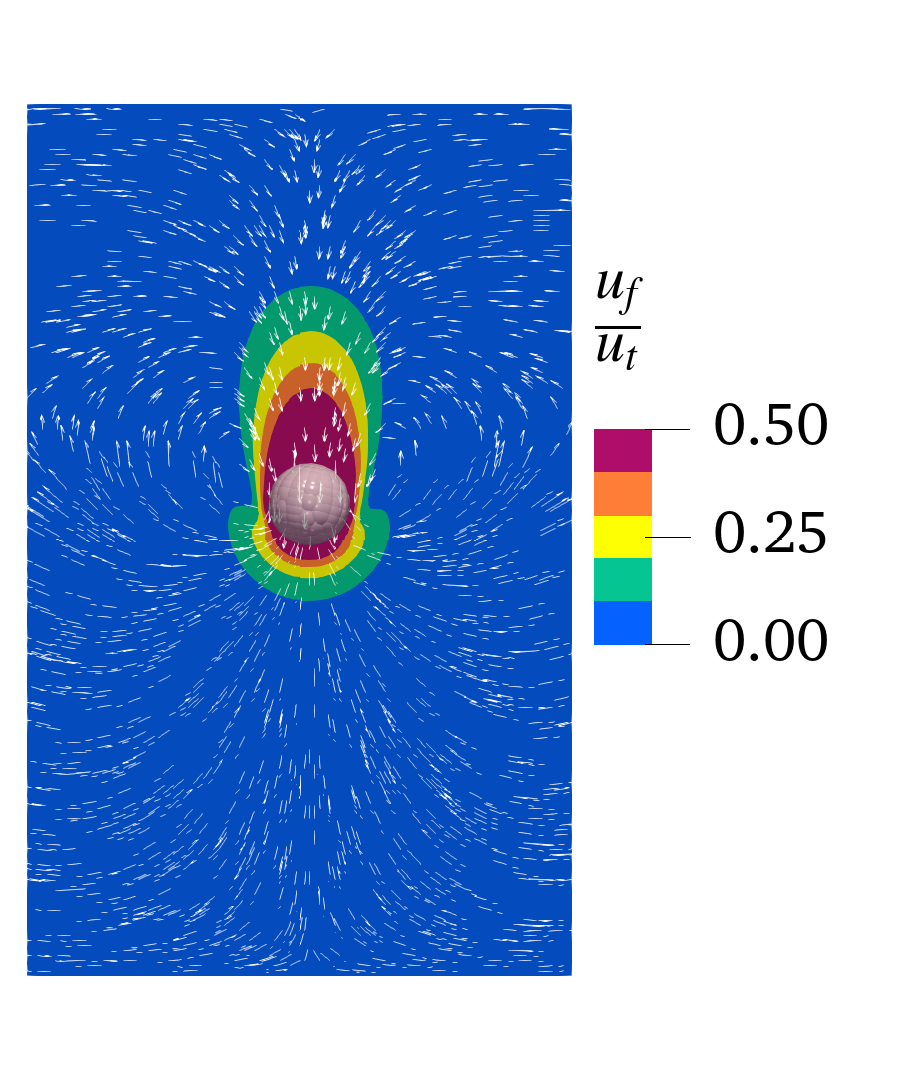}}
    \subfloat[]{\includegraphics[width = 0.25\textwidth]{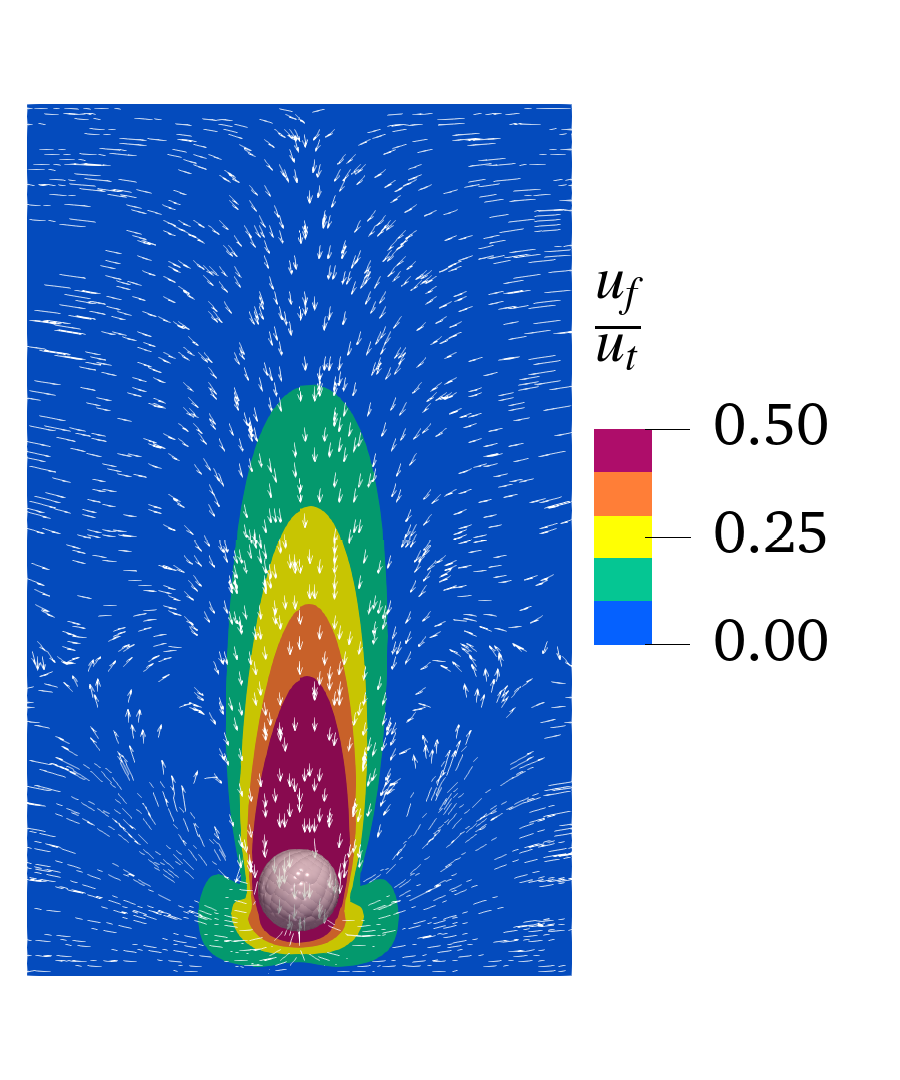}}
    \caption{Flow field around multisphere particle during sedimentation, showing wake asymmetries due to irregular geometry. (a) initial phase. (b) Steady settling at terminal velocity exhibiting local velocity gradient intensification near geometric irregularities. (c)  Near-wall interaction showing subtle flow perturbations induced by particle morphology.}
    \label{fig:figS4}
\end{figure}

\Cref{fig:figS4} illustrates the flow fields (vector glyphs) and normalized velocity contours\((u_f/u_t)\) for an idealized sphere approximated as a multisphere. The test particle comprises 206 overlapping sub-spheres arranged to approximate a spherical geometry while maintaining identical density (\(\rho_\text{s} = \SI{1120}{\kilo\gram\per\cubic\meter}\)) and volume, as the reference sphere from Case 2 (\(d = \SI{15}{\milli\meter}\)).

\newpage
\section{Drafting, Kissing, Tumbling}
\begin{figure}[H]
    \centering
    \subfloat[]{\includegraphics[width = 0.315\textwidth]{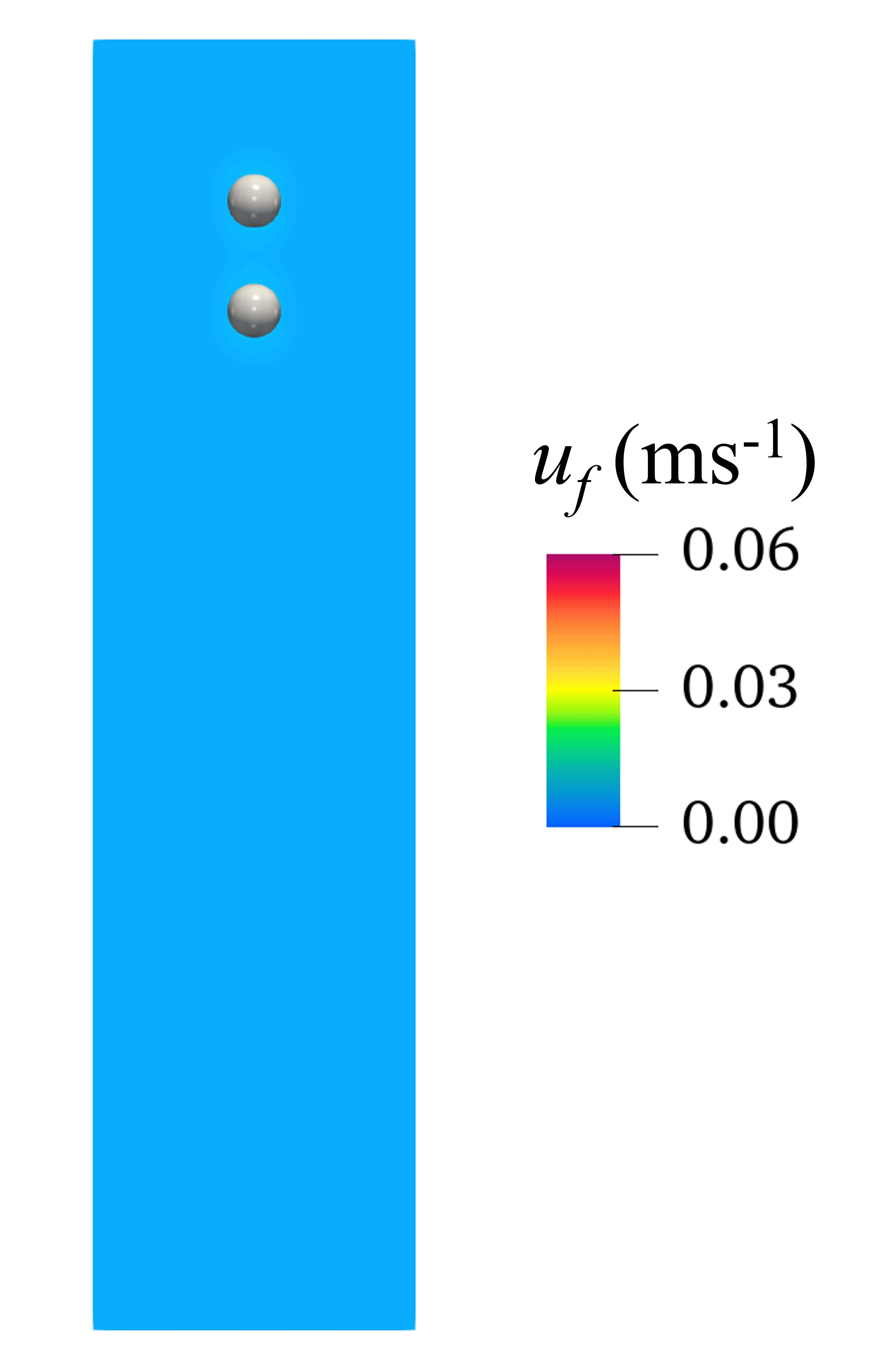}}
    \subfloat[]{\includegraphics[width = 0.185\textwidth]{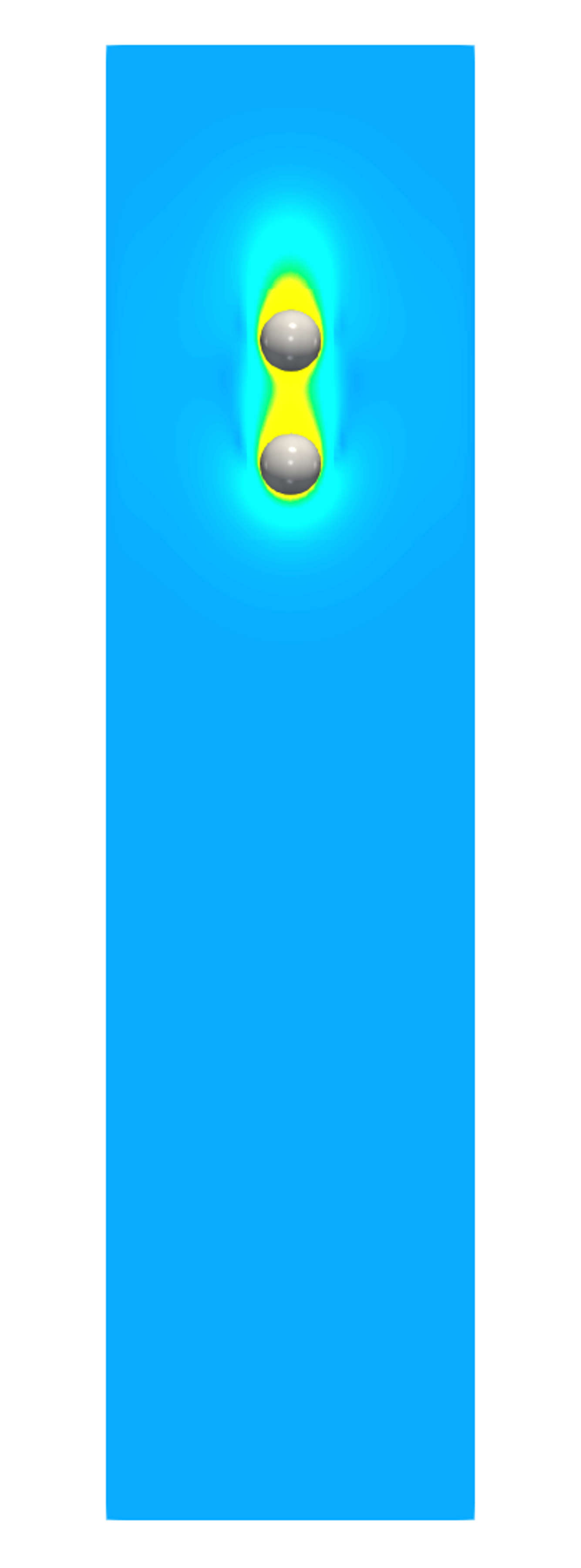}}
    \subfloat[]{\includegraphics[width = 0.185\textwidth]{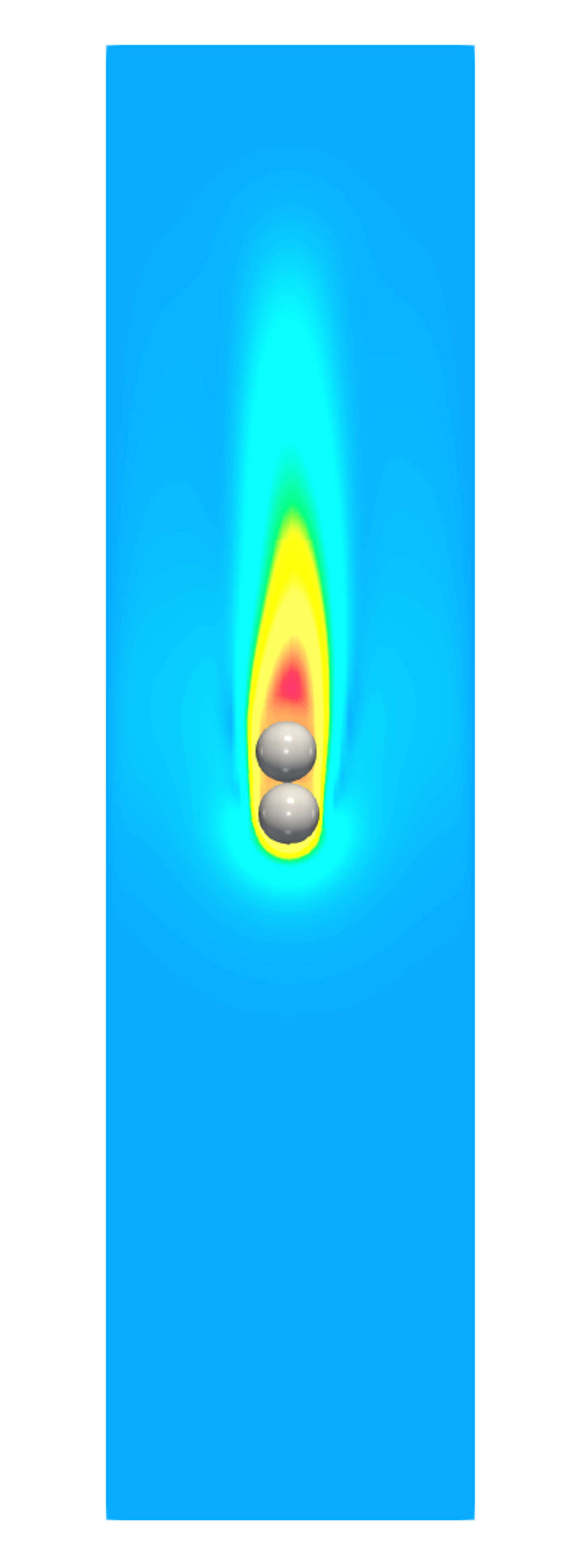}}
    \subfloat[]{\includegraphics[width = 0.185\textwidth]{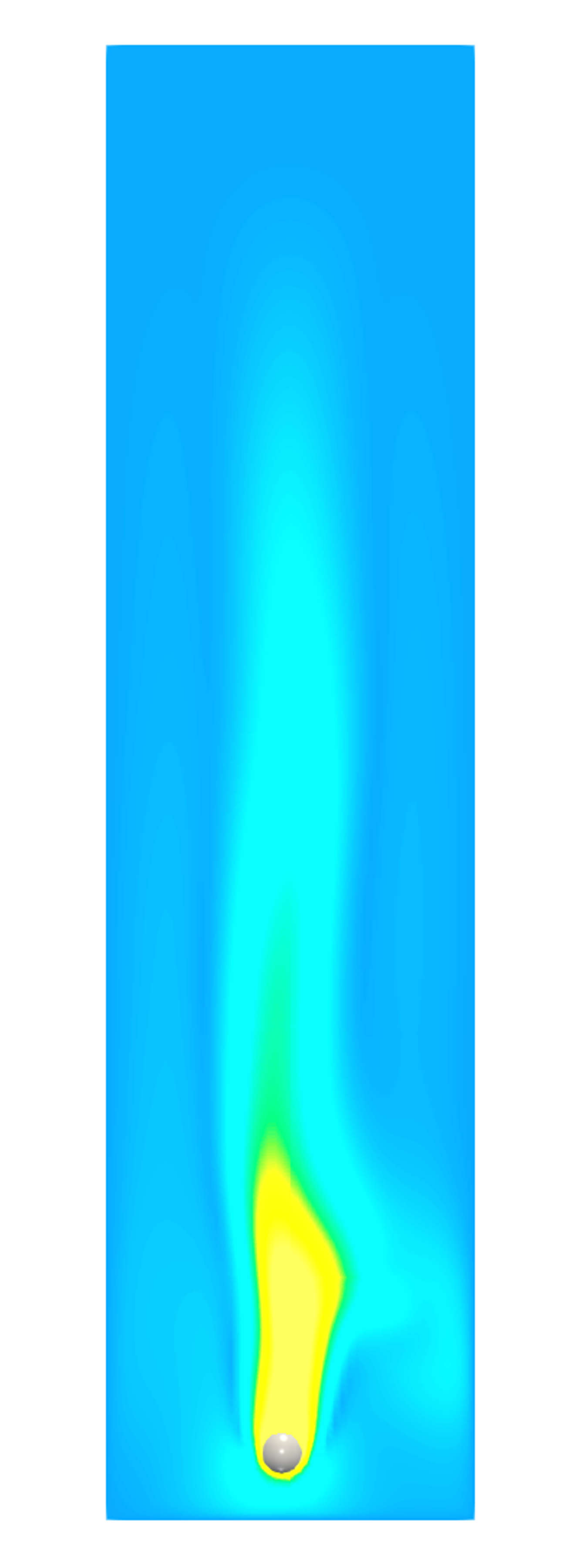}}
    \caption{Particle positions and fluid velocity magnitude contours at representative times during particle sedimentation, illustrating the drafting-kissing-tumbling phenomenon. (a) \SI{0}{\second}: initial configuration with particles vertically aligned. (b) \SIrange{0.14}{0.35}{\second}: onset of drafting phase as trailing particle enters leading particle's wake. (c) \SI{0.35}{\second}: kissing phase (particle collision). (d) \SIrange{0.35}{0.7}{\second}: post-collision tumbling with flow reorganization and momentum redistribution.}
    \label{fig:NS3}
\end{figure}

\Cref{fig:NS3} compares the flow fields for settling of the two spherical particles, revealing three characteristic phases of particle-particle interactions, viz., Drafting, Kissing, and Tumbling.

\newpage
\section{Statistical convergence}
\begin{figure}[H]
    \centering
    \includegraphics[width=0.5\linewidth]{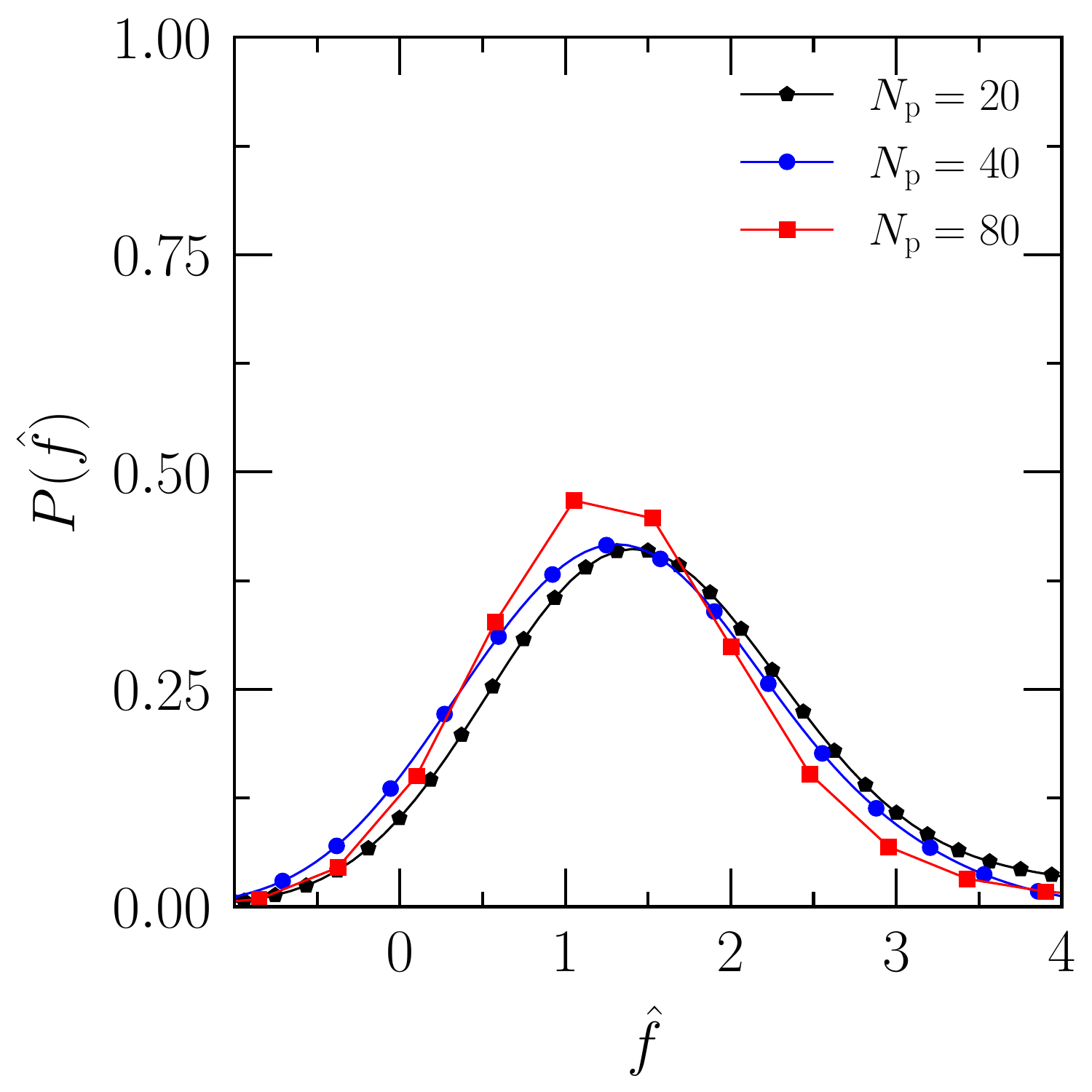}
    \caption{\revision{Statistical convergence for different ensemble sizes \(N_\text{p}\).}}
    \label{fig:NS4}
\end{figure}

\revision{We have performed convergence tests comparing drag-force distributions \(P(\hat{f})\) for ensemble sizes \(N_\text{p} = 20\), \(40\), and \(80\) for the large sphere case at \(u_\text{f} = 3.0u_\text{t}\), which has a higher \(Re_\text{p}\).~\Cref{fig:NS4} show that the \(N_\text{p} = 20\), \(N_\text{p} = 40\), and \(N_\text{p} = 80\) distributions agree closely in both peak location (\(\hat{f} \approx 1.2\)-\(1.4\)) and tail behavior (see~\Cref{fig:NS4}). This confirms that \(N_\text{p} = 40\) is sufficient for the first- and second-order statistics reported in the manuscript. We have added a sentence summarizing this convergence result.}

\section{Fluctuation Intensity}
\begin{figure}[H]
        \centering
        \subfloat[\(u_\text{f} = 1.0u_\text{t}\)]{\includegraphics[width=0.4\linewidth]{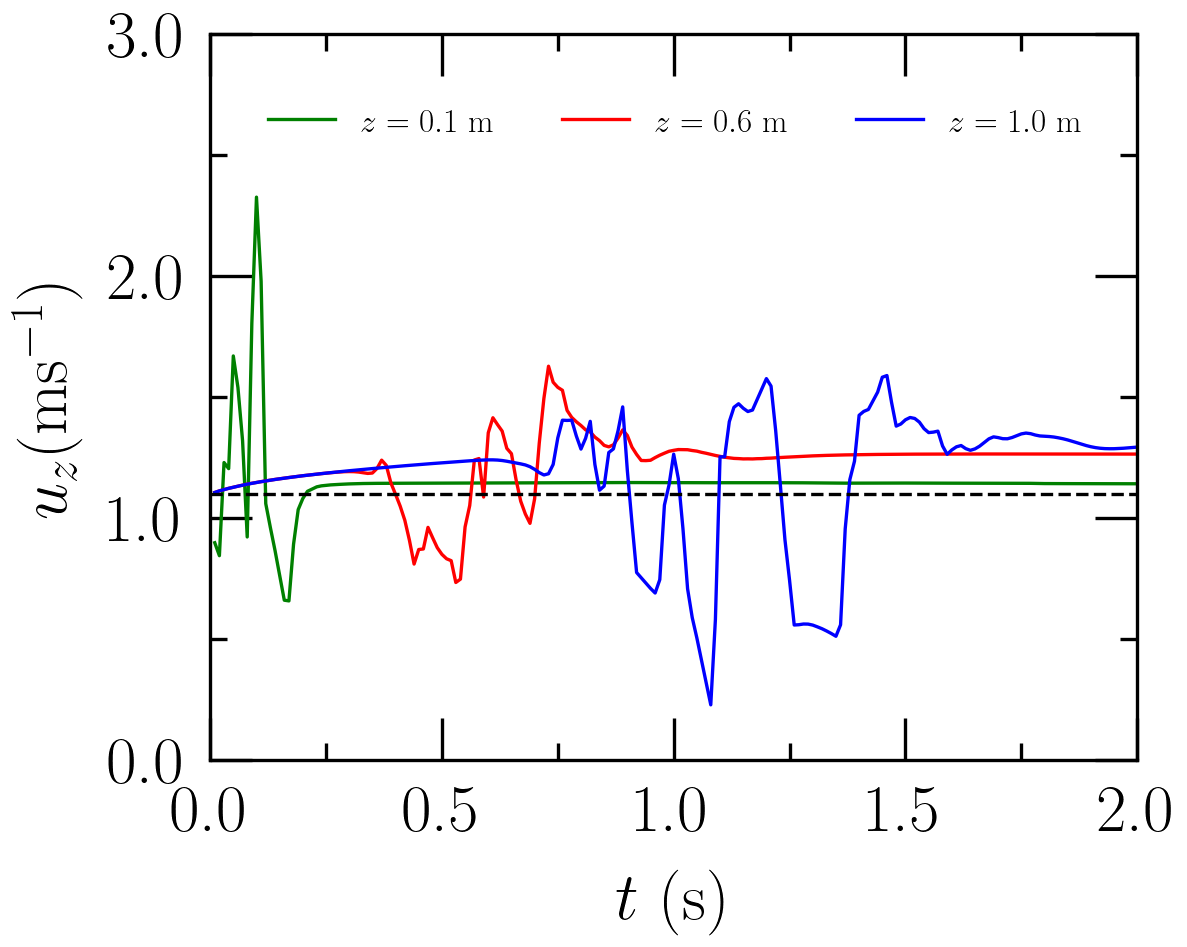}}
        \subfloat[\(u_\text{f} = 3.0u_\text{t}\)]{\includegraphics[width=0.4\linewidth]{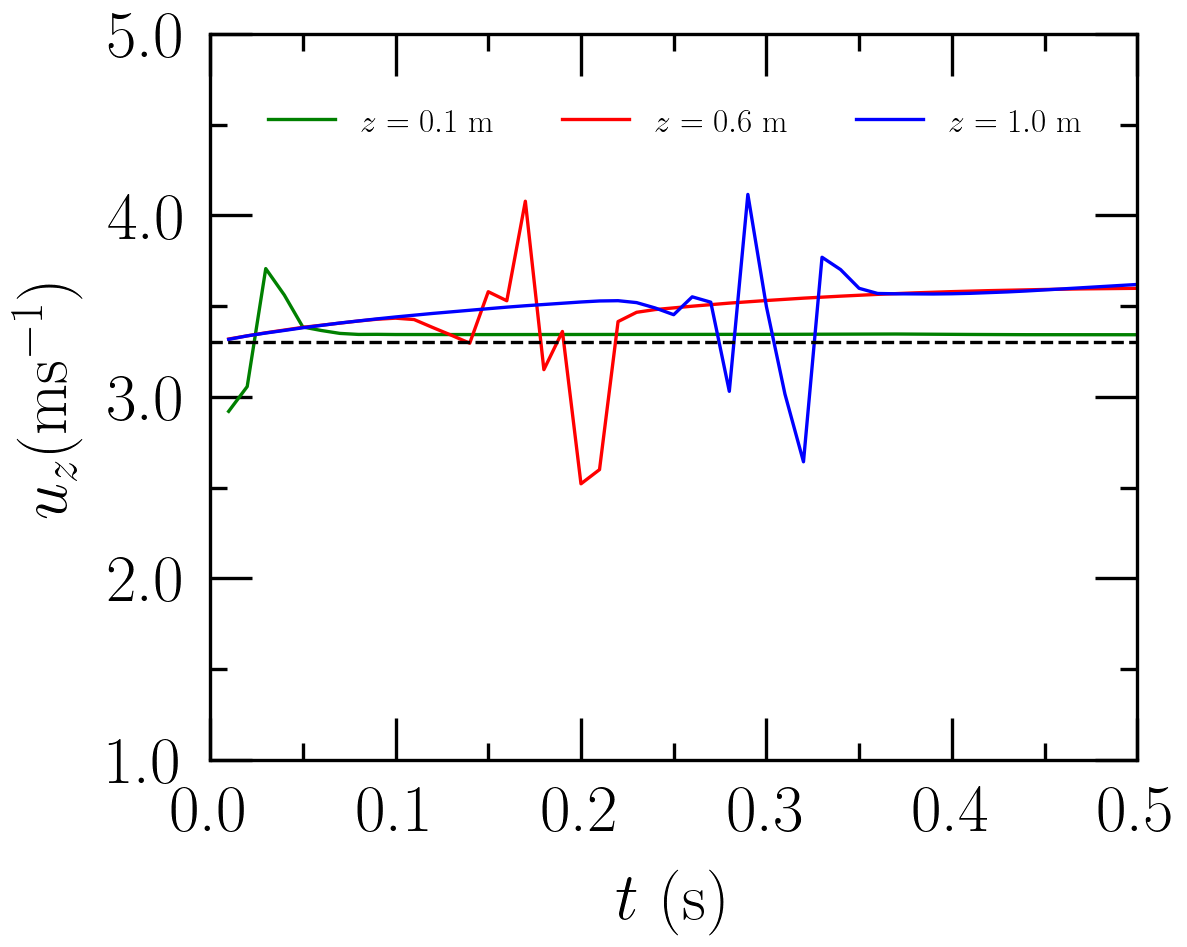}}
        \caption{\revision{Streamwise velocity fluctuation intensity \(I=u^{\prime}_\text{rms}/u_\text{f}\) measured at three axial locations for representative large-particle transport cases: (a) \(u_\text{f} = 1.0u_\text{t}\); (b) \(u_\text{f} = 3.0u_\text{t}\). At low velocity, fluctuation intensity increases downstream due to wake accumulation, whereas at high velocity the fluctuations become more spatially uniform, indicating a statistically homogeneous fluctuating state despite the absence of persistent coherent wake structures in instantaneous streamline visualizations.}}
        \label{fig:NS5}
    \end{figure}

\revision{we have computed the streamwise velocity fluctuation intensity \(I = u^{\prime}_\text{rms}/u_\text{f}\) at three axial locations for representative large-particle cases.~\Cref{fig:NS5} shows that \(I\) grows from \(\approx 11\%\) to \(\approx 23\%\) along the pipe at \(u_\text{f} = u_\text{t}\) (downstream accumulation of wake disturbances) and \(I \approx 13\%\) uniformly along the pipe at \(u_\text{f} = 3u_\text{t}\) (statistically homogeneous fluctuating state). This confirms that wake unsteadiness is active at high \(u_\text{f}\) even when not visible in streamline plots, and is consistent with the heavy-tailed drag-force distributions in Fig. 12b \& d.}

\end{document}